\documentclass[ALICE,manyauthors]{cernphprep}
\usepackage[utf8]{inputenc}
\usepackage[comma,square,numbers,sort&compress]{natbib}
\usepackage{hyperref}
\usepackage{lineno}
\usepackage{mathrsfs}
\usepackage{color}
\usepackage{multirow}
\usepackage{graphicx}

\newcommand{\pT}{\ensuremath{p_{\mbox{\tiny T}}}\xspace}
\newcommand{\mT}{\ensuremath{m_{\mbox{\tiny T}}}\xspace}
\newcommand{\GeVc}{\ensuremath{\mbox{GeV}/c}\xspace}
\newcommand{\MeVc}{\ensuremath{\mbox{MeV}/c}\xspace}
\newcommand{\dedx}{d$E$/d$x$\xspace}

\begin{document}%
\begin{titlepage}
\PHyear{2017}
\PHnumber{216}      
\PHdate{21 Aug}  
%

\title{$\pi^{0}$ and $\eta$ meson production in proton-proton collisions at $\sqrt{s}=8$ TeV}
\ShortTitle{$\pi^{0}$ and $\eta$ in pp collisions at $\sqrt{s}=8$ TeV}   

\Collaboration{ALICE Collaboration\thanks{See Appendix~\ref{app:collab} for the list of collaboration members}}
\ShortAuthor{ALICE Collaboration} 

\begin{abstract}
  An invariant differential cross section measurement of inclusive $\pi^{0}$ and $\eta$ meson production at mid-rapidity in pp collisions at $\sqrt{s}=8$~TeV was carried out by the ALICE experiment at the LHC. 
  The spectra of $\pi^{0}$ and $\eta$ mesons were measured in transverse momentum ranges of $0.3<\pT<35$~\GeVc and $0.5<\pT<35$~\GeVc, respectively.
  Next-to-leading order perturbative QCD calculations using fragmentation functions DSS14 for the $\pi^{0}$ and AESSS for the $\eta$ overestimate the cross sections of both neutral mesons, although such calculations agree with the measured $\eta/\pi^0$ ratio within uncertainties.
  The results were also compared with PYTHIA~8.2 predictions for which the Monash~2013 tune yields the best agreement with the measured neutral meson spectra.
  The measurements confirm a universal behavior of the $\eta/\pi^0$ ratio seen for NA27, PHENIX and ALICE data for pp collisions from $\sqrt{s}=27.5$~GeV to $\sqrt{s}=8$~TeV within experimental uncertainties.
  A relation between the $\pi^{0}$ and $\eta$ production cross sections for pp collisions at $\sqrt{s}=8$~TeV is given by \mT scaling for $\pT>3.5$~\GeVc.
  However, a deviation from this empirical scaling rule is observed for transverse momenta below $\pT<3.5$~\GeVc in the $\eta/\pi^0$ ratio with a significance of $6.2\sigma$.
\end{abstract}
\end{titlepage}
\setcounter{page}{2}
\section{Introduction}
\label{sec:introduction}

Measuring identified particle production in proton-proton (pp) collisions over wide kinematic ranges is considered an informative probe of strong interactions at high energies.
Quantum Chromodynamics (QCD) is the fundamental theory of the strong interaction \cite{Gross:1973ju}.
It succeeds in providing a qualitative description of a wide range of phenomena in hadronic collisions.  
At typical hadron collider energies its perturbative expansion (pQCD) permits a detailed quantitative comparison with experimental data.
However, it remains a challenge to provide a consistent description of hadron spectra at all collision energies reached experimentally.
In theoretical models, particle production is usually divided into two categories: the ``soft'' scattering regime describing particle production involving small momentum transfers and the ``hard'' scattering regime, responsible for producing particles with momenta of several \GeVc or more.


Only ``hard'' scattering processes with a sufficiently large transverse momentum transfer, $Q^{2}$, can be calculated using methods based on pQCD.
High-momentum particles originate from the fragmentation of partons produced in scattering processes with large $Q^{2}$.
The theoretical description of a ``hard'' scattering process can be factorized into parton distribution functions (PDFs), the QCD matrix element and fragmentation functions (FFs).
PDFs describe the fraction of the proton's longitudinal momentum carried by a scattered parton, $x$, and FFs describe the ratio of the observed hadron momentum to the final-state parton momentum, $z$, respectively.
Comprehensive parametrizations of PDFs and FFs are derived from global fits to the experimental data at various collision energies.
The energies reached at the LHC \cite{Evans:2008zzb} open up the domains in $x$ and $z$ not accessible at lower energy.
In the past, experiments at the LHC consequently found discrepancies between the measured $\pi^{0}$ and $\eta$ meson spectra \cite{Abelev:2012cn,Abelev:2014ypa,dEnterria:2013sgr} and pQCD calculations based on fragmentation functions, which include mostly data from experiments below the TeV scale \cite{deFlorian:2007aj}. 
Since the gluon contribution becomes more dominant with increased center of mass energy, $\sqrt{s}$, \cite{deFlorian:2007ekg}, $\pi^{0}$ and $\eta$ meson spectra at LHC energies provide new constraints on the gluon to light-flavor hadron fragmentation functions.
Recent progress in comprehensive global QCD analysis of parton-to-pion fragmentation functions at next-to-leading order (NLO) \cite{deFlorian:2014xna} derived from inclusive pion production in semi-inclusive electron-positron annihilation, deep-inelastic scattering and pp collisions over a wide energy range, including the LHC results \cite{Abelev:2012cn}, achieves a good and consistent description of pion spectra, including the latest measurements of $\pi^0$ and $\eta$ spectra in pp collisions at $\sqrt{s}=2.76$~TeV \cite{Acharya:2017hyu} and 7~TeV \cite{Abelev:2012cn}.
One of the conclusions of that analysis was that meson production from gluon fragmentation is reduced, which turns out to be at tension with previously available data obtained at RHIC \cite{dEnterria:2014tdl}.
In the quark model, the $\pi^{0}$ consists of light-flavor quark-antiquark pairs, $u\bar{u}$ and $d\bar{d}$, whereas the $\eta$ additionally contains hidden strangeness, $s\bar{s}$. 
Measurements of both neutral mesons are thus of particular interest due to their different quark content as they help to constrain the PDFs and FFs \cite{Aidala:2010bn} of the $s$ quark.

The majority of particles at low transverse momenta, \pT, are produced in ``soft'' processes involving a small $Q^{2}$.
In this regime, the pQCD calculations are not applicable for description of the production mechanisms and phenomenological models are based on previous measurements of neutral meson production cross sections or other light mesons by other experiments at lower collision energies.
Particle production measurements at transverse momenta down to a few hundred \MeVc, as reported here, are particularly important to further constrain such models.

The importance of precise identified particle production measurements is underlined by various empirical rules observed in relative particle yields which allow estimates of the hadronic background of rare probes such as direct photons, dileptons and heavy-quark production.
Almost all lower-energy experiments from ISR to RHIC report the observation of such an empricial rule, so-called \mT scaling, in particle production over wide \pT ranges \cite{BOURQUIN1976334, Khandai:2011cf}. 
The practical use of \mT scaling is the ability to derive the \pT-dependent differential yields of most of particles from the well measured light-flavor mesons, like pions and kaons, by assuming that the meson spectra can be described as a function of transverse mass \mT: $E{\rm d^{3}}\sigma/{\rm d}p^{3} = C^{h} f(\mT)$, where the function $f(\mT)$ is universal for all hadron species, so that their spectra share the same shape up to a normalization factor $C^{h}$ \cite{Altenkamper:2017qot}.
In the context of rare probes, this empirical relation is hence widely used to estimate the various background sources, for which no measurements are available.
However, phenomenological analyses of new data delivered by the LHC experiments show that \mT scaling is violated at higher \pT compared to lower collision energies \cite{Jiang:2013gxa,Altenkamper:2017qot}.
Therefore, precise measurements of identified hadron spectra over wide transverse momentum ranges at different LHC energies are of particular importance for the quantitative description of particle production at the LHC.

In this paper, the differential invariant production cross sections, $E{\rm d}^{3}\sigma/{\rm d}p^{3}$, of $\pi^{0}$ and $\eta$ mesons and the particle production ratio $\eta/\pi^0$ are presented, measured over wide \pT ranges at mid-rapidity in pp collisions at $\sqrt{s}=~8$~TeV by ALICE.
The new experimental results are compared with pQCD calculations using MSTW08 (PDF) \cite{Martin:2009iq} with DSS14 (FF) \cite{deFlorian:2014xna} for the $\pi^{0}$ and accordingly CTEQ6M5 (PDF) \cite{1126-6708-2007-02-053} with AESSS (FF) \cite{Aidala:2010bn} for the $\eta$, as well as the PYTHIA8.210 Monte Carlo (MC) event generator \cite{Sjostrand:2014zea} with the tunes Tune 4C \cite{Corke:2010yf} and Monash~2013 \cite{Skands:2014pea}.

This paper is organized as follows:
In Sec.~\ref{sec:detector}, the ALICE experiment is briefly described with the focus on the detectors used in this analysis, namely the calorimeters and the central tracking systems.
Sec.~\ref{sec:event} describes the datasets, the event selection and also introduces the calorimeter triggers used in this analysis.
In Sec.~\ref{sec:reconstruction}, the reconstruction principles for neutral mesons are introduced.
Furthermore, the determination of correction factors, which are used to calculate the differential invariant cross sections from the measured raw yields, is described.
Sec.~\ref{sec:systematics} discusses the various contributions to the statistical and systematic uncertainties of the measurements.
In Sec.~\ref{sec:result}, the \pT differential invariant cross sections for $\pi^{0}$ and $\eta$ meson production in pp collisions at $\sqrt{s}=~8$~TeV are presented and compared with pQCD calculations.
Subsequently, the measured ratio of $\eta/\pi^{0}$ is presented and compared to the same theoretical models.
Sec.~\ref{sec:conclusion} concludes the paper with a summary of the obtained results.

\section{Detector description}
\label{sec:detector}
Neutral mesons, $\pi^0$ and $\eta$, decay into photons, which are reconstructed via two fundamentally different detection methods.
The first method exploits the measurement of photons using electromagnetic calorimeters. 
Two such calorimeters are available in ALICE \cite{Aamodt:2008zz, Abelev:2014ffa}: the Electromagnetic Calorimeter (EMCal) \cite{Cortese:2008zza} and the Photon Spectrometer (PHOS) \cite{Dellacasa:1999kd}.
The second method of photon detection makes use of photons converted into $e^+e^-$ pairs within the inner detector material located between the interaction point and a radius which corresponds to the midpoint between the inner and outer field cage of the Time Projection Chamber (TPC)~\cite{Alme:2010ke}.
These electron-positron pairs, originating at secondary vertices (${\rm V^{0}}$), are reconstructed by the main tracking systems in ALICE centered at mid-rapidity and consisting of the Inner Tracking System (ITS) \cite{Aamodt:2010aa} and the TPC \cite{Alme:2010ke}.
The aforementioned detectors are described below, noting the detector configurations during pp data taking at $\sqrt{s}=~8$~TeV in 2012.

The EMCal detector \cite{Cortese:2008zza} is a sampling electromagnetic calorimeter.
Its active elements, called cells, are composed of 77 alternating layers of lead and plastic scintillator providing a radiation length of $20.1\,X_{0}$.
The scintillation light in each layer is collected by wavelength shifting fibers perpendicular to the face of each cell.
The fibers are connected to $5\times 5$~mm$^{2}$ active area Avalanche Photo Diodes (APDs) to detect the generated scintillation light.
Each cell has a size of $\Delta \eta \times \Delta \phi = 0.0143 \times 0.0143$ ($\approx6.0\times 6.0$~cm$^{2}$), corresponding to approximately twice the Molière radius. 
Groups of $2 \times 2$ cells are combined into modules, which are further combined into arrays of $12 \times 24$ modules called supermodules.
In total, there are ten active, full EMCal supermodules, covering $\Delta\phi=100^\circ$ in azimuth and $|\eta|<0.7$ in pseudorapidity with a total number of 11,520 cells.
The EMCal is located at a radial distance of 4.28~m at the closest point from the nominal collision vertex.
The intrinsic energy resolution of the EMCal is parametrized as
$\sigma_{E}/E = 4.8\%/E \oplus 11.3\%/\sqrt{E} \oplus 1.7\%$
with $E$ in units of GeV \cite{Abeysekara:2010ze}.
The relative energy calibration of the detector is performed by measuring, in each cell, the reconstructed $\pi^{0}$ mass in the invariant mass distribution of photon pairs built with one photon in the given cell.
The achieved calibration level is estimated to be 3\% and adds up quadratically to the constant term of the energy resolution.

The PHOS \cite{Dellacasa:1999kd,Aamodt:2008zz} is a homogeneous electromagnetic calorimeter composed of lead tungstate, $\rm{PbWO_{4}}$.
The size of its elementary active units, also called cells, is $\Delta \eta \times \Delta \phi = 0.004 \times 0.004$ ($\approx2.2\times 2.2$~cm$^{2}$).
Thus, the lateral dimensions of the cells are slightly larger than the $\rm{PbWO_{4}}$ Molière radius of 2~cm.
APDs with an active area of $5\times 5$~mm$^{2}$ detect the scintillation light generated within the detector cells.
The spectrometer covers $\Delta\phi=60^\circ$ in azimuth and $|\eta|<0.12$ in pseudorapidity and is located at a distance of 4.6~m from the interaction point.
It is operated at a temperature of $-25^{\circ}\rm{C}$, at which the light yield of $\rm{PbWO_{4}}$ increases by about a factor of three compared to room temperature. 
The energy resolution of the PHOS is $\sigma_{E}/E = 1.8\%/E \oplus 3.3\%/\sqrt{E}\oplus 1.1\%$, with $E$ in units of GeV.
The fine granularity of the detector enables the measurement of $\pi^{0}$ candidates up to $\pT\approx 50$~GeV/$c$.

The ITS \cite{Aamodt:2010aa} consists of three sub-detectors each with two layers to measure the trajectories of charged particles and to reconstruct primary vertices.
The two innermost layers are the Silicon Pixel Detectors (SPD) positioned at radial distances of $3.9$~cm and $7.6$~cm.
The middle two layers are Silicon Drift Detectors (SDD) located at $15.0$~cm and $23.9$~cm relative to the beam line. 
The outer two layers are Silicon Strip Detectors (SSD) located at radial distances of $38$~cm and $43$~cm.
The two layers of SPD cover pseudorapidity ranges of $|\eta|<2$ and $|\eta|<1.4$, respectively.
The SDD and SSD cover $|\eta|<0.9$ and $|\eta|<1.0$, accordingly.

The TPC \cite{Alme:2010ke} is a large (90~${\rm m}^{3})$ cylindrical drift detector filled with a gas mixture of ${\rm Ne}$-${\rm CO_{2}}$ ($90$\%-$10$\%).
It covers a pseudorapidity range of $|\eta|<0.9$ over full azimuth, providing up to 159 reconstructed space points per track.
A magnetic field of $B=0.5$~T is generated by a large solenoidal magnet surrounding the central barrel detectors.
Charged tracks originating from the primary vertex can be reconstructed down to $\pT\approx 100$~MeV/$c$ and charged secondaries down to $\pT\approx 50$~MeV/$c$ \cite{Abelev:2014ffa}.
The TPC provides particle identification via the measurement of energy loss, \dedx, with a resolution of $\approx5\%$ \cite{Alme:2010ke}.
Beyond the outer radius of the TPC, the Transition Radiation Detector (TRD) and the Time-Of-Flight detector (TOF) provide additional particle identification information, as well as allowing for improved momentum resolution and added triggering capability.
The detectors represent most of the material between the TPC and the EMCal and hence dominate the material budget in front of the EMCal.
These detectors are missing in front of PHOS in order to provide a minimal radiation length to profit from the high resolution of the spectrometer.

The V0 detector is made up of two scintillator arrays (V0A and V0C) \cite{Cortese:2004aa} covering $2.8<\eta<5.1$ and $-3.7<\eta<-1.7$. It is used to provide a minimum bias (MB) trigger \cite{Abelev:2012sea} and reduce background events \cite{Abelev:2014ffa}.
It is also involved in the definition of calorimeter triggers \cite{Kral:2012ae, Wang:2011zzd} and is used for luminosity determination as described in the next section.

In addition, the T0 detector \cite{Adam2017} was used for luminosity determination.
It consists of two arrays of Cherenkov counters, T0A and T0C, which respectively cover $4.61<\eta<4.92$ and $-3.28<\eta<-2.97$.
The T0 furthermore provides a precise timing signal to other detectors with a resolution of better than 50~ps, used as starting signal for the TOF detector for example. 

\section{Datasets and event selection}
\label{sec:event}

During the data taking period of pp collisions at $\sqrt{s}=~8$~TeV in 2012, the LHC operated at high beam intensities of approximately $2\,\times\,10^{14}$ protons per beam.
Collisions at the ALICE interaction point were realized using a so-called "main-satellite" bunch scheme, which involved proton collisions between the high intensity main bunches and low intensity satellite bunches.
The interaction probability per bunch-satellite crossing was about $0.01$, corresponding to an average instantaneous luminosity of about $5\times 10^{30}~\rm{cm^{-2}s^{-1}}$.
Background events caused by beam-gas interactions or detector noise are rejected in the analysis using the V0A and V0C timing information \cite{Abelev:2014ffa}.
Pileup events, with more than one pp collision per bunch crossing, are rejected based on SPD pileup identification algorithms looking for multiple primary vertices in a single event \cite{Abelev:2014ffa}.
Additionally, the SPD is used to reject background events by comparing the number of SPD clusters to the multiplicity of SPD track candidates found in the respective collision.
Only events with a $z$-vertex position of $|z|<10$~cm in the global ALICE coordinate system are accepted for the analysis.

Two different types of triggers were used during data taking to select the events to be recorded: the minimum bias (MB) trigger and the calorimeter triggers, which are provided by the EMCal and the PHOS, to enhance statistics at high \pT by selectively recording events with high energy deposits in the calorimeters.
The MB trigger is a hardware Level-0 (L0) trigger \cite{Abelev:2012sea}.
It requires at least one hit in each V0A and V0C \cite{Cortese:2004aa}.
Both calorimeters also provide L0 triggers: EMC-L0~\cite{Kral:2012ae} and PHOS-L0~\cite{Wang:2011zzd}.
These L0 calorimeter triggers are required to be in coincidence with the MB trigger and select events with a deposited energy exceeding a nominal threshold in $4\times4$ adjacent cells, which is set to $\overline{E}_{{\rm EMC}\text{-}\rm{L0}}\approx2$~GeV and $\overline{E}_{{\rm PHOS}\text{-}\rm{L0}}\approx4$~GeV, respectively.
A software Level-1 (L1) trigger is also deployed for the EMCal which inspects events preselected by the EMC-L0 trigger~\cite{1748-0221-5-12-C12048}.
The trigger algorithm is similar to the EMC-L0, but combines information from different trigger region units to enhance the trigger efficiency and overcome hardware boundary effects~\cite{1748-0221-5-12-C12048}.
Additionally, a larger trigger threshold of $\overline{E}_{{\rm EMC}\text{-}\rm{L1}}\approx8.4$~GeV is set to further obtain statistics at higher transverse momenta.

\begin{figure}[htb]
  \centering
  \includegraphics[width=0.66\hsize]{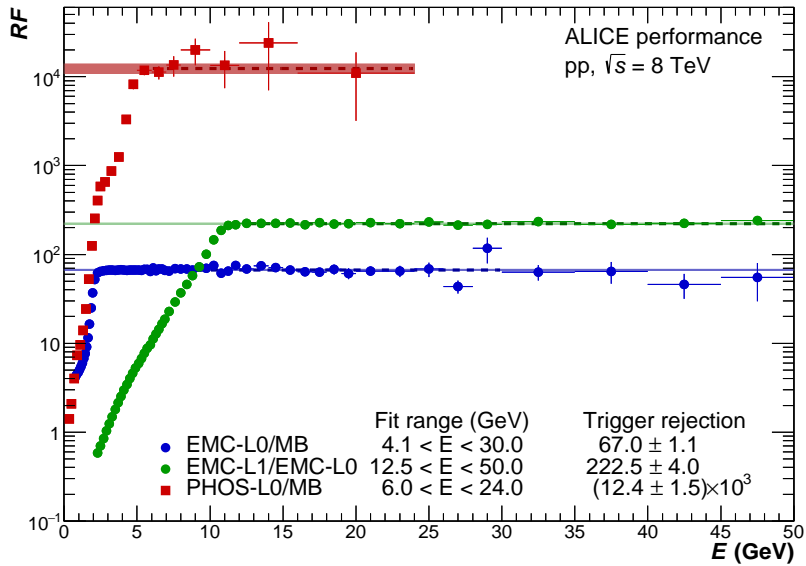}
  \caption{Determination of $RF$s for the PHOS-L0 and EMC-L0/L1 triggers. In the plateau region, the $RF$s are obtained by fits of constants in the given cluster energy ranges, illustrated by the dotted lines. The uncertainties of the determined $RF$s are indicated by light colored uncertainty bands, which are obtained by varying the fit ranges.}
  \label{fig:TrigEff}
  \bigskip
\end{figure}

In order to correctly normalize each trigger, the trigger rejection factors ($RF$) are determined by constructing the ratio of cluster energy spectra from MB and calorimeter triggered events as a function of the cluster energy, $E$, which are shown in Fig.~\ref{fig:TrigEff}.
The ratios are expected to follow a constant for high cluster energies, the so-called plateau region, assuming the triggers only enhance the rate of clusters but do not affect their reconstruction efficiency.
To reduce the statistical uncertainties, the $RF$s are always determined with respect to the next lower threshold trigger.
The cluster energy ratios have a steep turn-on near the respective trigger threshold energies.
Since the EMC-L0 trigger becomes fully efficient only above its triggering threshold of $\overline{E}_{{\rm EMC}\text{-}\rm{L0}}\approx2$~GeV, there is a change of slope visible in the turn-on region of the EMC-L1 trigger.
The turn-on curve of the PHOS-L0 trigger also changes its slope due to a non-uniformity of the channels hardware gains.
However, only the $RF$ plateau regions are mainly relevant for analysis, as they are needed to correctly normalize the triggered data, which are found to be: $RF_{\rm EMC\text{-}L0}=67.0\pm1.1$, $RF_{\rm PHOS\text{-}L0}=(12.4\pm 1.5)\times 10^3$ and $RF_{\rm EMC\text{-}L1}=(14.9\pm0.3)\times 10^3$.
The last factor is obtained by multiplying the two given rejection factors of the two EMCal triggers, see Fig.~\ref{fig:TrigEff}, as the $RF$ for EMC-L1 to MB trigger is of interest.

The luminosity determination is based on the cross-section of the MB trigger condition, $\sigma_{\rm MB_{AND}}$, measured in a van der Meer (vdM) scan \cite{vanderMeer:1968zz,ALICE-PUBLIC-2017-002}. 
The stability of the measured cross section throughout the whole data taking period is assessed by comparing the V0-based luminosity measurement with an independent luminosity signal, issued by the T0 detector. 
As discussed in Ref. \cite{ALICE-PUBLIC-2017-002}, this comparison results in an overall normalization uncertainty of 2.6\%, which includes contributions from both the vdM-based measurement and its stability over time.
The integrated luminosity of each triggered sample is calculated with the number of analyzed events, $N_{\text{events}}$, the respective rejection factors, $RF$, and the MB cross section, $\sigma_{\rm MB_{AND}}=55.80\pm1.45_{\text{(stat+sys)}}$~mb \cite{ALICE-PUBLIC-2017-002}, given by:
\begin{equation}
\mathscr{L_{\rm int}}=\frac{N_{\rm events}}{\sigma_{\rm MB_{AND}}}\times RF,
\end{equation}
for which $RF=1$ holds for the MB trigger.
As the good run lists for each detection method do not coincide, integrated luminosities are individually quoted for all cases in Tab.~\ref{tab:Luminosities}.

\renewcommand{\arraystretch}{1.2}
\begin{table}[h]
  \begin{center}
    \begin{tabular}{l||c|cc|c}        
    \cline{2-5}
    \multicolumn{1}{c|}{} & \multicolumn{4}{c}{$\mathscr{L_{\rm int}}$ (nb$^{-1}$)} \\ 
    \hline
    Reconstruction method & \multicolumn{2}{c|}{EMC \& PCM-EMC} & PHOS & PCM\\
    \hline
    \hline
    MB trigger & \multicolumn{2}{c|}{$1.94\pm0.05_{\text{norm}}$} & $1.25\pm0.04_{\text{norm}}$ & $2.17\pm0.06_{\text{norm}}$ \\
    EMC-/PHOS-L0 trigger & \multicolumn{2}{c|}{$40.9\pm0.7_{\text{sys}}\pm1.1_{\text{norm}}$} & $135.6\pm16.8_{\text{sys}}\pm3.6_{\text{norm}}$ & - \\
    EMC-L1 trigger & \multicolumn{2}{c|}{$615.0\pm15.0_{\text{sys}}\pm16.0_{\text{norm}}$} & - & - \\
    \hline 
    \end{tabular}
    \caption{The analyzed luminosities considering the individual statistics for the different reconstruction methods and triggers. The EMCal related measurements use the same list of good runs as indicated by the combined column. The uncertainties denoted with ``sys'' reflect the systematical uncertainty of $RF$ determination, whereas ``norm'' represents the uncertainties entering from the cross section determination of the MB trigger \cite{ALICE-PUBLIC-2017-002}.}
    \label{tab:Luminosities}
  \end{center}
\end{table}
\renewcommand{\arraystretch}{1.0}

\section{Neutral meson reconstruction}
\label{sec:reconstruction}

Both $\pi^{0}$ and $\eta$ mesons are reconstructed via their two-photon decay channels with branching ratios of $98.823\pm0.034\%$ and $39.31\pm0.20\%$ \cite{Olive:2016xmw} by means of an invariant mass analysis.
The neutral mesons are reconstructed using the two electromagnetic calorimeters, EMCal and PHOS, a photon conversion method (PCM) and a hybrid method, PCM-EMCal, which combines one photon candidate from the PCM and one from the EMCal, resulting in four (three) different methods for the reconstruction of $\pi^{0}$ ($\eta$) mesons.
The reconstruction of $\eta$ mesons is not accessible by PHOS due to the limited detector acceptance and, compared to the $\pi^{0}$, the wider opening angle of the decay photons.
The hybrid PCM-EMCal method benefits from the high momentum resolution of the PCM, a high reconstruction efficiency and, crucially, the triggering capabilities of the EMCal.
Moreover, an extended $p_{\rm T}$ coverage is achieved compared to the standalone EMCal measurement, as there is no limitation due to cluster merging effects, discussed later in this section. 

Photons and electrons/positrons generate electromagnetic showers when they enter an electromagnetic calorimeter.
They usually spread their energy over multiple adjacent calorimeter cells.
In order to reconstruct the full energy of impinging particles, those adjacent cells need to be grouped into clusters, which is realized by a clusterization algorithm. 
In the first step, the algorithm looks for the cell that recorded the highest energy in the event, exceeding the seed energy, $E_{\rm seed}$.
After the identification of such a seed cell, adjacent cells with recorded energies above a minimum energy, $E_{\rm min}$, are added to the cluster.
For the EMCal, the clusterization algorithm adds cells to the cluster as long as their recorded energy is smaller than the previous cell's energy and does not aggregate the respective cell, if it recorded a higher energy than the previous one.
The clusterization process continues in the same way with the remaining cells, until all cells above the energy thresholds are grouped into clusters.
Cluster energies are then calculated by $E=\sum^{N_{\rm cell}}_{i}e_{i}$, where $e_{i}$ stands for the energy recorded by the indicated cell.
The values of $E_{\rm seed}$ and $E_{\rm min}$ depend on the energy resolution and the noise level of the front-end electronics. 
For the EMCal, values of $E_{\rm seed}=500$~MeV and $E_{\rm min}=100$~MeV are chosen. 
For the PHOS, these parameters are set to $E_{\rm seed}=200$~MeV and $E_{\rm min}=15$~MeV.
Large clusters due to overlapping photon showers in the PHOS are separated into individual clusters by an unfolding method based on the knowledge of the lateral shape of the electromagnetic shower~\cite{Alessandro:2006yt}.

\begin{figure}[bht]
  \centering
  \includegraphics[width=0.49\hsize]{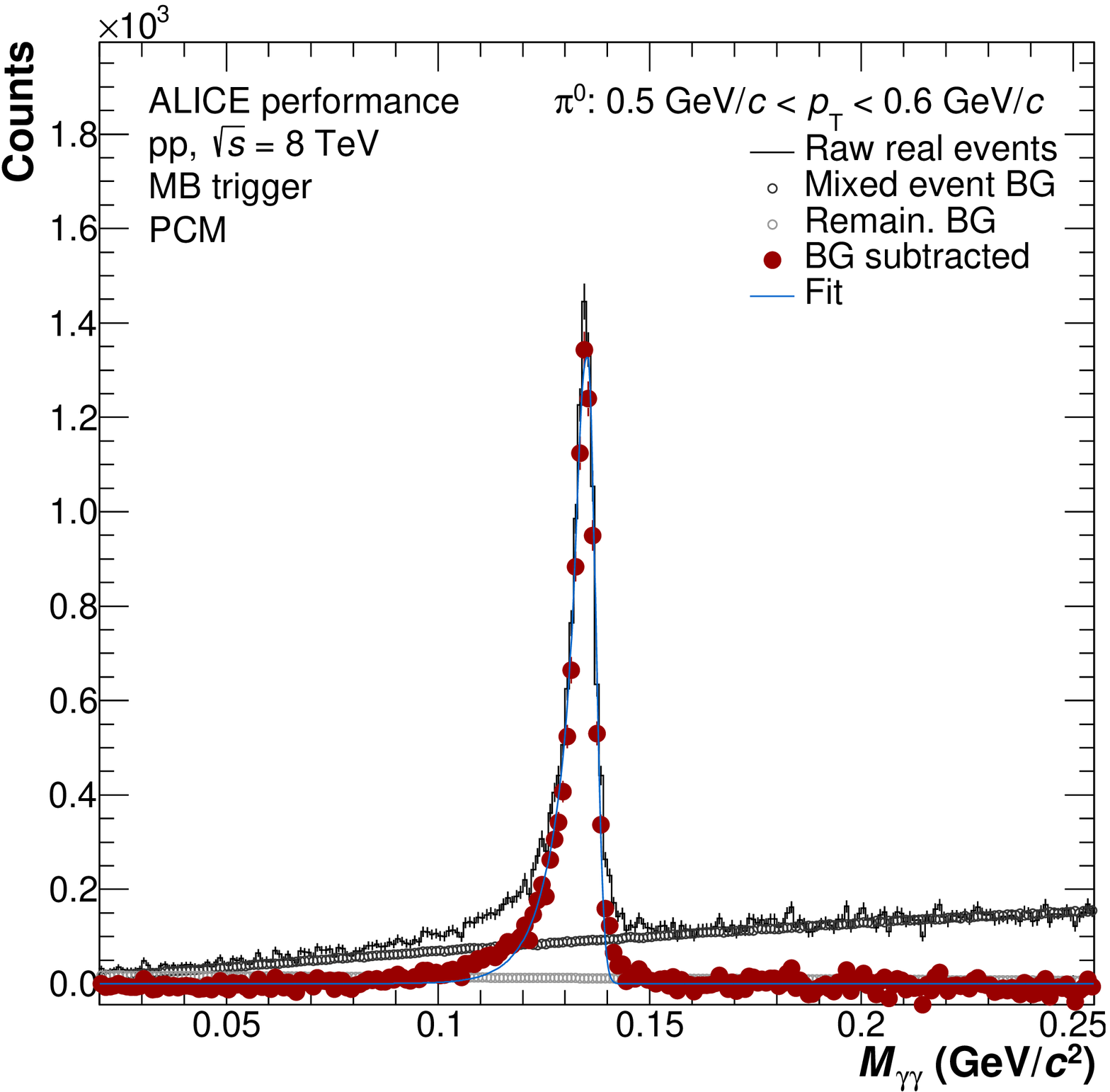}
  \hfil
  \includegraphics[width=0.49\hsize]{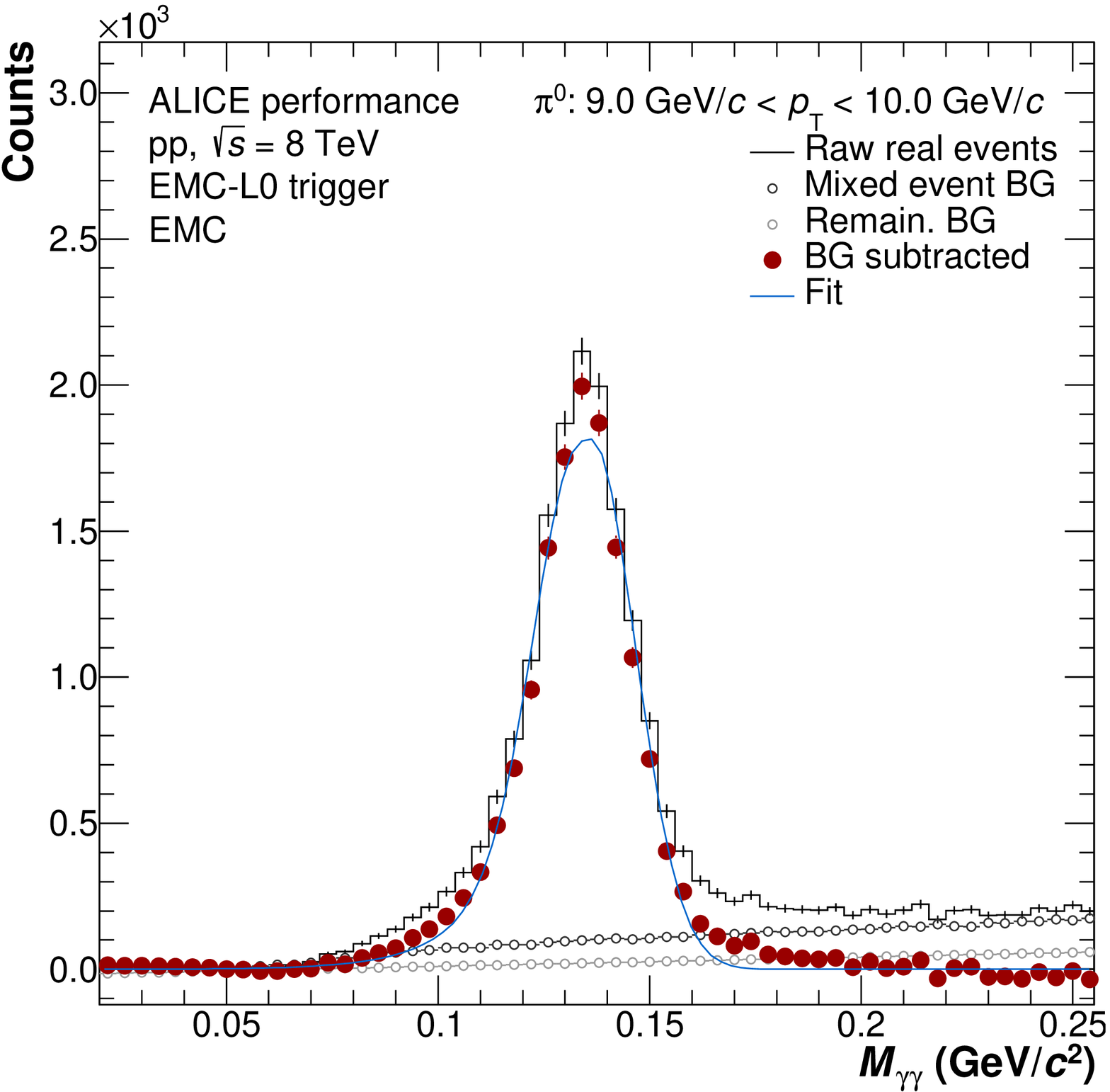}
  \includegraphics[width=0.49\hsize]{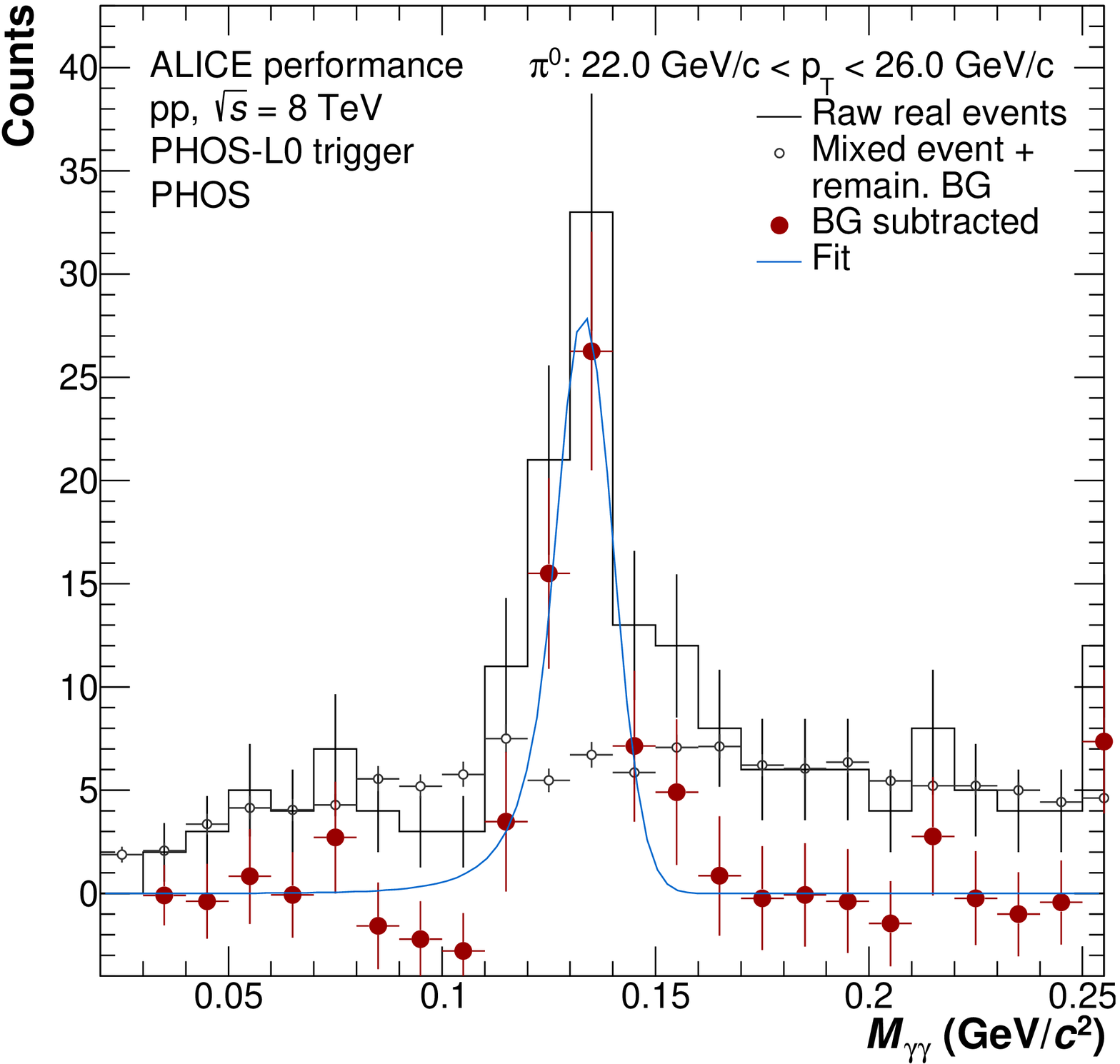}
  \hfil
  \includegraphics[width=0.49\hsize]{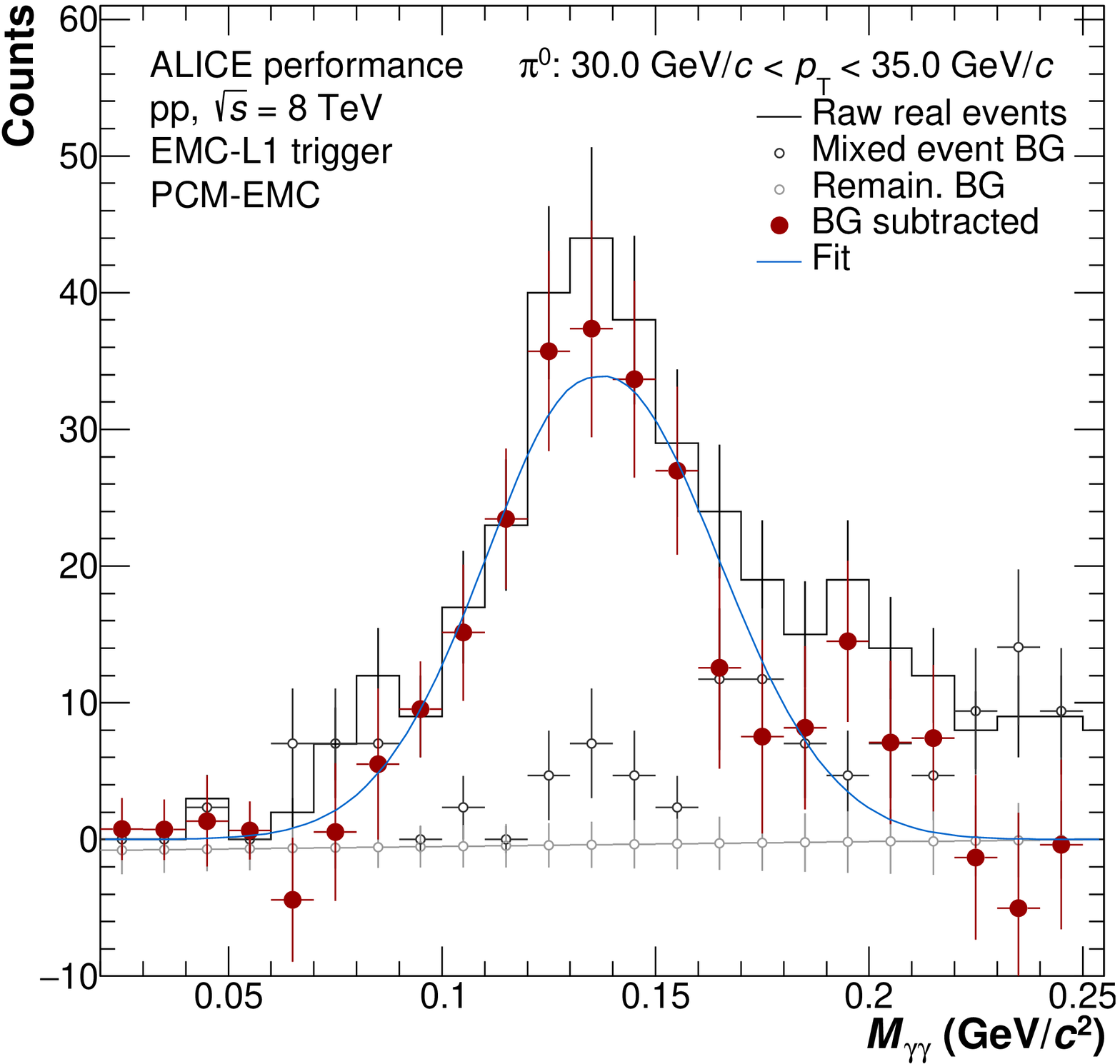}  
  \caption{Example invariant mass spectra in selected $p_{\rm T}$ slices for PCM (top left), PHOS (top right), EMC (bottom left) and PCM-EMC (bottom right) in the $\pi^{0}$ mass region.
  The black histograms show raw invariant mass distributions before any background subtraction.
  The grey points show mixed-event and residual correlated background contributions, which have been subtracted from raw real events to obtain the signal displayed with red data points.
  The blue curves represent fits to the background-subtracted invariant mass spectra.
  Additional examples of invariant mass distributions for the different methods are given in Ref. \cite{ALICE-PUBLIC-2017-009}.}
  \label{fig:InvMassPi0}
  \bigskip
\end{figure}

Cell energies are calibrated for both calorimeters to provide best estimates for the cluster energies.
After the cell-by-cell energy calibration of the EMCal \cite{Cortese:2008zza,Abeysekara:2010ze}, an improved correction for the relative energy scale as well as for the residual geometrical misalignment of the EMCal between data and MC simulations is derived by making use of the good momentum resolution of the PCM photon in the hybrid PCM-EMCal method. 
Using this method, the $\pi^{0}$ mass is evaluated as a function of EMCal cluster energy, $E_{\rm cluster}$, for data and MC.
Therefrom, a cluster energy correction is deduced for the simulation, for which the reconstructed $\pi^{0}$ masses are adjusted to the measured mass positions in data.
For $E_{\rm cluster}\approx 1$~GeV, the correction is of the order of 2\% and rises up to 4\% for higher energies.
Thus, a precise energy calibration scheme for the relevant energy regions is available which is found to be consistent for the EMCal and hybrid PCM-EMCal methods for $\pi^{0}$ as well as $\eta$ mesons at the same time, hence demonstrating the validity of the procedure. 
After applying this calibration in the analysis, the $\pi^{0}$ and $\eta$ mass values in data and MC are obtained for each \pT bin and their ratio is computed. 
Then, the ratios are plotted versus \pT and fitted with a constant, giving access to the residual miscalibration of the meson mass values between data and MC.
Such residual offsets of 0.005$\pm$0.043\% and 0.14$\pm$0.13\% are found for $\pi^{0}$ and $\eta$ mesons for the EMCal analysis, whereas 0.002$\pm$0.042\% and 0.02$\pm$0.14\% are obtained for PCM-EMCal, illustrating the performance of the calibration procedure.
For the PHOS, the energy deposition in each cell is calibrated by adjusting the $\pi^0$ peak position in the invariant mass spectra of photon pairs to the true mass of the $\pi^0$ meson. 
The accuracy of this calibration procedure is estimated to be better than 1\%.
It is evaluated from a comparison of the $\pi^0$ peak width in calibrated data and MC simulations by introducing random, normal-distributed decalibration parameters to the MC simulation. 

Photon identification criteria are applied to the sample of reconstructed clusters in order to primarily select clusters generated by true photon candidates.
For the photon reconstruction with PHOS, relatively loose identification cuts are applied because the shower overlap is negligible and the combinatorial background is found to be small in pp collisions.
A minimum cluster energy, $E_{\rm cluster}>0.3$~GeV, as well as a minimum number of cells forming a cluster, $N_{\rm cell}\geq3$, are required in order to reject electronic noise and minimum ionizing particles which deposit about $270$~MeV in the PHOS.
For the EMCal, a minimum energy cut of $E_{\rm cluster}>0.7$~GeV is applied and the minimum number of cells grouped in a cluster is set to $N_{\rm cell}\geq2$.
Furthermore, the selection criteria of $|\eta|<0.67$ and $1.40$~rad~$ < \varphi < 3.15$~rad are imposed for EMCal clusters.
Pileup from multiple events, which may occur within a readout interval of the front-end electronics, is rejected by applying a cluster timing cut relative to the collision time of $-25<t_{\rm cluster}<25$~ns for the PHOS and $-35<t_{\rm cluster}<25$~ns for the EMCal. 
Thus, photon candidates from different bunch crossings are removed with high efficiency of $\rm{>}99\%$.
For the EMCal, all clusters matched with a primary charged track are rejected.
This track matching procedure, referred to as general track matching, uses a track \pT-dependent matching in $\eta$ and $\varphi$, beginning from $|\Delta\eta|<0.04$ and $|\Delta\varphi|<0.09$ for very low track momenta of $\pT<0.5$~\GeVc and going down to $|\Delta\eta|<0.01$ and $|\Delta\varphi|<0.015$ for highest track momenta, using the \pT-dependent matching conditions $|\Delta\eta|<0.01+(\pT+4.07)^{-2.5}$ and $|\Delta\varphi|<0.015+(\pT+3.65)^{-2}$.
Applying these conditions, a primary track to cluster matching efficiency of more than 95\% is obtained over the full \pT range, rising above 98\% for the analyzed EMCal triggered datasets for \pT beyond 10~\GeVc. 
To further enhance the photon purity and to reject neutral hadrons, a cluster shape cut of $0.1\leq \sigma^{2}_{\rm long} \leq 0.7$ is applied for EMCal clusters, where $\sigma^{2}_{\rm long}$ stands for the smaller eigenvalue of the dispersion matrix of the shower shape ellipse defined by the responding cells and their energy contributions to the cluster \cite{Awes1992130,Acharya:2017hyu}.
The lower threshold of $\sigma^{2}_{\rm long}$ is chosen to remove contamination caused by neutrons hitting the APDs of the readout electronics.

Photons convert into lepton pairs within the detector material of ALICE with a probability of about 8.5\%.
The reconstruction of such photon conversion candidates using PCM may be divided into three major steps: (i) tracking of charged particles and secondary vertex (${\rm V^{0}}$) finding \cite{Alessandro:2006yt}; (ii) particle identification and (iii) photon candidate reconstruction and subsequent selection.
The ${\rm V^{0}}$s used in this analysis are obtained during data reconstruction using all available tracking information, recalculating the momenta of the daughter tracks under the assumption that both daughters are created with parallel momentum vectors at the ${\rm V^{0}}$.
The tracks associated with secondary vertices are required to have a minimum momentum of $p_{\rm T}^{\rm track}>50$~MeV/$c$ and at least 60\% of clusters from the maximum possible number of clusters, that a particle track can create in the TPC along its path, need to be found.
In order to reduce the contamination from Dalitz decays, conversion candidates are only considered with a vertex at a radial distance of at least $R>5$~cm.
In addition, a line-cut is applied to restrict the geometrical $\eta$ distribution of the ${\rm V^{0}}$s in order to remove photon candidates that would otherwise appear outside the angular dimensions of the detector. 
The condition $R_{\rm conv}>|Z_{\rm conv}| S_{\rm ZR}-7$~cm is applied with $S_{\rm ZR}=\tan\left(2\arctan(\exp(-\eta_{\rm cut}))\right)$ and $\eta_{\rm cut}=0.9$, where the coordinates $R_{\rm conv}$ and $Z_{\rm conv}$ are determined with respect to the nominal center of the detector.
Additional constraints are imposed on $R_{\rm conv}<180$~cm and $|Z_{\rm conv}|<240$~cm to ensure that the reconstruction of secondary tracks is performed inside the TPC.
Electrons and positrons from photon conversions are identified via their energy deposit, \dedx, in the TPC.
The difference of the measured \dedx value from the hypothesis of the electron/positron energy loss is used for particle identification.
The \dedx of measured charged tracks is required to be within $-3<{n\sigma_{e}<5}$ of the expected energy loss, which is a \pT-dependent observable defined by $n\sigma_{e}=(d$E$/d$x$-\langle $d$E/$d$x \rangle_{e})/\sigma_{e}$ with the average energy loss of the electron/positron, $\langle $d$E/$d$x \rangle_{e}$, and the Gaussian width of the fit to the measured \dedx distribution, $\sigma_{e}$.
To further reduce charged pion contamination as the pion \dedx-band begins to merge with the electron/positron \dedx-band above $p\gtrsim 4$~\GeVc, a cut based on the separation from the hypothesis of charged pion energy loss is applied in $n\sigma_{\pi}$, analog to the previous definition.
Tracks with energy losses closer to the pion line than $|n\sigma_{\pi}|<1$ are removed, which is done up to $3.5$~\GeVc.
The non-photon ${\rm V^{0}}$ candidate contamination is further suppressed by a triangular two-dimensional cut, $|\Psi_{\rm pair}|<\Psi_{\rm pair, max}(1-\chi^{2}_{\rm red}/\chi^{2}_{\rm red, max})$, with $\chi^{2}_{\rm red, max}=30$ and $\Psi_{\rm pair, max}=0.1$.
This cut is based on the reduced $\chi^{2}$ of the Kalman-Filter \cite{Fruhwirth:1987fm} hypothesis for the $e^{+}e^{-}$ pair and on the angle $\Psi_{\rm pair}$ between the plane perpendicular to the magnetic field of the ALICE magnet and the $e^{+}e^{-}$ pair plane.
Furthermore, a cut on the cosine of the pointing angle of $\cos(\theta_{\rm PA})>0.85$ is performed, where the pointing angle, $\theta_{\rm PA}$, is the angle between the reconstructed photon momentum vector and the vector joining the collision vertex.
The remaining ${\rm K^{0}_{S}}$, ${\rm \Lambda}$ and ${\rm \overline{\Lambda}}$ contamination is removed by selecting $q_{\rm T}<q_{\rm T, max}\sqrt{1-\alpha^{2}/\alpha^{2}_{\rm max}}$ on the Armenteros-Podolanski plot \cite{podolanski1954iii} with $q_{\rm T, max}=0.05$~\GeVc and $\alpha_{\rm max}=0.95$.
Additionally, the PCM measurement requires an out-of-bunch pileup correction which estimates the contamination of photon candidates from multiple events overlapping in the TPC.
The correction is based on a study of the distance of closest approach (DCA) of the conversion photon candidates which is the smallest distance in beam direction, $z$, between the primary vertex and the momentum vector of the photon candidate.
Photon candidates from different events generate a broad underlying Gaussian-like DCA distribution, which is fitted in order to estimate the out-of-bunch pileup contribution.
The correction is found to be \pT-dependent and ranges from 42\% at low \pT$\approx0.35$~\GeVc to 10\% at high~\pT~$\approx11$~\GeVc.

\begin{figure}[ht]
  \centering
  \includegraphics[width=0.49\hsize]{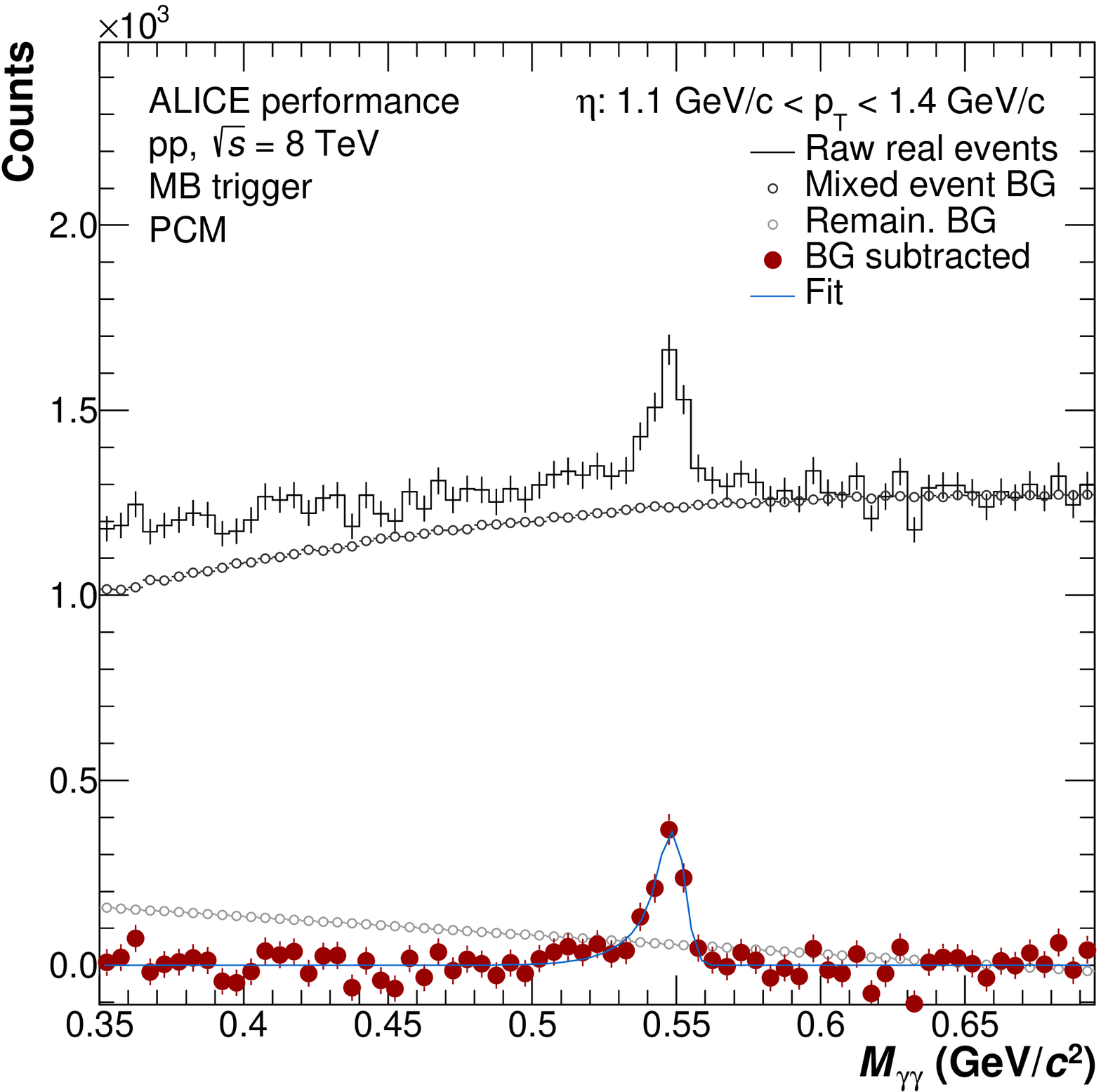}
  \hfil
  \includegraphics[width=0.49\hsize]{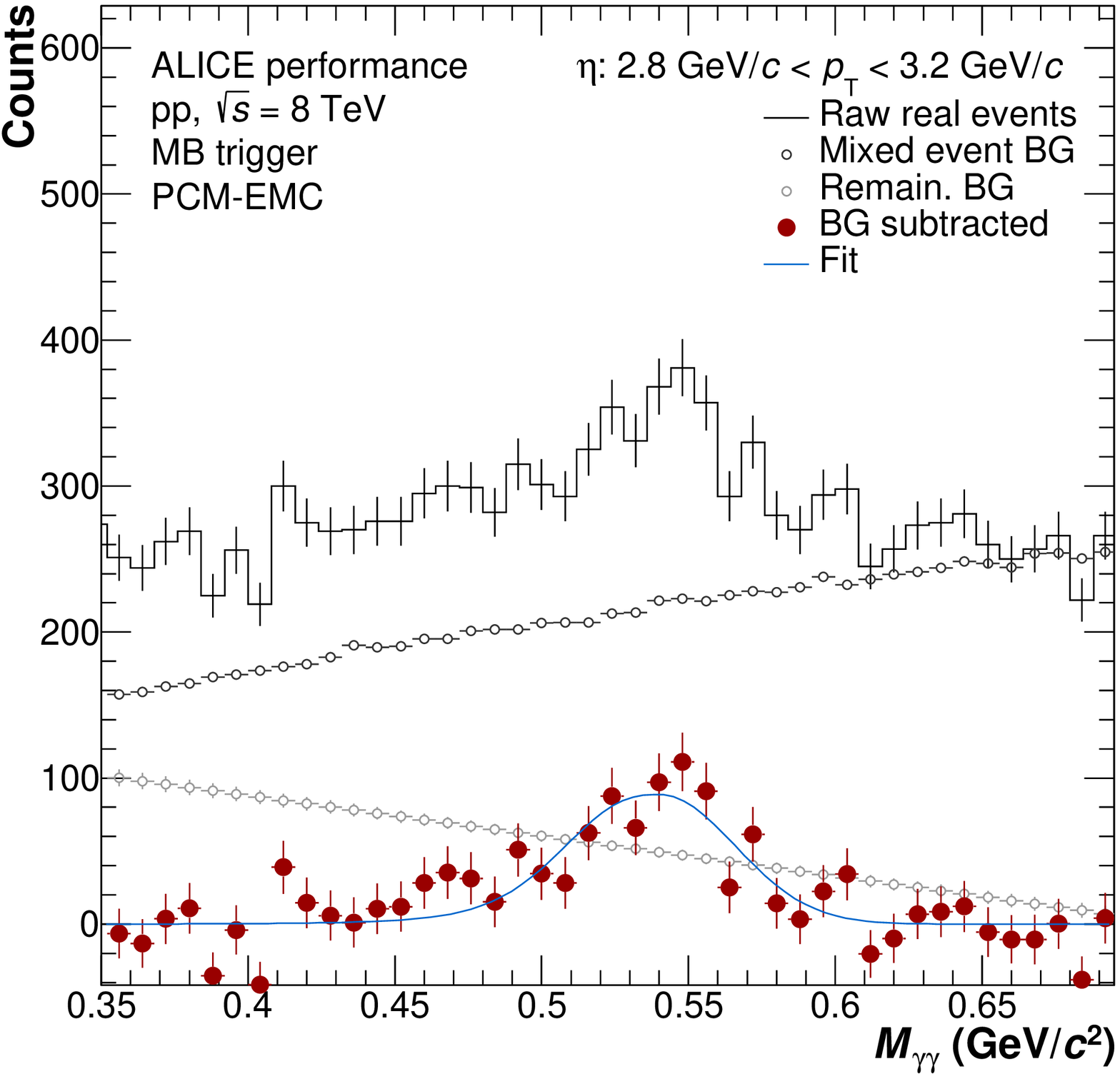}
  \includegraphics[width=0.49\hsize]{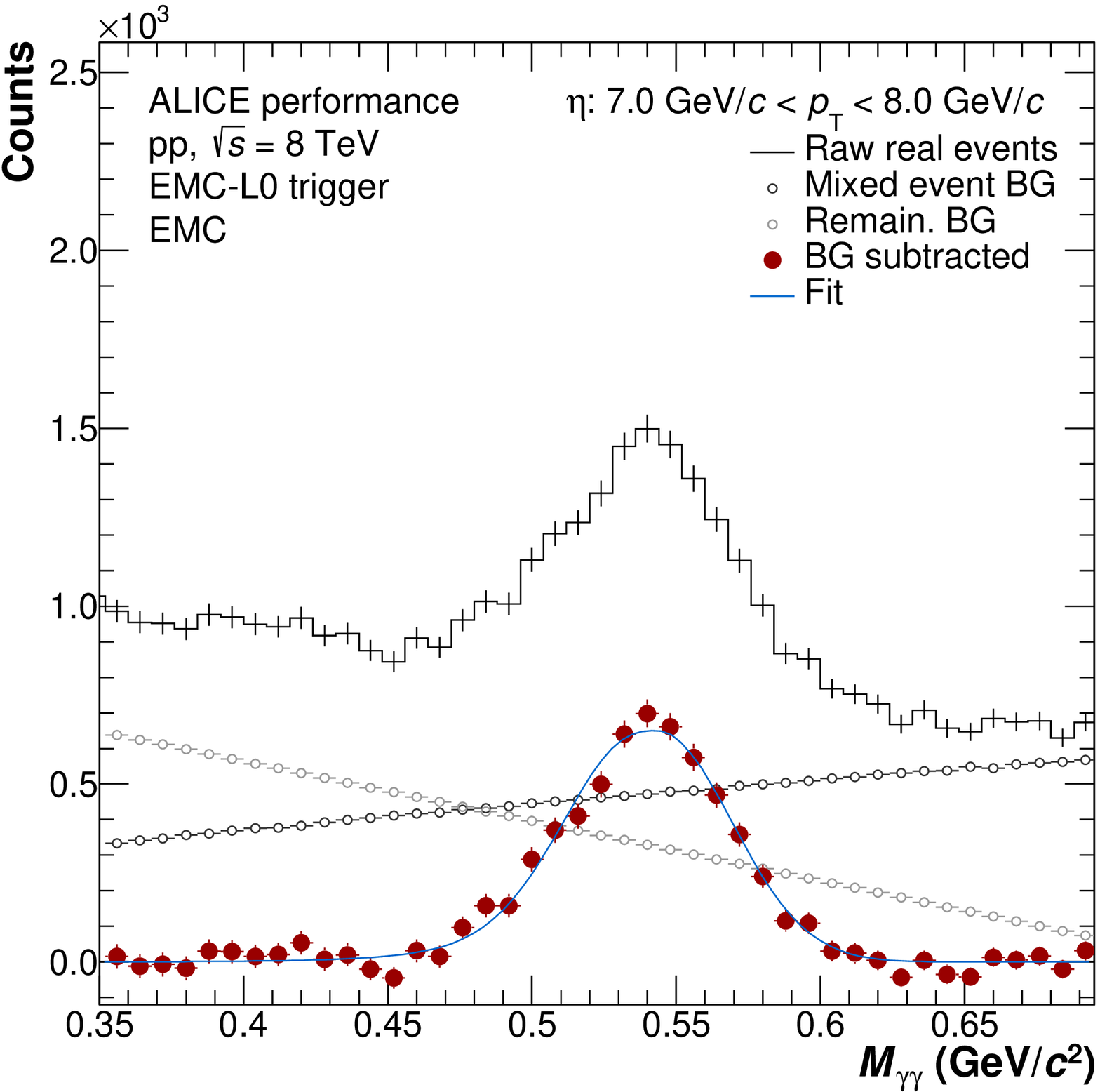}
  \hfil
  \includegraphics[width=0.49\hsize]{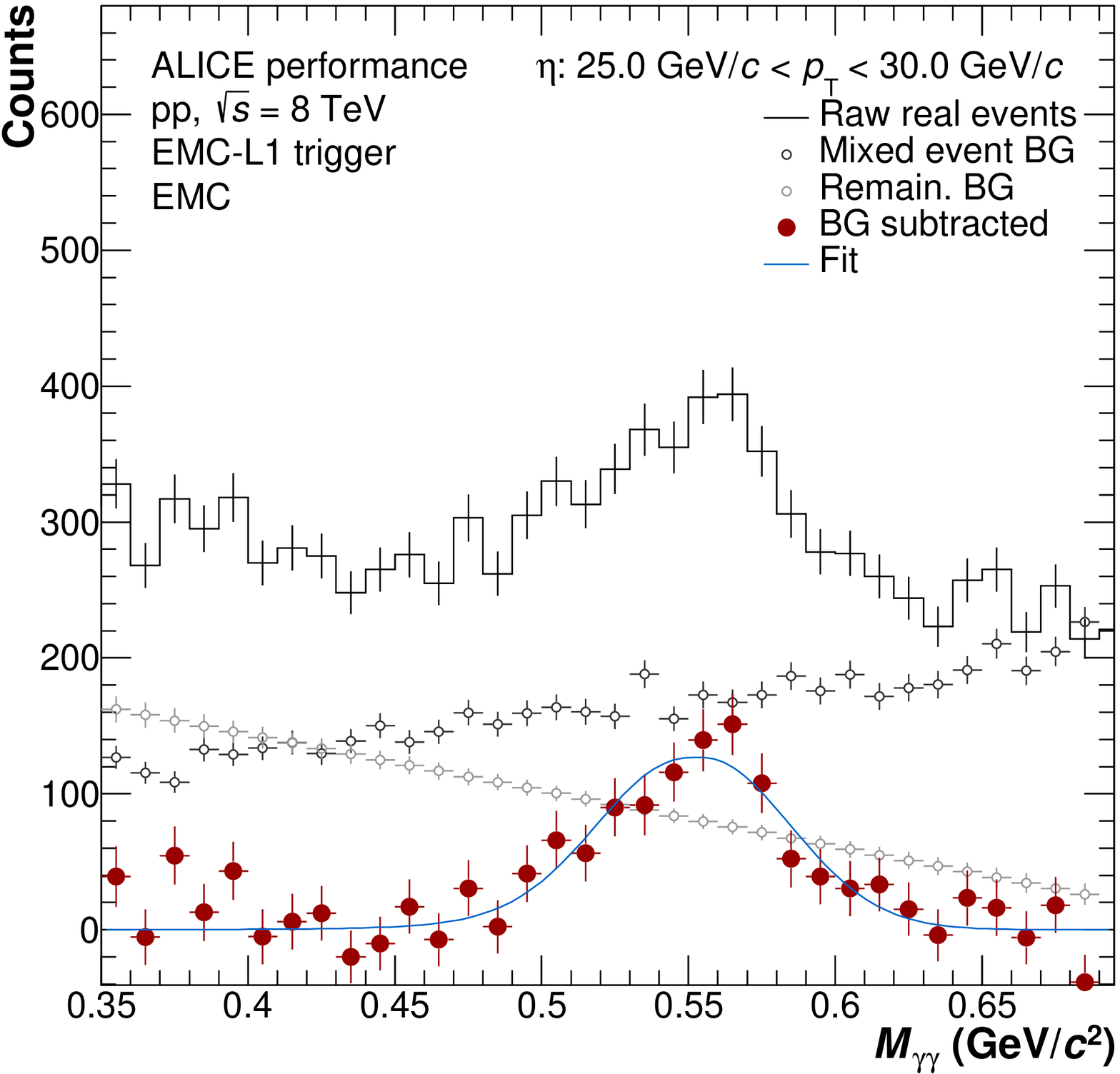}  
  \caption{Example invariant mass spectra in selected \pT slices in PCM (top left), PCM-EMCal (top right) and EMCal (bottom plots) in the $\eta$ mass region.
  The black histograms show raw invariant mass distributions before any background subtraction.
  The grey points show mixed-event and residual correlated background contributions, which have been subtracted from raw real events to obtain the signal displayed with red data points.
  The blue curves represent fits to the background-subtracted invariant mass spectra.
  Additional examples of invariant mass distributions for the different methods are given in Ref. \cite{ALICE-PUBLIC-2017-009}.}
  \label{fig:InvMassEta}
  \bigskip
\end{figure}

The hybrid PCM-EMCal method practically uses the same cuts on photon candidates as the respective standalone reconstruction methods. 
In context of the PCM, a wider cut of $-4<{\rm n\sigma_{e}<5}$ concerning the electron/positron energy loss hypothesis is used for the hybrid method and the \pT restriction of the charged pion \dedx cut is loosened.
Only the upper value of the cut on the short axis of the moment of the shower shape for the EMCal is changed and required to be $\sigma^{2}_{\rm long} \leq 0.5$ in order to further reject contamination of hadrons \cite{Acharya:2017hyu}. 
Due to the timing constraint of the EMCal restricting clusters to triggered bunch crossings, no DCA or additional out-of-bunch pileup rejection is needed for the hybrid method.
In addition to the general matching of primary charged particles to EMCal clusters already described, a dedicated track matching procedure for the two charged ${\rm V^{0}}$ daughters with respect to EMCal clusters is applied.
This cluster-${\rm V^{0}}$ track matching is the most important ingredient for the hybrid analysis, as pairing one leg of the ${\rm V^{0}}$ candidate with the EMCal cluster generated by one of these secondary charged tracks itself, leads to an auto-correlation and causes a broad peak between the masses of the $\pi^{0}$ and $\eta$ mesons at around 300~MeV/$c$.
The same parameters from the general track matching procedure are found to remove about 99\% of such candidates.

Invariant mass distributions of photon pairs, shown in Figs.~\ref{fig:InvMassPi0} and \ref{fig:InvMassEta}, include combinatorial background as well as the neutral meson signal for photon candidate pairs from the same, real event.
An opening angle cut of 17 mrad for the angle between the momentum vectors of the two paired photon candidates is applied for the EMCal measurement.
Requiring a minimum separation between such pairs is needed to ensure a proper background description by event mixing, in which two clusters from different events might otherwise be separated by an arbitrarily small distance.
In same events, such cluster configurations would overlap partially or even merge into single clusters, which has been explicitly considered
for event mixing by not allowing the cells with largest deposited energies of respective clusters to be direct neighbors on the EMCal surface.
For the PCM and hybrid PCM-EMCal methods, an opening angle cut of 5 mrad is further applied between the momentum vectors of the pair of conversion photon candidates and accordingly, the pair of PCM and EMCal photon candidates.
Furthermore, pairs are restricted to a rapidity of $|y|<0.12$ for the PHOS and $|y|<0.8$ for all other methods.

The uncorrelated combinatorial background is estimated by using an event mixing technique, in which photon candidates from different events are paired in order to prevent correlations between the candidates.
Different event pools are used for this purpose, binned by primary vertex position, photon candidate multiplicity and transverse momentum to ensure the mixing of similar events only.
In contrast to same-event combinations to extract the neutral meson signal, the mixed-event background is obtained with up to 80 different events, stored in each of the event pools, in order to minimize its statistical uncertainties.
Therefore, the mixed-event background needs to be scaled to match the integral of the raw signal in the vicinity of the right side of the neutral meson peak, just outside the peak integration interval, after which it is subtracted from the raw distribution.
The background-subtracted signal is then fitted to determine the mass peak position and width of $\pi^{0}$ and $\eta$ mesons for every \pT bin.
A function composed of a Gaussian modified by an exponential tail at the low mass side \cite{Matulewicz1990194} is used for this purpose.
The low mass tail accounts for late conversions of one or both photons for the EMCal method and for energy loss effects due to bremsstrahlung for the PCM and hybrid PCM-EMCal methods.
To reflect the residual correlated background components which remain after the subtraction of the mixed-event background, the fitting is performed by including an additional first order polynomial function (deduced from MC simulations), which is also shown in Figs.~\ref{fig:InvMassPi0} and \ref{fig:InvMassEta} and which is further being subtracted from the invariant mass distribution.
In contrast, a slightly different approach for the background description is followed for the PHOS as its limited acceptance results in a more complicated shape of the combinatorial background around the signal peak, especially at low \pT.
As both correlated and combinatorial backgrounds are influenced in the same manner, the ratio of the raw signal and mixed-event distributions is constructed and fitted with first or second order polynomial function outside the peak region. 
Then, the mixed-event distribution is scaled with the obtained polynomial function and subtracted from the raw signal, which can be followed in Fig.~\ref{fig:InvMassPi0}.
A Crystal Ball function \cite{CrystalBall:1980} is used as the main fit function for the PHOS method which also reproduces the tail at the low mass region to take into account the late conversion of photons in front of the calorimeter.
The signal distribution is then obtained by subtracting the scaled mixed-event background from the raw invariant mass distribution.
The resulting background-subtracted signal distributions as well as raw signals from real events, the normalized mixed-event and residual background distributions are shown in Figs.~\ref{fig:InvMassPi0}, \ref{fig:InvMassEta} and in Ref. \cite{ALICE-PUBLIC-2017-009} for the $\pi^{0}$ and $\eta$ meson mass region, respectively, for given example \pT bins, illustrating the meson reconstruction over the full reported \pT range.

\begin{figure}[h]
  \includegraphics[width=0.49\hsize]{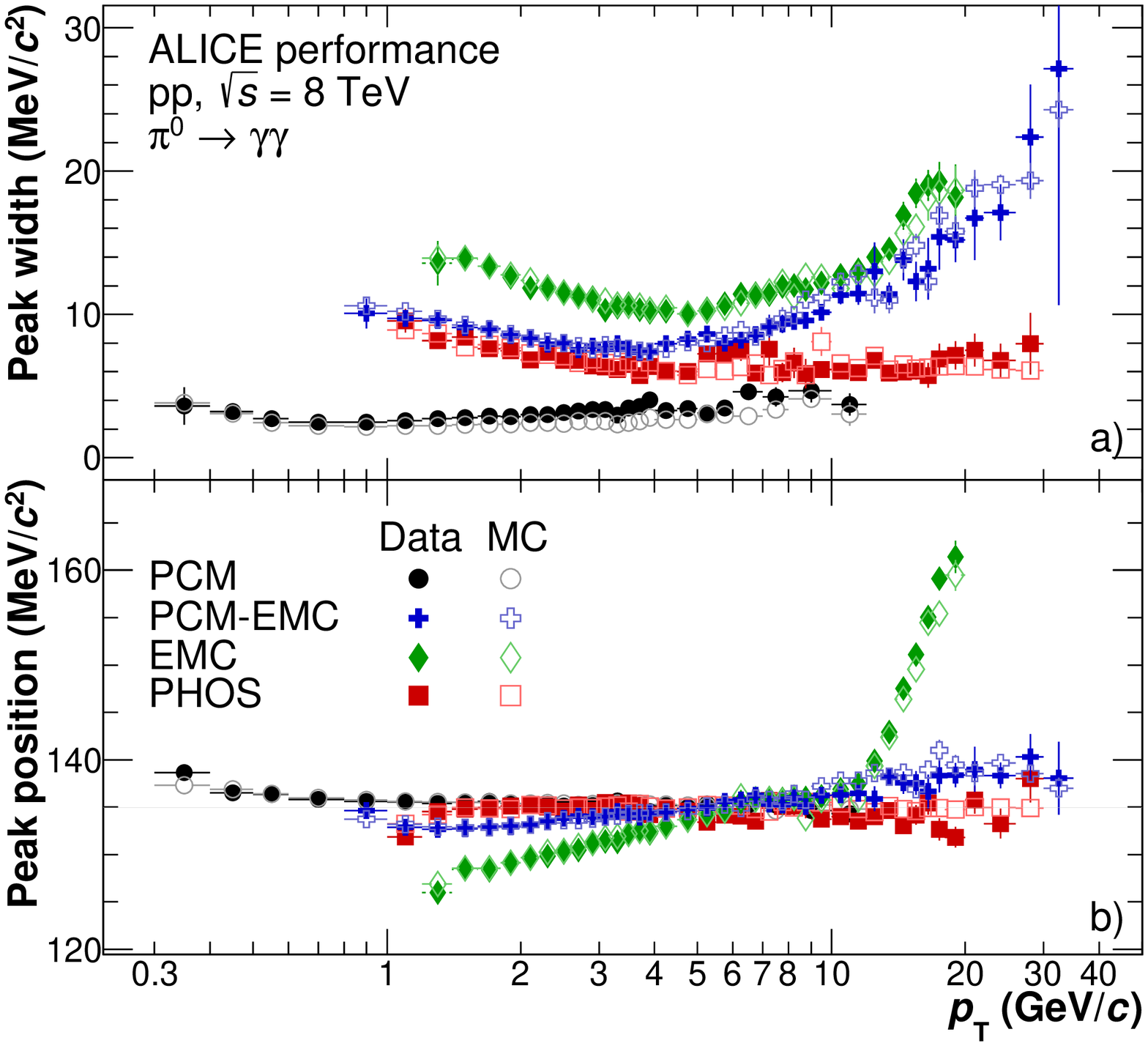}
  \hfil
  \includegraphics[width=0.49\hsize]{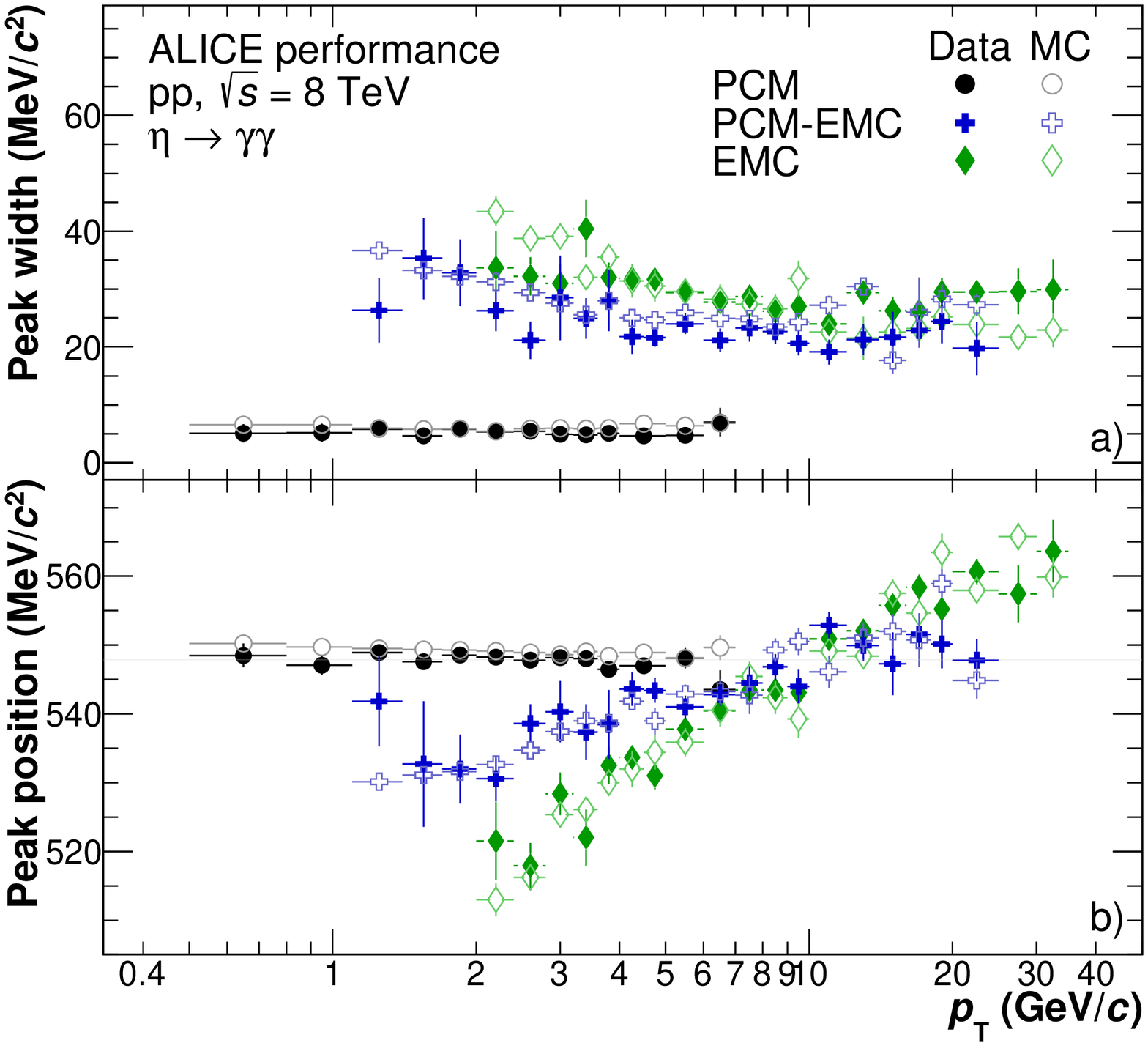}
  \caption{The left plots show reconstructed $\pi^{0}$ peak positions (left-bottom) and widths (left-top) of each reconstruction method compared to MC simulations for the transverse momentum bins used in the analysis. Corresponding plots for the $\eta$ meson are on the right for peak masses (right-bottom) and widths (right-top).}
  \label{fig:MassAndWidth}
  \bigskip
\end{figure}

The neutral meson raw yields are extracted by integrating the background-subtracted invariant mass distributions.
The integration windows are defined by the reconstructed mass position and width obtained by the respective fits of the signal distribution in a given \pT bin.
For the PHOS method, the integration range for $\pi^0$ is asymmetrically defined as $[-5\sigma,+3\sigma]$ around reconstructed peak position, where $\sigma$ is the standard deviation of the Gaussian part of the Crystal Ball function to take the asymmetric shape into account.
For the other methods, the integration windows for both neutral mesons are chosen to cover at least $[-3\sigma,+3\sigma]$ around the reconstructed peak position, where $\sigma$ is the standard deviation of the Gaussian part of the fit function.
For each reconstruction method, the peak position and width used for the signal extraction are shown in Fig.~\ref{fig:MassAndWidth} as a function of reconstructed \pT.

\begin{figure}[htb]
  \includegraphics[width=0.49\hsize]{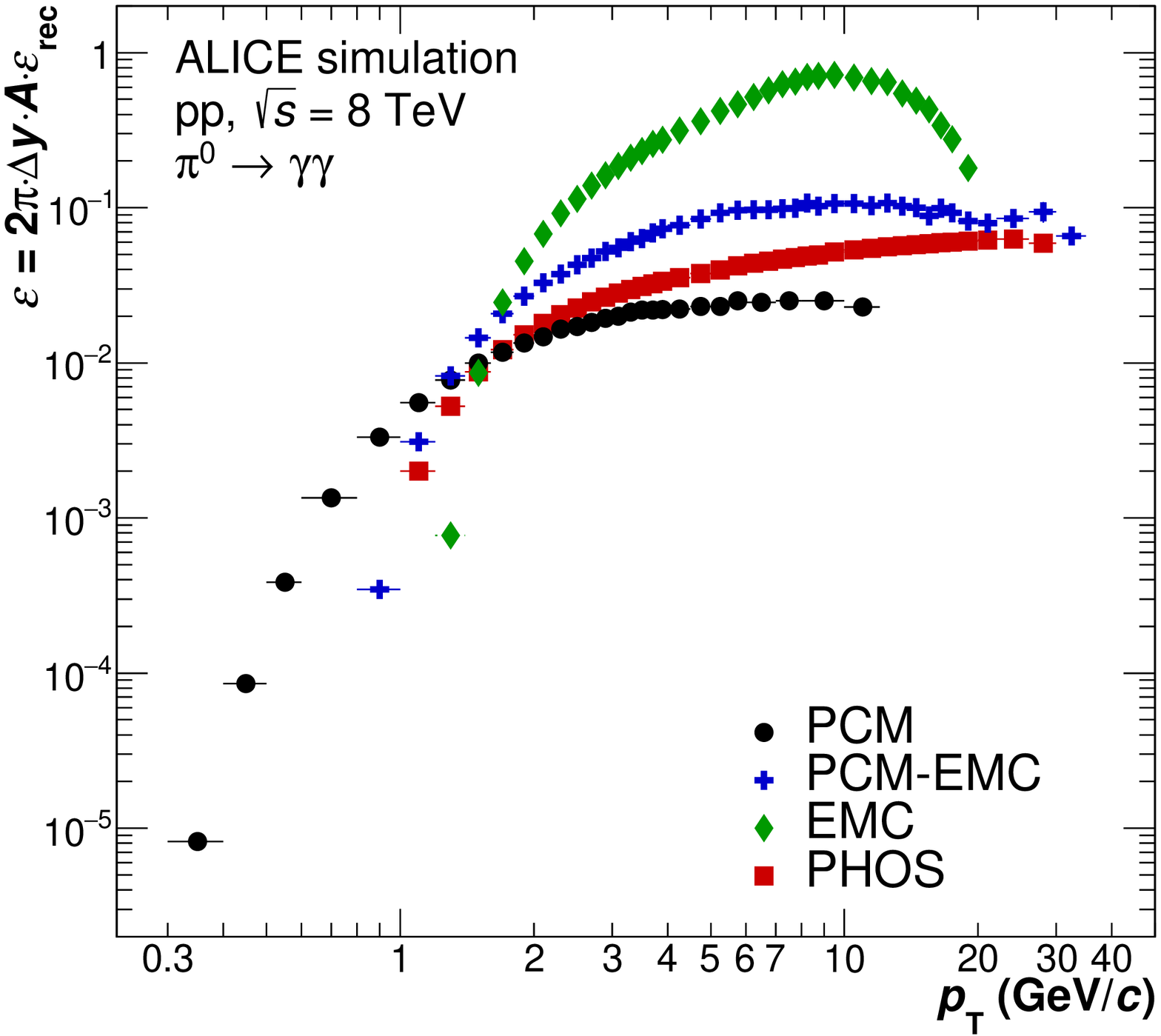}
  \hfil
  \includegraphics[width=0.49\hsize]{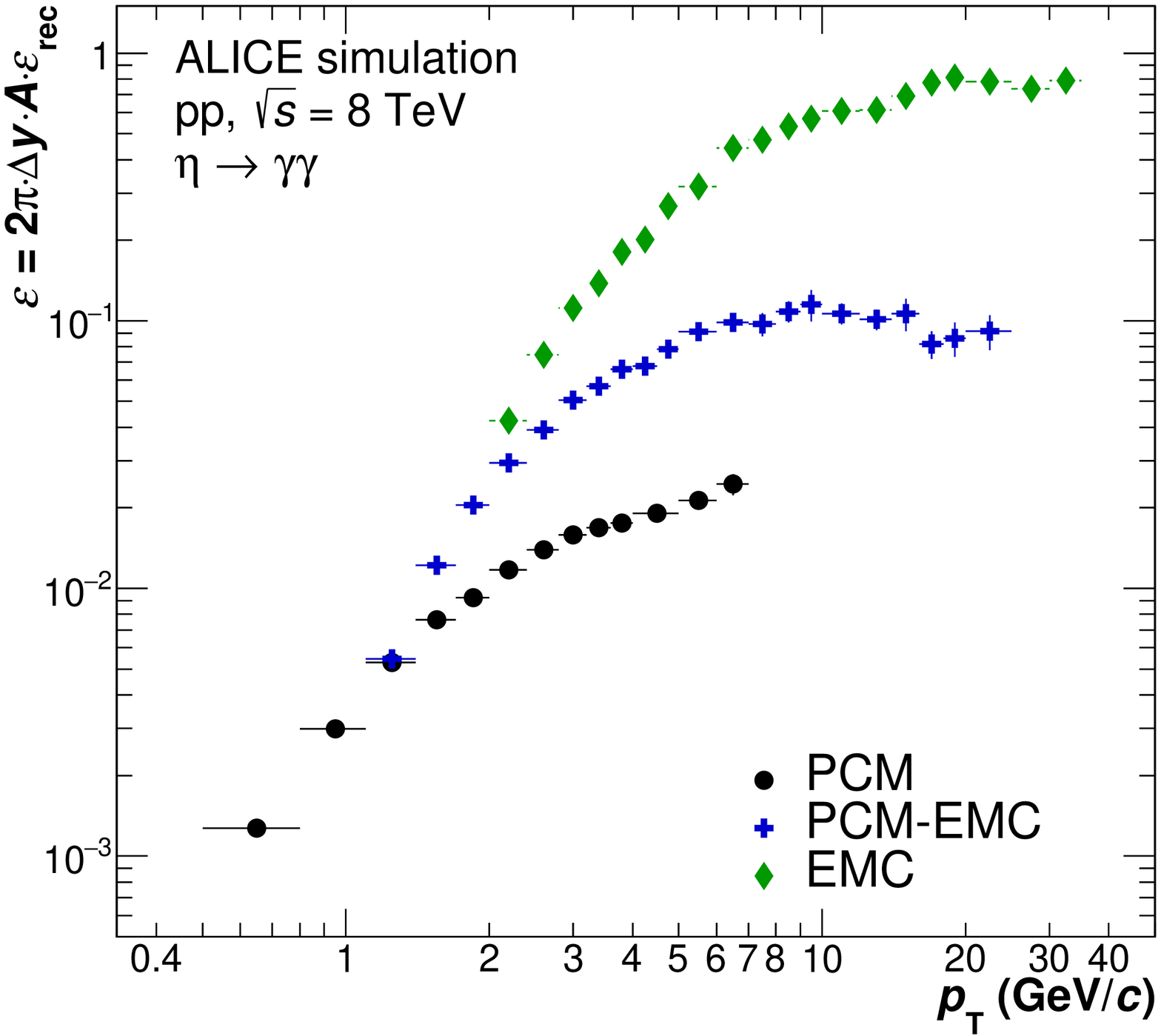}
  \caption{The normalized correction factors, $\epsilon$, for each reconstruction method for $\pi^{0}$ (left) and $\eta$ mesons (right) plotted versus \pT bins used in the analysis.
  The factors contain the detector acceptances and the respective reconstruction efficiencies, where acceptances are further normalized by the rapidity windows accessible with each method, $\Delta y$, and full azimuth coverage of $2\pi$, in order to enable a direct comparison between the different methods.}
  \label{fig:AcceptanceTimesEff}
  \bigskip
\end{figure}

Corrections for geometrical acceptance and reconstruction efficiencies are evaluated using MC simulations.
PYTHIA8 \cite{Sjostrand:2014zea} and PHOJET \cite{Engel:1995sb} event generators with minimum bias processes are used for this purpose.
The correction factors for both MC productions are found to be consistent and, hence, are combined.
To generate enough statistics for high meson momenta to be able to correct the raw yields obtained with triggered data, a PYTHIA8 simulation is used that is enriched with jets, generated in bins of hard scatterings, $p_{\mbox{\tiny T, hard}}$.
Particles generated by the event generators are propagated through the ALICE detector using GEANT3 \cite{Brun:1987ma} which realistically reproduces interactions between the particles and the detector material.
In the simulation, the same reconstruction algorithms and analysis cuts are applied as for real data.
In Fig.~\ref{fig:MassAndWidth}, the reconstructed $\pi^{0}$ and $\eta$ mass peak positions and widths are compared as a function of \pT between data and MC to confirm a proper detector response in the simulation. 
The normalized correction factors, $\epsilon$, for each method, containing the specific detector acceptances as well as the full reconstruction efficiencies,
are shown in Fig.~\ref{fig:AcceptanceTimesEff}.
For the EMCal analysis, the correction factor for the $\pi^0$ is observed to decrease for $\pT\gtrsim10$~\GeVc.
This is due to the effect of cluster merging, as due to the Lorentz boost the opening angles of $\pi^0$ mesons become too small to resolve adjacent clusters given the finite segmentation of the calorimeter.
While the dominant symmetric decays are first to merge, the asymmetric decay contributions become more relevant at higher momenta.
Above a certain limiting momentum, it is no longer possible to separate the two decay photons of the $\pi^0$, creating merged clusters that significantly reduce the reconstruction efficiency in the EMCal as seen in Fig.~\ref{fig:AcceptanceTimesEff}.
Thus, the natural upper limit for the $\pi^0$ reconstruction with the EMCal is of the order of $\pT^{\pi^{0}}\approx 20$~GeV/$c$.
In contrast, the PCM-EMCal hybrid approach overcomes the limitations of the EMCal cell segmentation and makes it possible to reconstruct $\pi^0$ mesons up to $\pT\approx35$~\GeVc as reported in this paper.
For the PHOS, such cluster merging effects are negligible for the reported \pT range owing to the high granularity of the calorimeter.
Since the opening angles of photons from $\eta$ meson decays are much larger compared to the $\pi^0$, merging effects are negligible  for all approaches over the full reported \pT range in this case.

The contributions of secondary $\pi^0$ from weak decays and hadronic interactions with the detector material are estimated and removed for the $\pi^0$ measurements.
Weak decays of ${\rm K_{S}^{0}}$ represent the main source of secondaries.
For all reconstruction methods, the spectra of the three main particles relevant for the secondary correction due to weak decays, ${\rm K_{S}^{0}}$, ${\rm K_{L}^{0}}$ and $\Lambda$, are obtained from Refs. \cite{Aamodt:2011zza, Abelev:2014laa, Adam:2016emw} with extrapolation of spectra to $\sqrt{s}=8$~TeV, assuming a power law for each \pT bin as function of $\sqrt{s}$.
These spectra are used as weights in a PYTHIA6.4 simulation, where the respective particle decays are simulated on generator level, taking into account the full decay kinematics.
Using this procedure, the invariant yields of secondary $\pi^0$s from weakly decaying particles are obtained.
From the full ALICE GEANT3 MC simulations, the acceptance and reconstruction efficiencies are calculated for these secondaries and multiplied with the respective invariant yields from the generator level MC simulation to arrive at the secondary $\pi^0$ raw yields from the different particles.
On the other hand, the $\pi^0$ raw yield from interactions with the detector material is purely obtained from the full MC simulation, which is the only viable approach.
All the estimated, secondary $\pi^0$ raw yields are subtracted from the reconstructed number of $\pi^0$s, as indicated in Eq. \ref{eq:InvCrossSection}.
The corrections are of the order of 1-3\% for ${\rm K_{S}^{0}}$, $<$0.5\% for ${\rm K_{L}^{0}}$, $\lesssim$0.02\% for $\Lambda$ and 0.1-2\% for material, varying within the given values for the different methods and triggers used.

As there are three different triggers available for the EMCal and hybrid PCM-EMCal methods, and two different ones for the PHOS measurement, each with its own statistical and systematic uncertainties, as well as correlations between the different systematical uncertainties, the results from each trigger class are properly combined in order to obtain the final result for each reconstruction method.
Statistical uncertainties are ensured to be uncorrelated since different triggers use non-overlapping data samples.
For the systematic uncertainties, the \pT-dependent correlation coefficients are determined.
Only a few systematic uncertainties are found to be uncorrelated, such as the uncertainty of signal extraction and partly ``efficiency'' and ``trigger'' related uncertainties, for which further details are contained in Sec. \ref{sec:systematics}.
The correlation coefficients are found to be generally above 0.8.
The respective \pT-dependent weights are calculated according to the BLUE algorithm \cite{Lyons1988rp,Valassi2003mu,Lyons1986em,barlow1989statistics,Valassi2013bga}, which are used to combine the spectra from each method.

\section{Systematic uncertainties}
\label{sec:systematics}

Systematic uncertainties are respectively summarized in Tabs.~\ref{tab:table_system_Pi0}, \ref{tab:table_system_Eta} and \ref{tab:table_system_EtaToPi0} for the neutral mesons $\pi^{0}$, $\eta$ and their ratio $\eta/\pi^{0}$. 
The values are given in percent and refer to relative systematic uncertainties of the measured values.
Three different example \pT bins are listed for each reconstruction method in order to illustrate their relative strengths.
An additional, more detailed description of the systematic sources and the determination of uncertainties for all methods except PHOS may be found in Ref. \cite{Acharya:2017hyu}, which is fully applicable to this paper.

\begin{table}[thb]
  \begin{center}
  \scalebox{0.88}{
  \hspace*{-0.5cm}
    \begin{tabular}{l||c|c|c|c||c|c|c|c||c|c|c}
      \hline
      \pT interval & \multicolumn{4}{c||}{$1.4-1.6$ GeV/$c$} & \multicolumn{4}{c||}{$5.0-5.5$ GeV/$c$} & \multicolumn{3}{c}{$15.0-16.0$ GeV/$c$} \\ \hline
      \multirow{2}{*}{Method} & \multirow{2}{*}{PCM} & PCM- & \multirow{2}{*}{EMC} & \multirow{2}{*}{PHOS} & \multirow{2}{*}{PCM} & PCM- & \multirow{2}{*}{EMC} & \multirow{2}{*}{PHOS} & PCM- & \multirow{2}{*}{EMC} & \multirow{2}{*}{PHOS} \\ 
             &  & EMC & & & & EMC & & & EMC & & \\ \hline\hline
      Signal extraction      & $4.8$ & $1.9$ & $2.3$ & $3.0$ & $5.4$ & $2.4$ & $1.5$ & $1.8$ & $3.3$ & $4.6$ & $1.0$  \\
      Inner material         & $9.0$ & $4.5$ &     - &     - & $9.0$ & $4.5$ &     - &     - & $4.5$ &     - &     -  \\
      Outer material         &     - & $2.1$ & $4.2$ & $3.5$ &     - & $2.1$ & $4.2$ & $3.5$ & $2.1$ & $4.2$ & $3.5$  \\
      PCM track rec.         & $1.0$ & $0.5$ &     - &     - & $1.0$ & $0.9$ &     - &     - & $2.1$ &     - &     -  \\
      PCM electron PID       & $1.8$ & $0.6$ &     - &     - & $1.1$ & $1.3$ &     - &     - & $3.1$ &     - &     -  \\
      PCM photon PID         & $1.7$ & $0.5$ &     - &     - & $2.1$ & $1.1$ &     - &     - & $3.5$ &     - &     -  \\
      Cluster description    &     - & $2.5$ & $4.4$ &     - &     - & $2.5$ & $3.7$ &     - & $4.3$ & $4.0$ &     -  \\
      Cluster energy calib.  &     - & $1.8$ & $2.5$ & $2.6$ &     - & $1.9$ & $1.8$ & $0.6$ & $2.8$ & $2.0$ & $0.6$  \\
      Track match to cluster &     - & $0.2$ & $3.1$ &     - &     - & $0.5$ & $2.0$ &     - & $3.3$ & $3.7$ &     -  \\
      Efficiency             &     - & $2.0$ & $2.5$ & $7.0$ &     - & $2.8$ & $2.7$ & $7.0$ & $2.7$ & $3.7$ & $7.5$  \\
      Trigg. norm.\&pileup   & $3.4$ & $0.1$ & $0.1$ & $1.2$ & $2.2$ & $0.7$ & $0.3$ & $1.2$ & $2.3$ & $2.4$ & $12.5$ \\ \hline
      Total syst. uncertainty& $11.1$& $6.5$ & $8.0$ & $8.9$ & $11.0$& $7.3$ & $6.9$ & $8.2$ & $10.6$& $9.6$ & $15.0$ \\ 
      \hline\hline
      Statistical uncertainty& $1.5$ & $1.5$ & $3.4$ & $7.2$ & $7.5$ & $3.3$ & $2.2$ & $8.2$ & $7.9$ & $4.4$ & $10.6$ \\
      \hline\hline
      Combined stat. unc.    & \multicolumn{4}{c||}{2.1} & \multicolumn{4}{c||}{2.2} & \multicolumn{3}{c}{4.0} \\
      \hline
      Combined syst. unc.    & \multicolumn{4}{c||}{5.1} & \multicolumn{4}{c||}{5.1} & \multicolumn{3}{c}{7.6} \\ 
      \hline
    \end{tabular}}
    \caption{Summary of relative systematic uncertainties in percent for selected \pT bins for the reconstruction of $\pi^{0}$ mesons. The statistical uncertainties are given in addition to the total systematic uncertainties for each bin. Moreover, the combined statistical and systematic uncertainties are also listed, obtained by applying the BLUE method \cite{Lyons1988rp,Valassi2003mu,Lyons1986em,barlow1989statistics,Valassi2013bga} for all reconstruction methods available in the given \pT bin, considering the uncertainty correlations for the different methods (see Sec. \ref{sec:result} for further details). The uncertainty from $\sigma_{\rm MB_{AND}}$ determination of 2.6\%, see Ref. \cite{ALICE-PUBLIC-2017-002}, is independent from the reported measurements and is separately indicated in the following plots below.}
    \label{tab:table_system_Pi0}
  \end{center}
\end{table}

For the $\pi^0$ measurement by PHOS, the systematic uncertainty related to the signal extraction is evaluated by varying the fitting range and the assumptions about the mass peak and background shapes.
The systematic uncertainty related to the material budget is taken from Ref. \cite{Abelev:2012cn}, which is estimated by comparing the results of the analysis with and without magnetic field in the ALICE solenoid.
Photons, which converted to $e^+e^-$ pairs within the detector material, are most likely being reconstructed as two clusters in the presence of a magnetic field. 
Without a field, the secondary tracks from photon conversions are less separated and can be dominantly detected as single clusters, building the correct invariant masses for $\pi^0$s in a di-cluster analysis.
Therefore, comparing the $\pi^0$ spectra from data and MC with nominal and zero magnetic fields is a straightforward method to evaluate the uncertainty of the material budget description in simulations.
Systematic uncertainties due to the cluster energy calibration are decomposed into the uncertainty of the energy scale of clusters and non-linearity effects.
The energy scale uncertainty of $0.1\%$ is estimated from a comparison of the $\pi^{0}$ mass peak position for the two-photon invariant mass spectra in data and MC. 
This energy uncertainty is translated to an uncertainty of the $\pi^{0}$ yield by convolution with the shape of the \pT spectrum.
The systematic uncertainty due to the non-linearity correction is evaluated by introducing different non-linearity correction schemes and calibration parameters for the MC simulation, whereas the \pT dependence of the $\pi^{0}$ peak position and width is always kept consistent with data.
The efficiency uncertainty consists of acceptance variations and differences between MC event generators. 
The acceptance uncertainty is estimated by changing the good cluster selection criteria, and the MC generator-dependent uncertainty is evaluated by comparing efficiencies of MB MC generators and single particle MC simulation which generates events containing single neutral mesons with realistic transverse momentum and rapidity distributions.
Moreover, it includes the trigger efficiency uncertainty for the high energy photon trigger analysis, which is estimated by comparing the trigger turn-on curve from data with MC simulations.  
``Trigger normalization \& pileup'' summarizes systematic uncertainties due to the trigger normalization factor and pileup effects.
The uncertainty related to the trigger normalization factor is estimated by changing the range of the fit to determine the rejection factor ($RF$).
Furthermore, the out-of-bunch pileup contribution is evaluated by varying the timing cut to accept clusters.

For the PCM measurement, the main source of systematic uncertainty is the material budget, for which the same value is used as previously calculated in Ref. \cite{Abelev:2012cn}.
The signal extraction uncertainty is estimated by changing the integration window around the invariant mass peak, the normalization range of the mixed-event background and by using different order polynomials as well as other fit functions to evaluate the remaining background contribution.
``Track reconstruction'' summarizes the systematic uncertainties found by requiring different numbers of TPC clusters and by applying different minimum transverse momentum cuts on tracks.
The systematic uncertainties due to the electron identification (``electron PID'' and ``PCM photon PID'') are determined by varying the PID cuts, which are elaborated in Sec.~\ref{sec:reconstruction}, and by comparing the respective results.
For PCM, the ``trigger normalization \& pileup'' uncertainty is dominated by the uncertainty of the DCA$z$ background description for the out-of-bunch pileup estimation.
Furthermore, it contains the systematic uncertainty due to the pileup rejection by the SPD due to its finite efficiency to remove pileup events.

\begin{table}[thb]
  \begin{center}
  \scalebox{0.88}{
    \begin{tabular}{l||c|c|c||c|c|c||c|c}
      \hline
      \pT interval & \multicolumn{3}{c||}{$2.0-2.4$ GeV/$c$} & \multicolumn{3}{c||}{$5.0-6.0$ GeV/$c$} & \multicolumn{2}{c}{$18.0-20.0$ GeV/$c$} \\ \hline
      \multirow{2}{*}{Method} & \multirow{2}{*}{PCM} & PCM- & \multirow{2}{*}{EMC} & \multirow{2}{*}{PCM} & PCM- & \multirow{2}{*}{EMC} & PCM- & \multirow{2}{*}{EMC} \\ 
             &  & EMC & & & EMC & & EMC & \\ \hline\hline
      Signal extraction      & $5.1$ & $9.0$ & $9.3$ & $7.3$ & $7.2$ & $6.0$ & $10.6$ & $8.1$  \\
      Inner material         & $9.0$ & $4.5$ &     - & $9.0$ & $4.5$ &     - & $4.5$  &     -  \\
      Outer material         &     - & $2.1$ & $4.2$ &     - & $2.1$ & $4.2$ & $2.1$  & $4.2$  \\
      PCM track rec.         & $1.5$ & $1.8$ &     - & $2.0$ & $2.4$ &     - & $3.3$  &     -  \\
      PCM electron PID       & $2.4$ & $1.8$ &     - & $2.2$ & $2.9$ &     - & $6.5$  &     -  \\
      PCM photon PID         & $3.6$ & $2.9$ &     - & $6.3$ & $3.0$ &     - & $7.9$  &     -  \\
      Cluster description    &     - & $3.1$ & $4.6$ &     - & $4.0$ & $4.9$ & $6.0$  & $4.9$  \\
      Cluster energy calib.  &     - & $3.2$ & $3.5$ &     - & $3.9$ & $3.4$ & $4.5$  & $3.5$  \\
      Track match to cluster &     - & $1.5$ & $4.0$ &     - & $1.7$ & $3.2$ & $4.2$  & $3.3$  \\
      Efficiency             &     - & $5.0$ & $4.3$ &     - & $9.7$ & $5.5$ & $10.0$ & $6.3$  \\
      Trigg. norm.\&pileup   & $2.1$ & $0.1$ & $0.1$ & $1.4$ & $1.4$ & $1.5$ & $3.0$  & $2.8$  \\ \hline
      Total syst. uncertainty& $11.5$& $13.0$& $13.1$& $13.6$& $15.2$& $11.5$& $20.9$ & $13.3$ \\ 
      \hline\hline
      Statistical uncertainty& $10.1$ & $12.1$& $16.8$& $18.3$& $6.8$ & $5.4$ & $21.3$ & $8.2$ \\
      \hline\hline
      Combined stat. unc.    & \multicolumn{3}{c||}{7.4} & \multicolumn{3}{c||}{5.0} & \multicolumn{2}{c}{7.9} \\
      \hline
      Combined syst. unc.    & \multicolumn{3}{c||}{8.7} & \multicolumn{3}{c||}{9.0} & \multicolumn{2}{c}{12.3} \\ 
      \hline
    \end{tabular}}
    \caption{Summary of relative systematic uncertainties in percent for selected \pT bins for the reconstruction of $\eta$ mesons, see Tab.~\ref{tab:table_system_Pi0} for further explanations which also apply here.}
    \label{tab:table_system_Eta}
  \end{center}
\end{table}

For the EMCal, one main systematic uncertainty arises from the knowledge of the outer material budget, which is defined by all detector components from the radial center of the TPC up to the EMCal.
The uncertainty is assessed by running the analysis only with/without TRD modules in front of the EMCal, since part of the data taking in 2012 occurred with the EMCal only partially obscured by the TRD.
Since TRD and TOF have similar material budgets, the same uncertainty is assigned to the TOF as well, which covered the full polar angle so that a similar assessment as for the TRD is not feasible.
Both uncertainties are quadratically combined to arrive at the given uncertainties which are listed in the tables.
The signal extraction uncertainty contains the systematic uncertainties obtained from variations of the background normalization region, the choice of the background fit function and integration intervals, analog to the PCM method, as well as from variations of the minimum opening angle cut on the meson level.
The systematic uncertainty denoted as ``cluster description'' reflects the mismatch of the description of the clusterization process between data and MC simulations, giving rise to modified reconstruction efficiencies, which includes the following cluster related quantities: minimum energy, shower shape, number of cells and timing as well as variations of the energy thresholds used for the clusterization process.
Moreover, cell timing cut variations are also included in this category.
``Cluster energy calibration'' considers the systematic uncertainties due to non-linearity effects and the energy scale of clusters.
Different non-linearity schemes are used in this analysis from which this uncertainty is obtained.
Moreover, the energy scale uncertainty is determined by obtaining the residual differences of reconstructed meson mass values from data and MC simulations.
The systematic uncertainty induced by the charged particle veto on cluster level, introduced as ``general track matching'' in Sec. \ref{sec:reconstruction}, is determined by variations of the matching residuals.
The ``efficiency'' uncertainty reflects differences between MB MC generators for the calculation of reconstruction efficiencies. 
Moreover, it contains the uncertainty of the actual trigger turn-on, obtained by comparing the turn-on curves in data and MC.
The uncertainties from the determination of trigger rejection factors ($RF$) as well as from the pileup rejection by the SPD, which has a finite efficiency for pileup removal, are summarized with ``trigger normalization \& pileup''.

\begin{table}[bht]
  \begin{center}
   \scalebox{0.88}{
    \begin{tabular}{l||c|c|c||c|c|c||c|c}
      \hline
      \pT interval & \multicolumn{3}{c||}{$2.0-2.4$ GeV/$c$} & \multicolumn{3}{c||}{$5.0-6.0$ GeV/$c$} & \multicolumn{2}{c}{$18.0-20.0$ GeV/$c$} \\ \hline
      \multirow{2}{*}{Method} & \multirow{2}{*}{PCM} & PCM- & \multirow{2}{*}{EMC} & \multirow{2}{*}{PCM} & PCM- & \multirow{2}{*}{EMC} & PCM- & \multirow{2}{*}{EMC} \\ 
             &  & EMC & & & EMC & & EMC & \\ \hline\hline
      Signal extraction      & $5.9$ & $9.0$ & $9.3$ & $8.2$ & $7.5$ & $6.6$ & $11.2$ & $12.8$ \\
      PCM track rec.         & $1.5$ & $1.9$ &     - & $2.0$ & $2.4$ &     - & $3.8$  &     -  \\
      PCM electron PID       & $2.4$ & $1.9$ &     - & $2.2$ & $3.5$ &     - & $7.4$  &     -  \\
      PCM photon PID         & $3.6$ & $3.2$ &     - & $6.3$ & $3.6$ &     - & $9.0$  &     -  \\
      Cluster description    &     - & $3.5$ & $4.9$ &     - & $4.1$ & $5.1$ & $8.9$  & $5.5$  \\
      Cluster energy calib.  &     - & $3.4$ & $4.2$ &     - & $4.6$ & $4.2$ & $5.5$  & $4.5$  \\
      Track match to cluster &     - & $1.5$ & $3.9$ &     - & $1.8$ & $3.2$ & $6.1$  & $3.3$  \\
      Efficiency             &     - & $5.4$ & $4.5$ &     - & $9.8$ & $5.8$ & $10.5$ & $7.5$  \\ \hline
      Total syst. uncertainty& $7.5$ & $12.4$& $12.8$& $10.8$& $15.0$& $11.6$& $23.1$ & $16.8$ \\ 
      \hline\hline
      Statistical uncertainty& $10.2$& $12.2$& $5.4$ & $19.2$& $7.4$ & $2.7$ & $23.3$ & $19.0$ \\
      \hline\hline
      Combined stat. unc.    & \multicolumn{3}{c||}{5.5} & \multicolumn{3}{c||}{3.9} & \multicolumn{2}{c}{15.1} \\
      \hline
      Combined syst. unc.    & \multicolumn{3}{c||}{7.1} & \multicolumn{3}{c||}{8.7} & \multicolumn{2}{c}{13.0} \\ 
      \hline
    \end{tabular}}
    \caption{Summary of relative systematic uncertainties in percent for selected \pT bins for the determination of the $\eta/\pi^{0}$ ratio. The statistical uncertainties are given in addition to the total systematic uncertainties for each bin. Moreover, the combined statistical and systematic uncertainties are listed as well, see also explanations in caption of Tab. \ref{tab:table_system_Pi0}.}
    \label{tab:table_system_EtaToPi0}
  \end{center}
\end{table}
\renewcommand{\arraystretch}{1.0}

For the hybrid method PCM-EMCal, the same cut variations are performed as for the standalone methods.
However, given the fact that only one photon candidate of each system is used, most systematic uncertainties are found to be of different size or behavior, e.g. the minimum opening angle cut variations.
The ``track matching to cluster'' uncertainty reflects the ${\rm V^{0}}$-track to cluster matching, which is assessed by varying the matching residuals.

As indicated in Tab.~\ref{tab:table_system_EtaToPi0}, many uncertainties cancel for the $\eta/\pi^{0}$ ratio, such as the material-related systematics.
For the remaining categories, the respective uncertainties of the $\pi^0$ and $\eta$ measurements are added quadratically and canceled partially beforehand, if applicable.

\section{Results}
\label{sec:result}

The invariant differential cross sections of $\pi^{0}$ and $\eta$ production are obtained by
\begin{eqnarray}
  E\frac{{\rm d^{3}}\sigma^{pp\rightarrow\pi^{0}(\eta)+X}}{{\rm d}p^{3}} =
  \frac{1}{2\pi \pT}\, \frac{1}{\mathcal{L}_{\rm int}}\, \frac{1}{A\cdot \varepsilon_{\rm rec}}\,
  \frac{1}{Br_{\pi^{0}(\eta)\rightarrow\gamma\gamma}}\, \frac{N^{\pi^{0}(\eta)}-N^{\pi^{0}}_{\rm sec}}{\Delta y \Delta \pT},
  \label{eq:InvCrossSection}
\end{eqnarray}
where $N^{\pi^{0}(\eta)}$ is the number of reconstructed $\pi^{0}(\eta)$ meson in a given $\pT$ bin, 
$N^{\pi^{0}}_{\rm sec}$ represents the estimated number of secondary $\pi^{0}$ mesons,
$\mathcal{L}_{\rm int}$ is the integrated luminosity,
$A\cdot\varepsilon_{\rm rec}$ is the product of the geometrical acceptance and reconstruction efficiency, also referred to as $\epsilon$ in Fig.~\ref{fig:AcceptanceTimesEff},
$Br_{\pi^{0}(\eta)\rightarrow\gamma\gamma}$ is the branching ratio for the two-gamma decay channel and $\Delta y \Delta \pT$ is the bin width in rapidity and transverse momentum.
For the measurement of $\pi^{0}$ mesons by PCM, the out-of-bunch pileup correction has to be noted for completeness and to be applied as well.

The invariant differential cross sections are independently calculated for each method.
The final spectra are obtained by combining the results in the overlap regions using the BLUE method \cite{Lyons1988rp,Valassi2003mu,Lyons1986em,barlow1989statistics,Valassi2013bga},
properly taking into account the correlations of the systematic uncertainties of the different methods.
Possible statistical correlations between the measurements, for instance due to the conversions at small distances relative to the beam axis, are negligible due to the small conversion probability and the small likelihood of reconstructing the respective electron in the calorimeters leading to a meson candidate which finally ends up in the respective integration window.
As there are no common uncertainties present for PCM, EMCal and PHOS, all systematic uncertainties are considered to be completely uncorrelated in those cases. On the other hand, the correlations introduced by including the hybrid PCM-EMCal measurement have to be taken into account.
By construction, there are different numbers of conversion photons entering the two methods.
Thus, all systematic uncertainty sources from PCM are found to be partially correlated in the PCM-EMCal method.
Half of the size of the material budget uncertainty, for example, is assumed to be uncorrelated.
Furthermore, the uncorrelated systematic uncertainties from PCM-EMCal with respect to PCM are, with full size, all the calorimeter related uncertainties as well as trigger and efficiency uncertainties.

Due to finite bin widths of the measured production cross sections, the neutral meson spectra are shifted along the horizontal axis \cite{Lafferty:1995}.
All bin width corrections are of the order of $1\%$ and below.
In contrast, the reported $\eta/\pi^{0}$ ratios are shifted along the vertical axis, as otherwise the ratio could not be computed and the different measurements could not be combined.
The correction is below $1\%$ for $\pT>2$~\GeVc, but becomes significant for smaller momenta and rises to $8\%$ for the lowest bin.

\begin{figure}[htbp]
  \includegraphics[width=0.49\hsize]{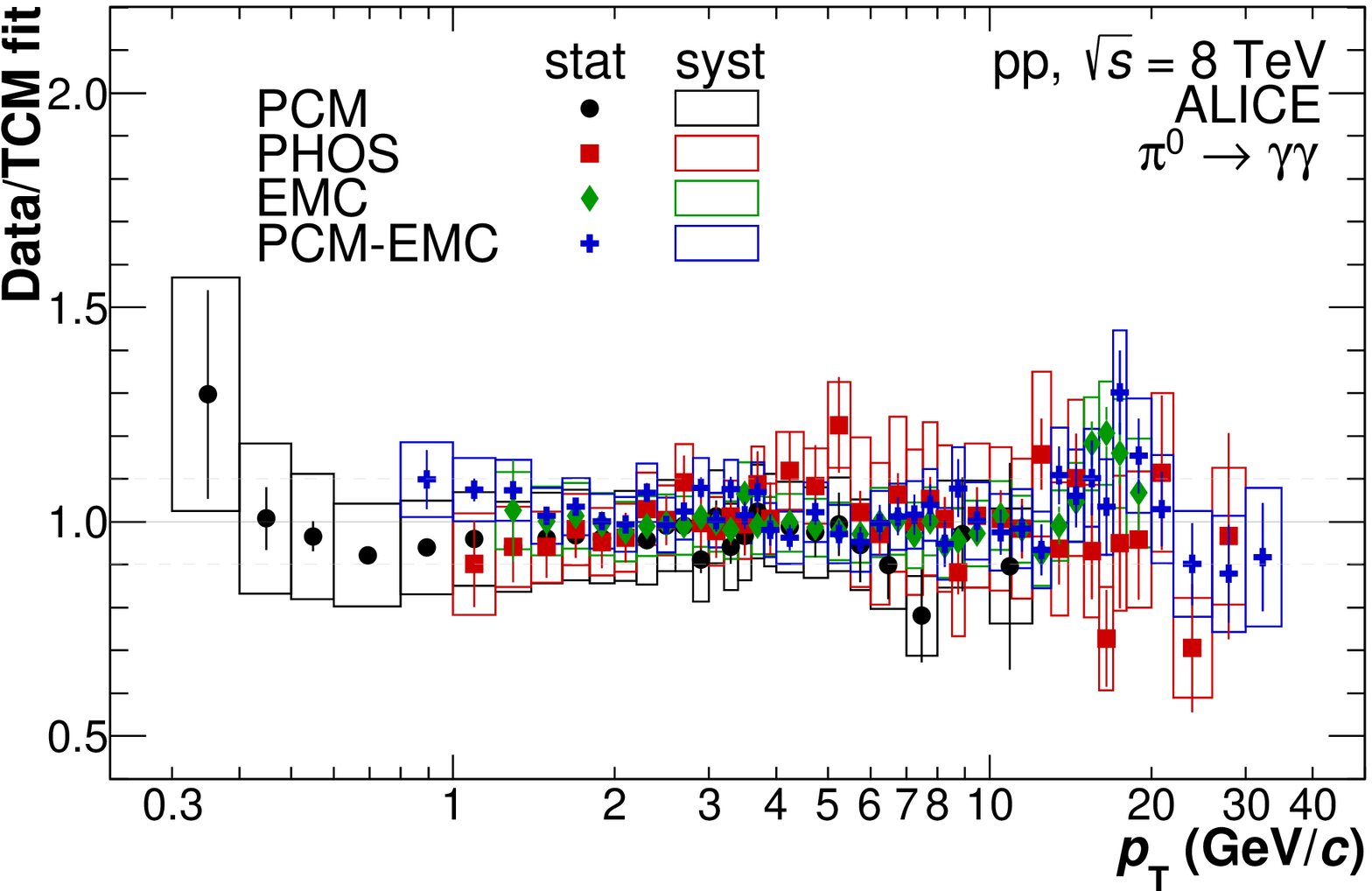}
  \hfil
  \includegraphics[width=0.49\hsize]{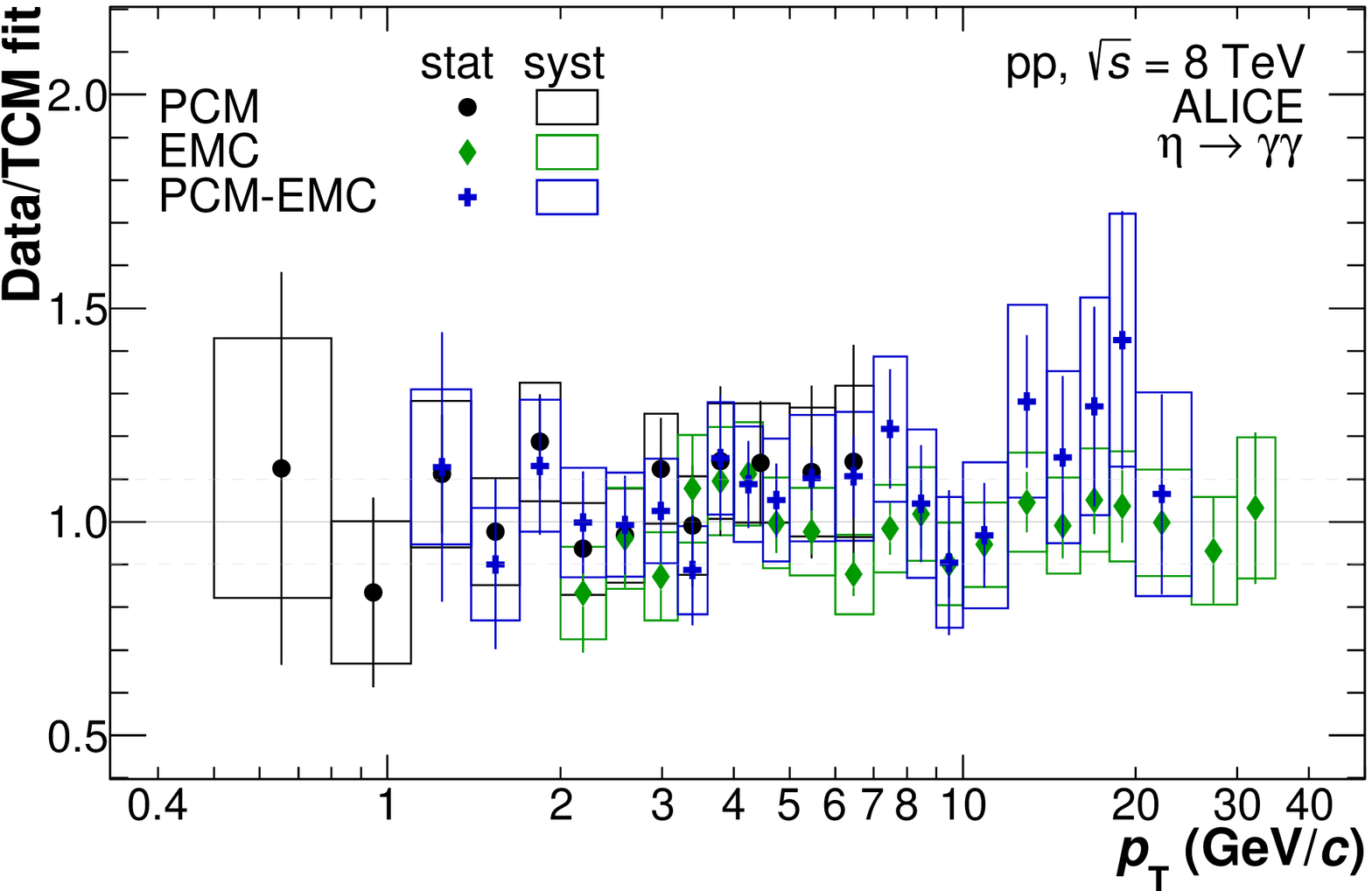}
  \caption{Ratios of the fully corrected $\pi^{0}$ (left) and $\eta$ (right) spectra for each reconstruction method to the TCM fit of the combined spectrum.}
  \label{fig:RatioOfIndividualMeasToCombFit_PP}
  \bigskip
\end{figure}

The combined invariant cross sections of inclusive $\pi^{0}$ and $\eta$ meson production cover transverse momentum ranges of $0.3<\pT<35$~\GeVc and $0.5<\pT<35$~\GeVc, respectively.
The total uncertainties of the measurements, obtained by quadratically adding the combined statistical and systematic uncertainties, are of the order of $5$\% for the $\pi^{0}$ and $10$\% for the $\eta$ meson for most of the \pT bins covered, increasing for lowest and highest momenta due to statistical limitations as well as systematic effects.
Both combined neutral meson spectra are fitted with a two-component model (TCM), proposed in Ref. \cite{Bylinkin:2015xya}, by using the total uncertainties for each \pT bin.
The functional form of the TCM is a combination of a Boltzmann component and a power-law part, which, in general, should be the dominant components at low and high \pT, respectively.
The fit function is able to reproduce the spectra over the full \pT range and is described as:
\begin{eqnarray}
  E\frac{{\rm d}^{3}\sigma}{{\rm d}p^{3}} = A_{\rm e}\,\exp\left(-E_{\rm T,kin}/T_{\rm e}\right)+A\left( 1+\frac{\pT^{2}}{T^{2}n}\right)^{-n},
  \label{eq:Bylinkin}
\end{eqnarray}
where $E_{\rm T,kin}=\sqrt{\pT^{2}+m^{2}}-m$ is the transverse kinematic energy with the meson rest mass $m$ and $A_{\rm e}$, $A$, $T_{e}$, $T$ as well as $n$ are the free parameters.
To compare the different methods, the ratios of spectra measured by each reconstruction method to the TCM fit of the combined spectrum are shown in Fig.~\ref{fig:RatioOfIndividualMeasToCombFit_PP}.
The vertical error bars represent the statistical uncertainties, whereas the boxes quantify the bin widths in horizontal direction and the systematic uncertainties in vertical direction.
All measurements agree within uncertainties over the full $\pT$ range.

\begin{figure}[ht!]
  \includegraphics[width=0.49\hsize]{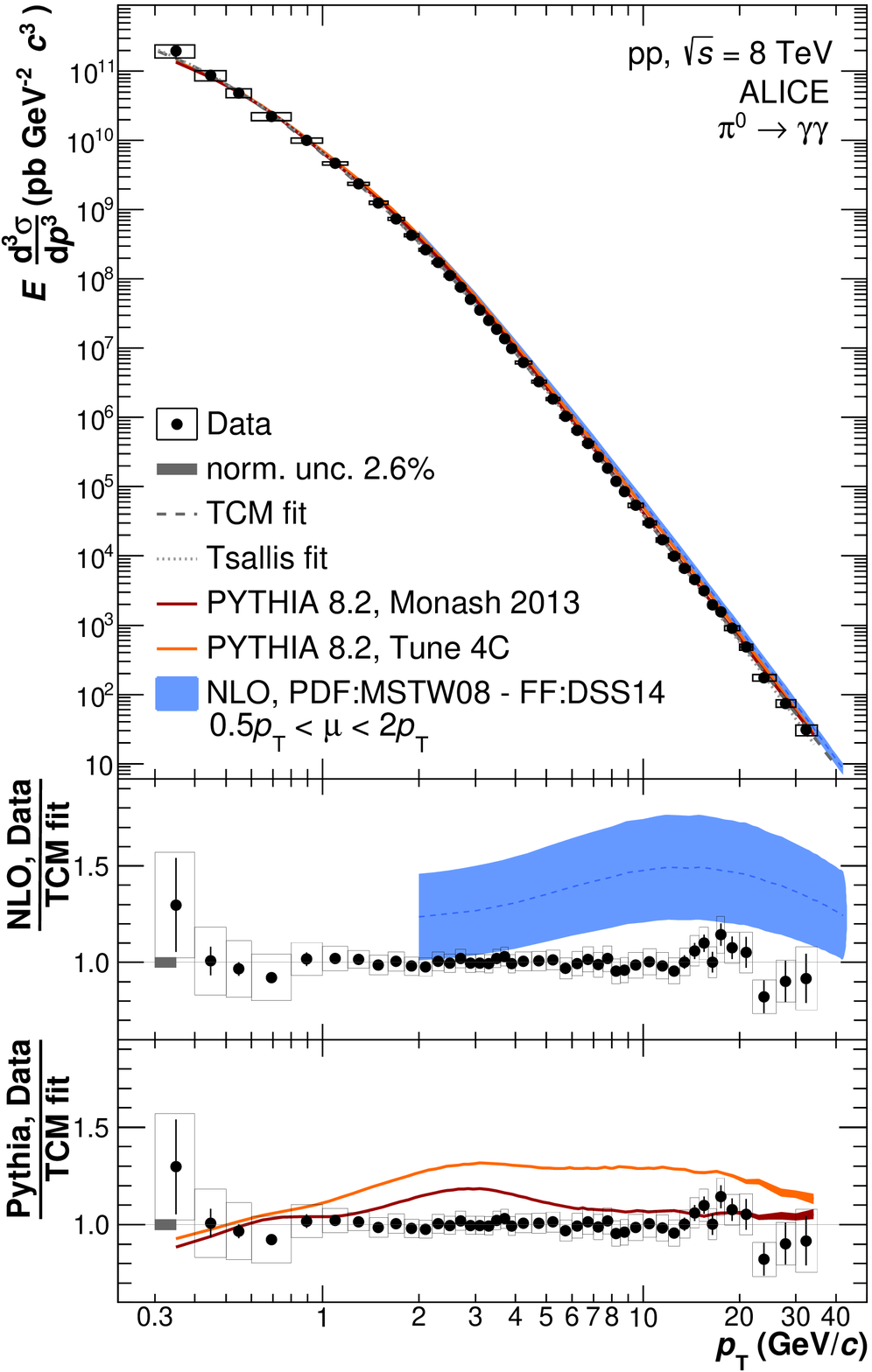}
  \hfil
  \includegraphics[width=0.49\hsize]{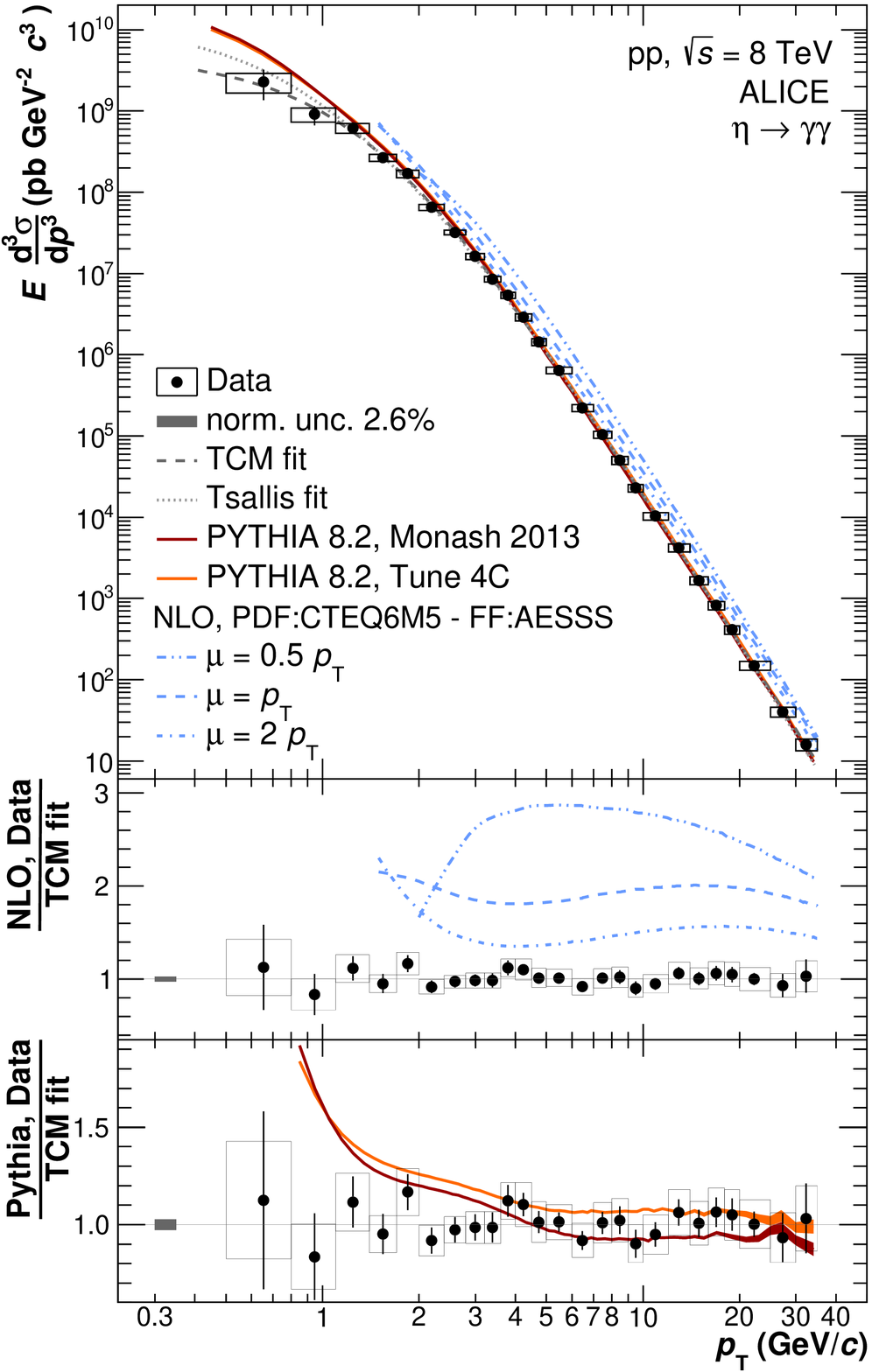}
  \caption{Invariant cross sections for neutral meson production are shown together with NLO pQCD predictions using PDFs MSTW08 (CTEQ6M5) with FFs DSS14 (AESSS) for $\pi^{0}$ ($\eta$) as well as PYTHIA8.210 calculations, for which two different tunes are available. The data points are fitted using a TCM fit, Eq.~\ref{eq:Bylinkin}, and a Tsallis fit, Eq.~\ref{eq:Tsallis}.}
  \label{fig:InvXSectionWithRatios_Paper}
  \bigskip
\end{figure}

The $\pi^{0}$ and $\eta$ meson cross sections are also fitted with a Tsallis function \cite{Tsallis1987eu}, which has been used in previous measurements of $\pi^0$ and $\eta$ meson production in pp collisions reported by ALICE \cite{Abelev:2012cn,Abelev:2014ypa}:
\begin{eqnarray}
  E\frac{{\rm d}^{3}\sigma}{{\rm d}p^{3}}
  =
  \frac{C}{2\pi}\,
  \frac{(n-1)(n-2)} {n T(n T+m(n-2))}  \left(1+\frac{\mT-m}{nT}\right)^{-n},
  \label{eq:Tsallis}
\end{eqnarray}
where $C$, $n$ and $T$ are free parameters of the fit with $m$ and $\mT$ being the rest as well as the transverse mass of the meson.
The fit parameters extracted from both the TCM and Tsallis fits are summarized in Tab.~\ref{tab:FitParam}.
The TCM is chosen as the standard fit function, since it better describes the spectra at low and high \pT than the Tsallis counterpart \cite{ALICE-PUBLIC-2017-009}.
This is also reflected in the smaller values obtained for the reduced $\chi^2_{\rm red}$ of the respective fits, which are also recorded in Tab.~\ref{tab:FitParam}.
These values are calculated without assuming any correlation of systematic uncertainties and are found to be rather small for both fits, as the total uncertainties of meson spectra are used for their calculation.
A direct comparison of TCM and Tsallis fits can be found in Fig.~\ref{fig:InvXSectionWithRatios_Paper}, where both fits are plotted, in addition to the measured spectra and theory calculations.

\renewcommand{\arraystretch}{1.3}
\begin{table}[h!]
  \begin{center}
  \scalebox{0.92}{
    \begin{tabular}{c||cccccc}
     \hline
     TCM & $A_{e}$ (pb~GeV$^{-2}c^{3}$) & $T_{e}$ (GeV) & $A$ (pb~GeV$^{-2}c^{3}$) & $T$ (GeV) & $n$ & $\chi^2_{\rm red}$ \\
     \hline\hline
     $\pi^{0}$ & (6.84$\pm$2.79)$\times10^{11}$ & 0.142$\pm$0.020 & (3.68$\pm$0.89)$\times10^{10}$ & 0.597$\pm$0.030 & 3.028$\pm$0.018 & 0.28 \\
     \hline
     $\eta$ & (1.62$\pm$4.35)$\times10^{9}$ & 0.229$\pm$0.203 & (2.89$\pm$1.81)$\times10^{9}$ & 0.810$\pm$0.103 & 3.043$\pm$0.045 & 0.33\\
     \hline
     \multicolumn{6}{c}{}\\
     \hline
     Tsallis & \multicolumn{2}{c}{$C$ (pb)} & $T$ (GeV) & \multicolumn{2}{c}{$n$} & $\chi^2_{\rm red}$ \\
     \hline\hline
     $\pi^{0}$ & \multicolumn{2}{c}{(2.46$\pm$0.18)$\times10^{11}$} & 0.121$\pm$0.004 & \multicolumn{2}{c}{6.465$\pm$0.042} & 0.57 \\
     \hline
     $\eta$ & \multicolumn{2}{c}{(1.56$\pm$0.19)$\times10^{10}$} & 0.221$\pm$0.012 & \multicolumn{2}{c}{6.560$\pm$0.113} & 0.59\\
     \hline
    \end{tabular}}
    \caption{Parameters of the fits to the $\pi^{0}$ and $\eta$ invariant differential cross sections using the TCM fit \cite{Bylinkin:2015xya} from Eq. \ref{eq:Bylinkin} as well as using a Tsallis fit \cite{Tsallis1987eu} from Eq.~\ref{eq:Tsallis}.}
    \label{tab:FitParam}
  \end{center}
\end{table}
\renewcommand{\arraystretch}{1.0}

The measured invariant differential cross sections are compared with NLO pQCD calculations \cite{deFlorian:2014xna,Aidala:2010bn} using MSTW08 (PDF) \cite{Martin:2009iq} with DSS14 (FF) \cite{deFlorian:2014xna} for $\pi^{0}$ and
CTEQ6M5 (PDF) \cite{1126-6708-2007-02-053} with AESSS (FF) \cite{Aidala:2010bn} for the $\eta$ meson.
The same factorization scale value, $\mu$, ($0.5\pT < \mu < 2\pT$) is chosen for the factorization, renormalisation and fragmentation scales used in the NLO pQCD calculations.
For the $\pi^{0}$ the NLO PDF, pQCD and FF combination describes the RHIC data rather well \cite{Adare:2015ozj}, whereas for $\sqrt{s}=2.76$~TeV pQCD overpredicts ALICE data by 30\% at moderate \pT and agrees at higher \pT \cite{Acharya:2017hyu}. 
The ratios of data and NLO pQCD predictions to the TCM fits of neutral meson spectra are shown in Fig.~\ref{fig:InvXSectionWithRatios_Paper}. 
The largest uncertainty of the NLO pQCD calculation is due to the choice of $\mu$. 
For all $\mu$ values, these calculations overestimate the measured data for both $\pi^{0}$ and $\eta$ mesons.
This is also observed for meson measurements at $\sqrt{s} = 2.76$~TeV by ALICE \cite{Acharya:2017hyu}, although better description of data is achieved for $\mu=2\pT$, for which calculations are above data by $10-40\%$ depending on \pT.
It has to be noted that FF uncertainties of NLO pQCD calculations have been considerably reduced after including the published $\pi^{0}$ measurement at $\sqrt{s}=7$~TeV \cite{Abelev:2012cn} for DSS14.
Including precise new data for $\eta$ meson production measured at $\sqrt{s}=2.76$, $7$ and $8$~TeV \cite{Acharya:2017hyu,Abelev:2012cn} will also help to considerably reduce NLO pQCD uncertainty bands in that case.
In addition, the reported neutral meson measurements at $\sqrt{s}=8$~TeV are compared to PYTHIA8.210 \cite{Sjostrand:2014zea} references; Tune 4C \cite{Corke:2010yf} and Monash~2013 tune \cite{Skands:2014pea}.
To enable a proper comparison of the PYTHIA tunes with the measured neutral meson spectra, $\pi^{0}$ from decays of long-living strange particles ($K^0_S$, $\Lambda$, $\Sigma$ and $\Xi$) are excluded.
The Tune 4C calculation is about $30\%$ above the $\pi^{0}$ measurement for $\pT>1.5$~\GeVc.
In contrast, the Monash~2013 tune reproduces the $\pi^{0}$ spectrum within 10\% for almost the complete transverse momentum range, although both tunes are not able to describe the shape of the measured spectrum indicated by the bump at approximately 3~\GeVc.
Concerning the $\eta$ meson, both tunes reproduce the measured spectrum for $\pT>1.5$~\GeVc within uncertainties.
At lower momenta $\pT<1.5$~\GeVc, both tunes follow the same trend and deviate significantly in magnitude and shape from data.
The tuning parameters of the soft QCD part of PYTHIA apparently fail to describe the measured $\eta$ meson spectrum below $\pT<1.5$~\GeVc, whereas there is further tension up to $\pT\approx3.5$~\GeVc.
On the other hand, both PYTHIA tunes are consistent within uncertainties with the measured $\pi^{0}$ spectrum for the low transverse momentum interval $0.3<\pT<1.5$~\GeVc.

The mean transverse momenta, $\langle\pT\rangle$, are determined for the neutral meson spectra shown in Fig.~\ref{fig:InvXSectionWithRatios_Paper}.
Three different fit functions are used in this context: a TCM, Eq. \ref{eq:Bylinkin}, a Tsallis, Eq. \ref{eq:Tsallis}, and a modified Hagedorn \cite{PhysRevC.81.034911} fit that is used as the default fit function, since it yields the best agreement with data at lowest \pT measured \cite{ALICE-PUBLIC-2017-009}.
The obtained values for $\pi^{0}$ and $\eta$ mesons are listed in Tab. \ref{tab:meanPtAndYields}, where statistical and systematic uncertainties are quoted.
The additional uncertainty term denoted with ``fit sys'' reflects the choice of the fitting function.
Moreover, the introduced fit functions are also used to calculate the integrated yields, $\text{d}N/\text{d}y|_{y\,\,\approx\,\,0}$, for both neutral mesons in inelastic events.
The cross section for inelastic pp collisions at $\sqrt{s}=8$~TeV, $\sigma_{\rm INEL} = 74.7 \pm 1.7$~mb \cite{PhysRevLett.111.012001}, is used for this purpose.
The obtained yields are given in Tab. \ref{tab:meanPtAndYields}, which are based on extrapolation fractions, $F_{\rm extpol}$, of about 45\% for the $\pi^{0}$ and about 34\% for the $\eta$ meson.
Additionally, the integrated $\eta/\pi^{0}$ ratio is estimated and can be found in Tab. \ref{tab:meanPtAndYields} as well.
For the recent paper by ALICE on neutral meson production in pp collisions at $\sqrt{s}=2.76$~TeV \cite{Acharya:2017hyu}, the mean \pT as well as the integrated yields are also calculated for the reported spectra, which are furthermore added to Tab. \ref{tab:meanPtAndYields}.
The inelastic pp cross section at $\sqrt{s}=2.76$~TeV, quoted in Ref. \cite{Acharya:2017hyu} as well, is used to calculate the integrated yields which include extrapolation fractions of about 59\% for the $\pi^{0}$ and about 52\% for the $\eta$ meson.
The obtained values for $\langle\pT\rangle$ and $\text{d}N/\text{d}y|_{y\,\,\approx\,\,0}$ for both neutral mesons are compared with measurements of average transverse momenta of charged particles \cite{Abelev:2013bla} and with results concerning charged-particle multiplicity \cite{Adam:2015gka}.
Due to a large extrapolation fraction of the $\pi^{0}$ and $\eta$ meson spectra with respect to charged particles and the given systematics for the lowest transverse momenta, the uncertainties of $\langle\pT\rangle$ and $\text{d}N/\text{d}y|_{y\,\,\approx\,\,0}$ are found to be larger.
Hence, the integrated $\eta/\pi^{0}$ ratios are also affected.
Nevertheless, all values quoted in this paragraph are consistent within experimental uncertainties with the results from charged particle measurements~\cite{Abelev:2014laa, Adam:2015qaa}.
Within their substantial uncertainties, the $\eta/\pi^{0}$ ratios at both pp energies are found to be consistent as well.

\renewcommand{\arraystretch}{1.3}
\begin{table}[h]
  \begin{center}
  \hspace*{-0.35cm}
  \scalebox{0.93}{
    \begin{tabular}{c||c|c|c}
     \hline
     $\sqrt{s}=8$~TeV & $\langle\pT\rangle$~(\GeVc) & $\text{d}N/\text{d}y|_{y\,\,\approx\,\,0}$ & $F_{\rm extpol}$ \\
     \hline\hline
     $\pi^{0}$ & $0.431\pm0.006_{\mbox{\tiny (stat)}}\pm0.020_{\mbox{\tiny (sys)}}\pm0.012_{\mbox{\tiny (fit sys)}}$ & $3.252\pm0.128_{\mbox{\tiny (stat)}}\pm0.918_{\mbox{\tiny (sys)}}\pm0.146_{\mbox{\tiny (fit sys)}}$ & 45\%\\
     \hline
     $\eta$ & $0.929\pm0.110_{\mbox{\tiny (stat)}}\pm0.126_{\mbox{\tiny (sys)}}\pm0.085_{\mbox{\tiny (fit sys)}}$ & $0.164\pm0.033_{\mbox{\tiny (stat)}}\pm0.052_{\mbox{\tiny (sys)}}\pm0.023_{\mbox{\tiny (fit sys)}}$ & 34\% \\
     \hline
     \hline
     $\eta/\pi^{0}$ & \multicolumn{2}{c|}{$0.050\pm0.010_{\mbox{\tiny (stat)}}\pm0.022_{\mbox{\tiny (sys)}}\pm0.008_{\mbox{\tiny (fit sys)}}$} & \\
     \cline{1-3}
     \multicolumn{4}{c}{}\\
     \hline
     $\sqrt{s}=2.76$~TeV & $\langle\pT\rangle$~(\GeVc) & $\text{d}N/\text{d}y|_{y\,\,\approx\,\,0}$ & $F_{\rm extpol}$ \\
     \hline\hline
     $\pi^{0}$ & $0.451\pm0.008_{\mbox{\tiny (stat)}}\pm0.014_{\mbox{\tiny (sys)}}\pm0.152_{\mbox{\tiny (fit sys)}}$ & $1.803\pm0.058_{\mbox{\tiny (stat)}}\pm0.352_{\mbox{\tiny (sys)}}\pm0.646_{\mbox{\tiny (fit sys)}}$ & 59\%\\
     \hline
     $\eta$ & $0.647\pm0.068_{\mbox{\tiny (stat)}}\pm0.040_{\mbox{\tiny (sys)}}\pm0.140_{\mbox{\tiny (fit sys)}}$ & $0.250\pm0.050_{\mbox{\tiny (stat)}}\pm0.052_{\mbox{\tiny (sys)}}\pm0.063_{\mbox{\tiny (fit sys)}}$ & 52\% \\
     \hline
     \hline
     $\eta/\pi^{0}$ & \multicolumn{2}{c|}{$0.139\pm0.028_{\mbox{\tiny (stat)}}\pm0.040_{\mbox{\tiny (sys)}}\pm0.061_{\mbox{\tiny (fit sys)}}$} & \\
     \cline{1-3}
    \end{tabular}}
    \caption{The mean transverse momenta, $\langle\pT\rangle$, and integrated yields, $\text{d}N/\text{d}y|_{y\,\,\approx\,\,0}$, for ALICE measurements of $\pi^{0}$ and $\eta$ mesons at $\sqrt{s}=2.76$ and 8~TeV are summarized \cite{ALICE-PUBLIC-2017-009}. It has to be noted that the uncertainties from the measurements of the inelastic cross sections are not included for the given numbers, which are $^{+3.9\%}_{-6.4\%}(model)\pm2.0(lumi)$\% for $\sqrt{s}=2.76$~TeV \cite{Abelev:2012sea} and $\pm2.3$\% for 8~TeV \cite{PhysRevLett.111.012001}. Moreover, the integrated $\eta/\pi^{0}$ ratios are quoted for the different energies.}
    \label{tab:meanPtAndYields}
  \end{center}
\end{table}
\renewcommand{\arraystretch}{1.0}

\begin{figure}[hbt]
  \includegraphics[width=0.49\hsize]{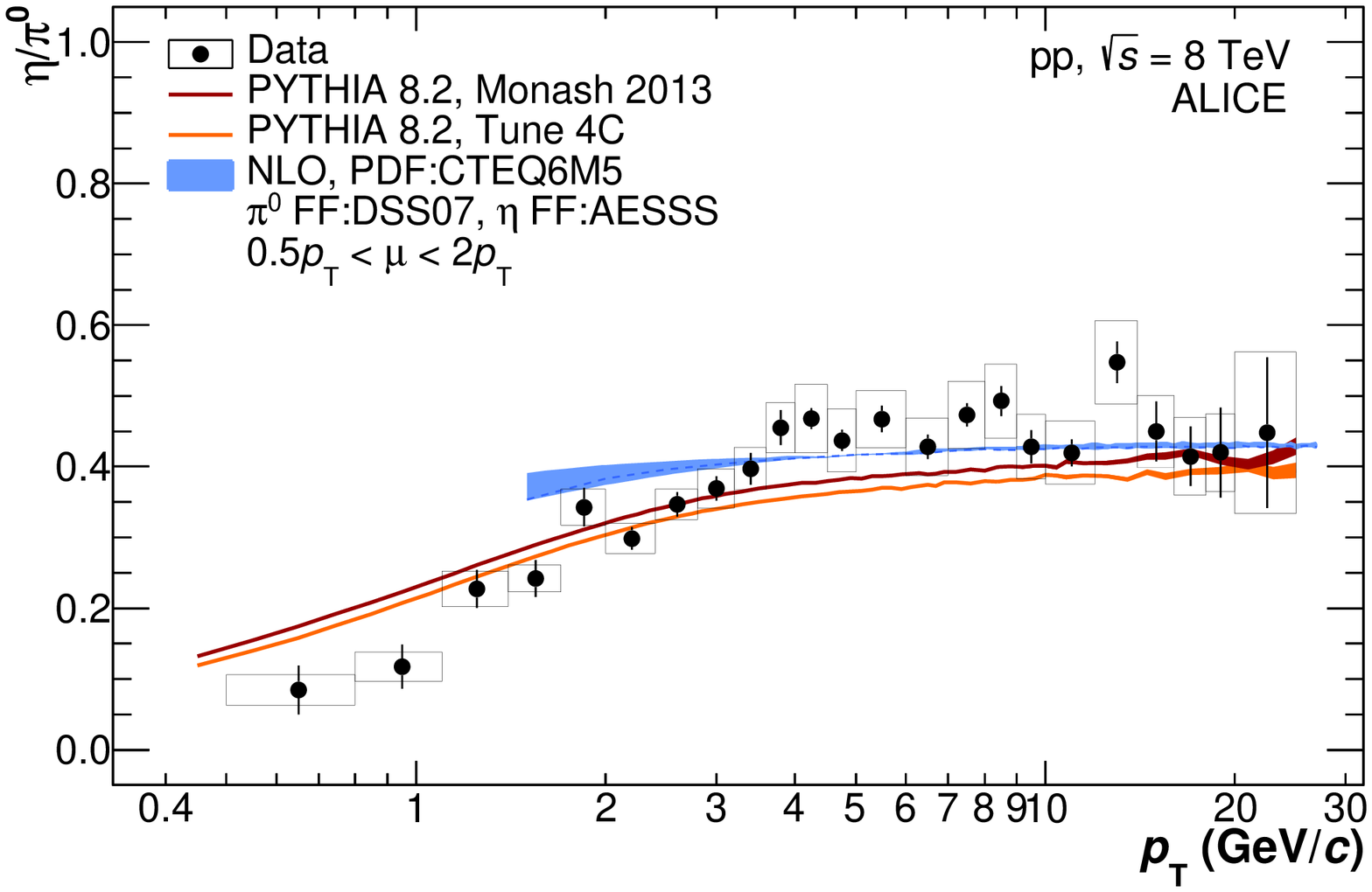}
  \hfil
  \includegraphics[width=0.49\hsize]{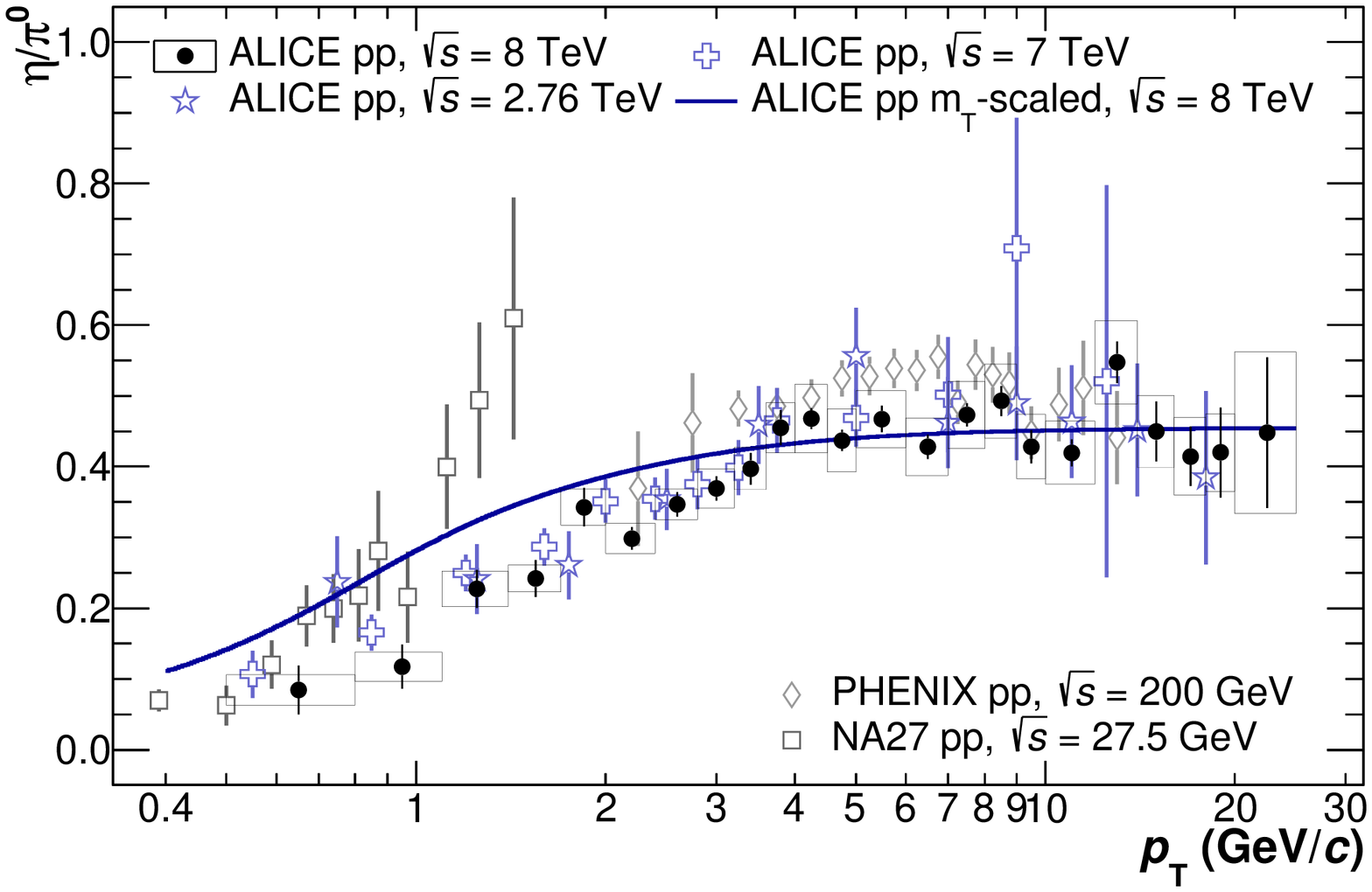}
  \caption{Left: $\eta/\pi^{0}$ ratio compared to NLO pQCD predictions using PDF CTEQ6M5 and FFs DSS07 for the $\pi^0$ and AESSS for the $\eta$, and PYTHIA8.210 calculations using Tune~4C and Monash~2013 tune. The total uncertainties of the measured $\eta/\pi^{0}$ ratio are of the order of $10$\% for most of the \pT bins covered, increasing for lower and higher momenta due to limited statistics as well as systematic effects. Right: Comparison of the $\eta/\pi^{0}$ ratio to related, previous ALICE measurements as well as other experiments at lower collision energies, for which total uncertainties are drawn. Furthermore, a comparison to the $\eta/\pi^{0}$ ratio obtained with \mT scaling is added.}
  \label{fig:EtaToPi0_Theory_Paper}
  \bigskip
\end{figure}

Both meson spectra, which are shown in Fig.~\ref{fig:InvXSectionWithRatios_Paper}, exhibit a similar power-law behavior, $E{\rm d}^3\sigma/{\rm d}p^3 \propto \pT^{-n}$, with $n_{\pi^{0}} = 5.936\pm0.012\mbox{(stat)}\pm0.023\mbox{(sys)}$ and
$n_{\eta} = 5.930\pm0.029\mbox{(stat)}\pm0.044\mbox{(sys)}$ for high momenta of $\pT>3.5$~\GeVc.
This is also reflected in the $\eta/\pi^{0}$ ratio which is shown in Fig.~\ref{fig:EtaToPi0_Theory_Paper}.
The ratio is flat for $\pT>3.5$~\GeVc with a constant value of $C^{\eta/\pi^0}=0.455\pm0.006\mbox{(stat)}\pm0.014\mbox{(sys)}$. 
Despite of the inability of NLO calculations to describe individual $\pi^0$ and $\eta$ meson spectra, the $\eta/\pi^{0}$ ratio is reproduced fairly well, as it can be followed from left part of Fig.~\ref{fig:EtaToPi0_Theory_Paper}.
It has to be noted that a different FF for the $\pi^0$ is used to compile the theory curve, namely DSS07, since there is no recent calculation for the $\eta$ meson available which could be compared to the recent DSS14 $\pi^0$ prediction.
The agreement of pQCD calculations with the data can be viewed as an indication that the $\eta/\pi^{0}$ ratio is driven by the $\pi^{0}$ and $\eta$ meson FFs in the factorized picture of pQCD.
A comparison of the reported $\eta/\pi^{0}$ ratio to the different PYTHIA tunes indicates an agreement within uncertainties down to $\pT\approx 1.5$~\GeVc, although the shape, as well as the ratio, cannot be fully reproduced below $\pT<1.5$~\GeVc due to the already mentioned deviations of PYTHIA tunes from data in this region.

The validity of \mT scaling is tested by means of the $\eta/\pi^{0}$ ratio.
For this purpose, the TCM parameterization of the measured $\pi^0$ spectrum, given in Tab. \ref{tab:FitParam}, is used to obtain the $\eta$ spectrum via the application of \mT scaling by replacing the $\pi^0$ mass with the $\eta$ mass and using the normalization ratio $C^{\eta}/C^{\pi^0}=0.455$.
From these two spectra, the $\eta/\pi^{0}$ ratio is constructed, plotted as blue curve in the right part of Fig.~\ref{fig:EtaToPi0_Theory_Paper}.
The measured $\eta/\pi^0$ ratio is consistent with the \mT scaling prediction (blue curve) above $\pT>3.5$~\GeVc.
However, for smaller transverse momenta of $\pT<3.5$~\GeVc the ratio of the measured $\eta/\pi^0$ ratio over the $\eta/\pi^0$ ratio obtained with \mT scaling, which can be found in Ref. \cite{ALICE-PUBLIC-2017-009}, constantly decreases and reaches about 45\% at around $1$~\GeVc.
For the region below $3.5$~\GeVc, \mT scaling is observed to be clearly broken with a significance of 6.2$\sigma$.
Given this observation, the measured $\eta/\pi^0$ ratios in pp collisions at $\sqrt{s}=2.76$~TeV and $7$~TeV, previously reported by ALICE \cite{Acharya:2017hyu,Abelev:2012cn}, are re-evaluated.
Whereas there is indication for a \mT scaling violation with 2.1$\sigma$ for $2.76$~TeV, we also observe a significant disagreement of 5.7$\sigma$ for $7$~TeV.
Hence, both ratios are found to be consistent with our observation at $8$~TeV.
Furthermore, both $\eta/\pi^0$ ratios are fitted with a constant for $\pT>3.5$~\GeVc, yielding values of $C^{\eta/\pi^0}=0.474\pm0.015\mbox{(stat)}\pm0.024\mbox{(sys)}$ for $2.76$~TeV and $C^{\eta/\pi^0}=0.476\pm0.020\mbox{(stat)}\pm0.020\mbox{(sys)}$ for $7$~TeV. 
They are consistent within uncertainties with the measured $\eta/\pi^0$ ratio at $8$~TeV for the given \pT range.
Therefore, all three ALICE measurements are simultaneously fitted with a constant for $\pT>3.5$~\GeVc in order to obtain a combined value of  $C^{\eta/\pi^0}=0.459\pm0.006\mbox{(stat)}\pm0.011\mbox{(sys)}$.
For the region $\pT<3.5$~\GeVc, all collision energies covered by ALICE also agree within experimental uncertainties, so that $\eta/\pi^0$ ratios may be claimed to be consistent within accuracy for ALICE measurements in pp collisions at $\sqrt{s}=2.76$, $7$ and $8$~TeV.

Before the LHC era, the precision of $\eta/\pi^0$ measurements was not sufficient to probe \mT scaling over broad ranges of \pT with high statistics.
PHENIX and NA27 provide the $\eta/\pi^0$ ratio with highest accuracy at high and low \pT and therefore are compared to the reported measurement.
PHENIX measurements for pp collisions at $\sqrt{s}=200$~GeV are available only for \pT region $>2.25$~\GeVc \cite{Adare:2010cy}, where $\pi^0$ and $\eta$ spectra are already described by \mT scaling.
However, PHENIX notably does not apply any secondary $\pi^0$ correction concerning weak decays, which further has to be taken into account when comparing with data points from ALICE.
Measurements of $\pi^0$ and $\eta$ spectra in pp collisions at $\sqrt{s}=27.5$~GeV from NA27 \cite{AguilarBenitez:1991yy} are used to obtain the $\eta/\pi^0$ ratio in the \pT range of $0.4<\pT<1.6$~\GeVc.
The paper does not mention a secondary correction of $\pi^{0}$ spectrum; however, it cannot significantly change the conclusions to be drawn from the measurement.
Although the NA27 measurement provides the world's most precise published data points for the $\eta/\pi^0$ ratio at low $\pT<2.0$~\GeVc in the pre-LHC era for pp collisions, it is not conclusive concerning \mT scaling violation.
The first NA27 points at $\pT<1$~\GeVc are consistent with both the \mT scaling curve and the new data from pp collisions at $\sqrt{s}=2.76$, $7$ and $8$~TeV within uncertainties, whereas for $\pT>1$~\GeVc the results of NA27 show a tendency to be above the \mT scaling prediction, although uncertainties become significant.
A clearer confirmation of the \mT scaling at low \pT is observed for other particle species, such as kaons, $\phi$ and $J/\psi$ in pp collisions at $\sqrt{s}=200$~GeV \cite{Khandai:2011cf}.
Whether the magnitude of \mT scaling violation depends on the collision energy can be clarified in future by ongoing analysis of hadron spectra measurements in pp collisions at $\sqrt{s}=13$~TeV delivered by the LHC.




\section{Conclusion}
\label{sec:conclusion}
The invariant differential cross sections of $\pi^{0}$ and $\eta$ mesons in pp collisions at $\sqrt{s}$~=~8~TeV have been measured at mid-rapidity over a wide $\pT$ range by the ALICE experiment, using four different reconstruction methods for the $\pi^{0}$, and three for the $\eta$ meson.  
NLO pQCD calculations with MSTW08 (PDF) with DSS14 (FF) for the $\pi^{0}$ and CTEQ6M5 (PDF) with AESSS (FF) for the $\eta$ meson overestimate the measured spectra of both neutral mesons.
This discrepancy is also reported for pp collisions at $\sqrt{s}=2.76$~TeV by ALICE.
However, the ratio of $\eta/\pi^{0}$ is reproduced by NLO pQCD calculations within uncertainties, which is a test for the FFs in the factorized picture of pQCD.
The prediction from PYTHIA8.2 Tune 4C describes the $\eta$ spectrum within uncertainties for $\pT>1.5$~\GeVc, but it is about $30\%$ larger than the measured $\pi^{0}$ production cross section.
On the other hand, the Monash~2013 tune agrees with the reported neutral meson measurements within 10\% for $\pT>1.5$~\GeVc.
Both PYTHIA8.2 tunes are able to reproduce the $\pi^{0}$ spectrum below $\pT<1.5$~\GeVc within uncertainties, but fail to describe the $\eta$ spectrum in this region.
The $\eta/\pi^0$ ratio is described by \mT scaling for $\pT>3.5$~\GeVc, whereas a deviation from this empirical scaling law is found for $\pT<3.5$~\GeVc with a significance of $6.2\sigma$.
Within experimental uncertainties, the $\eta/\pi^0$ ratios measured by NA27, PHENIX and ALICE are in agreement for the covered transverse momentum intervals of each measurement, representing pp collisions starting at center of mass energies of $\sqrt{s}=27.5$~GeV up to $\sqrt{s}=8$~TeV.


%
%

\newenvironment{acknowledgement}{\relax}{\relax}
\begin{acknowledgement}
\section*{Acknowledgements}
We thank Werner Vogelsang and Marco Stratmann for providing the pQCD calculations.

The ALICE Collaboration would like to thank all its engineers and technicians for their invaluable contributions to the construction of the experiment and the CERN accelerator teams for the outstanding performance of the LHC complex.
The ALICE Collaboration gratefully acknowledges the resources and support provided by all Grid centres and the Worldwide LHC Computing Grid (WLCG) collaboration.
The ALICE Collaboration acknowledges the following funding agencies for their support in building and running the ALICE detector:
A. I. Alikhanyan National Science Laboratory (Yerevan Physics Institute) Foundation (ANSL), State Committee of Science and World Federation of Scientists (WFS), Armenia;
Austrian Academy of Sciences and Nationalstiftung f\"{u}r Forschung, Technologie und Entwicklung, Austria;
Ministry of Communications and High Technologies, National Nuclear Research Center, Azerbaijan;
Conselho Nacional de Desenvolvimento Cient\'{\i}fico e Tecnol\'{o}gico (CNPq), Universidade Federal do Rio Grande do Sul (UFRGS), Financiadora de Estudos e Projetos (Finep) and Funda\c{c}\~{a}o de Amparo \`{a} Pesquisa do Estado de S\~{a}o Paulo (FAPESP), Brazil;
Ministry of Science \& Technology of China (MSTC), National Natural Science Foundation of China (NSFC) and Ministry of Education of China (MOEC) , China;
Ministry of Science, Education and Sport and Croatian Science Foundation, Croatia;
Ministry of Education, Youth and Sports of the Czech Republic, Czech Republic;
The Danish Council for Independent Research | Natural Sciences, the Carlsberg Foundation and Danish National Research Foundation (DNRF), Denmark;
Helsinki Institute of Physics (HIP), Finland;
Commissariat \`{a} l'Energie Atomique (CEA) and Institut National de Physique Nucl\'{e}aire et de Physique des Particules (IN2P3) and Centre National de la Recherche Scientifique (CNRS), France;
Bundesministerium f\"{u}r Bildung, Wissenschaft, Forschung und Technologie (BMBF) and GSI Helmholtzzentrum f\"{u}r Schwerionenforschung GmbH, Germany;
General Secretariat for Research and Technology, Ministry of Education, Research and Religions, Greece;
National Research, Development and Innovation Office, Hungary;
Department of Atomic Energy Government of India (DAE) and Council of Scientific and Industrial Research (CSIR), New Delhi, India;
Indonesian Institute of Science, Indonesia;
Centro Fermi - Museo Storico della Fisica e Centro Studi e Ricerche Enrico Fermi and Istituto Nazionale di Fisica Nucleare (INFN), Italy;
Institute for Innovative Science and Technology , Nagasaki Institute of Applied Science (IIST), Japan Society for the Promotion of Science (JSPS) KAKENHI and Japanese Ministry of Education, Culture, Sports, Science and Technology (MEXT), Japan;
Consejo Nacional de Ciencia (CONACYT) y Tecnolog\'{i}a, through Fondo de Cooperaci\'{o}n Internacional en Ciencia y Tecnolog\'{i}a (FONCICYT) and Direcci\'{o}n General de Asuntos del Personal Academico (DGAPA), Mexico;
Nederlandse Organisatie voor Wetenschappelijk Onderzoek (NWO), Netherlands;
The Research Council of Norway, Norway;
Commission on Science and Technology for Sustainable Development in the South (COMSATS), Pakistan;
Pontificia Universidad Cat\'{o}lica del Per\'{u}, Peru;
Ministry of Science and Higher Education and National Science Centre, Poland;
Korea Institute of Science and Technology Information and National Research Foundation of Korea (NRF), Republic of Korea;
Ministry of Education and Scientific Research, Institute of Atomic Physics and Romanian National Agency for Science, Technology and Innovation, Romania;
Joint Institute for Nuclear Research (JINR), Ministry of Education and Science of the Russian Federation and National Research Centre Kurchatov Institute, Russia;
Ministry of Education, Science, Research and Sport of the Slovak Republic, Slovakia;
National Research Foundation of South Africa, South Africa;
Centro de Aplicaciones Tecnol\'{o}gicas y Desarrollo Nuclear (CEADEN), Cubaenerg\'{\i}a, Cuba, Ministerio de Ciencia e Innovacion and Centro de Investigaciones Energ\'{e}ticas, Medioambientales y Tecnol\'{o}gicas (CIEMAT), Spain;
Swedish Research Council (VR) and Knut \& Alice Wallenberg Foundation (KAW), Sweden;
European Organization for Nuclear Research, Switzerland;
National Science and Technology Development Agency (NSDTA), Suranaree University of Technology (SUT) and Office of the Higher Education Commission under NRU project of Thailand, Thailand;
Turkish Atomic Energy Agency (TAEK), Turkey;
National Academy of  Sciences of Ukraine, Ukraine;
Science and Technology Facilities Council (STFC), United Kingdom;
National Science Foundation of the United States of America (NSF) and United States Department of Energy, Office of Nuclear Physics (DOE NP), United States of America.    
\end{acknowledgement}

\bibliographystyle{utphys}   
\bibliography{biblio}

\providecommand{\href}[2]{#2}\begingroup\raggedright\begin{thebibliography}{10}

\bibitem{Gross:1973ju}
D.~J. Gross and F.~Wilczek, ``{Asymptotically Free Gauge Theories. 1},''
\href{http://dx.doi.org/10.1103/PhysRevD.8.3633}{{\em Phys. Rev.} {\bfseries
  D8} (1973) 3633--3652}.

\bibitem{Evans:2008zzb}
L.~Evans and P.~Bryant, ``{LHC Machine},''
\href{http://dx.doi.org/10.1088/1748-0221/3/08/S08001}{{\em JINST} {\bfseries
  3} (2008) S08001}.

\bibitem{Abelev:2012cn}
{\bfseries ALICE} Collaboration, B.~Abelev {\em et~al.}, ``{Neutral pion and
  $\eta$ meson production in proton-proton collisions at $\sqrt{s}=0.9$ TeV and
  $\sqrt{s}=7$ TeV},''
  \href{http://dx.doi.org/10.1016/j.physletb.2012.09.015}{{\em Phys. Lett.}
  {\bfseries B717} (2012) 162--172},
\href{http://arxiv.org/abs/1205.5724}{{\ttfamily arXiv:1205.5724 [hep-ex]}}.

\bibitem{Abelev:2014ypa}
{\bfseries ALICE} Collaboration, B.~B. Abelev {\em et~al.}, ``{Neutral pion
  production at midrapidity in pp and Pb-Pb collisions at $\sqrt{s}$= 2.76
  TeV},'' \href{http://dx.doi.org/10.1140/epjc/s10052-014-3108-8}{{\em Eur.
  Phys. J.} {\bfseries C74} (2014) 3108},
\href{http://arxiv.org/abs/1405.3794}{{\ttfamily arXiv:1405.3794 [nucl-ex]}}.

\bibitem{dEnterria:2013sgr}
D.~d'Enterria, K.~J. Eskola, I.~Helenius, and H.~Paukkunen, ``{Confronting
  current NLO parton fragmentation functions with inclusive charged-particle
  spectra at hadron colliders},''
  \href{http://dx.doi.org/10.1016/j.nuclphysb.2014.04.006}{{\em Nucl. Phys.}
  {\bfseries B883} (2014) 615--628},
\href{http://arxiv.org/abs/1311.1415}{{\ttfamily arXiv:1311.1415 [hep-ph]}}.

\bibitem{deFlorian:2007aj}
D.~de~Florian, R.~Sassot, and M.~Stratmann, ``{Global analysis of fragmentation
  functions for pions and kaons and their uncertainties},''
  \href{http://dx.doi.org/10.1103/PhysRevD.75.114010}{{\em Phys. Rev.}
  {\bfseries D75} (2007) 114010},
\href{http://arxiv.org/abs/hep-ph/0703242}{{\ttfamily arXiv:hep-ph/0703242
  [hep-ph]}}.

\bibitem{deFlorian:2007ekg}
D.~de~Florian, R.~Sassot, and M.~Stratmann, ``{Global analysis of fragmentation
  functions for protons and charged hadrons},''
  \href{http://dx.doi.org/10.1103/PhysRevD.76.074033}{{\em Phys. Rev.}
  {\bfseries D76} (2007) 074033},
\href{http://arxiv.org/abs/0707.1506}{{\ttfamily arXiv:0707.1506 [hep-ph]}}.

\bibitem{deFlorian:2014xna}
D.~de~Florian, R.~Sassot, M.~Epele, R.~J. Hernández-Pinto, and M.~Stratmann,
  ``{Parton-to-Pion Fragmentation Reloaded},''
  \href{http://dx.doi.org/10.1103/PhysRevD.91.014035}{{\em Phys. Rev.}
  {\bfseries D91} (2015) 014035},
\href{http://arxiv.org/abs/1410.6027}{{\ttfamily arXiv:1410.6027 [hep-ph]}}.

\bibitem{Acharya:2017hyu}
{\bfseries ALICE} Collaboration, S.~Acharya {\em et~al.}, ``{Production of
  ${\pi ^0}$ and $\eta $ mesons up to high transverse momentum in pp collisions
  at 2.76 TeV},'' \href{http://dx.doi.org/10.1140/epjc/s10052-017-4890-x}{{\em
  Eur. Phys. J.} {\bfseries C77} (2017) 339},
\href{http://arxiv.org/abs/1702.00917}{{\ttfamily arXiv:1702.00917 [hep-ex]}}.

\bibitem{dEnterria:2014tdl}
D.~d'Enterria, K.~J. Eskola, I.~Helenius, and H.~Paukkunen, ``{LHC data
  challenges the contemporary parton-to-hadron fragmentation functions},'' {\em
  PoS} {\bfseries DIS2014} (2014) 148,
\href{http://arxiv.org/abs/1408.4659}{{\ttfamily arXiv:1408.4659 [hep-ph]}}.

\bibitem{Aidala:2010bn}
C.~A. Aidala, F.~Ellinghaus, R.~Sassot, J.~P. Seele, and M.~Stratmann,
  ``{Global Analysis of Fragmentation Functions for Eta Mesons},''
  \href{http://dx.doi.org/10.1103/PhysRevD.83.034002}{{\em Phys. Rev.}
  {\bfseries D83} (2011) 034002},
\href{http://arxiv.org/abs/1009.6145}{{\ttfamily arXiv:1009.6145 [hep-ph]}}.

\bibitem{BOURQUIN1976334}
M.~Bourquin and J.-M. Gaillard, ``{A simple phenomenological description of
  hadron production},''
  \href{http://dx.doi.org/10.1016/0550-3213(76)90592-7}{{\em Nucl. Phys.}
  {\bfseries B114} (1976) 334 -- 364}.

\bibitem{Khandai:2011cf}
P.~K. Khandai, P.~Shukla, and V.~Singh, ``{Meson spectra and $m_T$ scaling in
  $p + p$, $d + $Au, and Au + Au collisions at $\sqrt{s_{NN}}=200$ GeV},''
  \href{http://dx.doi.org/10.1103/PhysRevC.84.054904}{{\em Phys. Rev.}
  {\bfseries C84} (2011) 054904},
\href{http://arxiv.org/abs/1110.3929}{{\ttfamily arXiv:1110.3929 [hep-ph]}}.

\bibitem{Altenkamper:2017qot}
L.~Altenkämper, F.~Bock, C.~Loizides, and N.~Schmidt, ``{Applicability of
  transverse mass scaling in hadronic collisions at the LHC},''
\href{http://arxiv.org/abs/1710.01933}{{\ttfamily arXiv:1710.01933 [hep-ph]}}.

\bibitem{Jiang:2013gxa}
K.~Jiang, Y.~Zhu, W.~Liu, H.~Chen, C.~Li, L.~Ruan, Z.~Tang, Z.~Xu, and Z.~Xu,
  ``{Onset of radial flow in p+p collisions},''
  \href{http://dx.doi.org/10.1103/PhysRevC.91.024910}{{\em Phys. Rev.}
  {\bfseries C91} (2015) 024910},
\href{http://arxiv.org/abs/1312.4230}{{\ttfamily arXiv:1312.4230 [nucl-ex]}}.

\bibitem{Martin:2009iq}
A.~D. Martin, W.~J. Stirling, R.~S. Thorne, and G.~Watt, ``{Parton
  distributions for the LHC},''
  \href{http://dx.doi.org/10.1140/epjc/s10052-009-1072-5}{{\em Eur. Phys. J.}
  {\bfseries C63} (2009) 189--285},
\href{http://arxiv.org/abs/0901.0002}{{\ttfamily arXiv:0901.0002 [hep-ph]}}.

\bibitem{1126-6708-2007-02-053}
W.~Tung, H.~Lai, A.~Belyaev, J.~Pumplin, D.~Stump, and C.-P. Yuan, ``{Heavy
  quark mass effects in deep inelastic scattering and global QCD analysis},''
  \href{http://dx.doi.org/10.1088/1126-6708/2007/02/053}{{\em JHEP} {\bfseries
  02} (2007) 053}, \href{http://arxiv.org/abs/hep-ph/0611254}{{\ttfamily
  arXiv:hep-ph/0611254 [hep-ph]}}.

\bibitem{Sjostrand:2014zea}
T.~Sjöstrand, S.~Ask, J.~R. Christiansen, R.~Corke, N.~Desai, P.~Ilten,
  S.~Mrenna, S.~Prestel, C.~O. Rasmussen, and P.~Z. Skands, ``{An Introduction
  to PYTHIA 8.2},'' \href{http://dx.doi.org/10.1016/j.cpc.2015.01.024}{{\em
  Comput. Phys. Commun.} {\bfseries 191} (2015) 159--177},
\href{http://arxiv.org/abs/1410.3012}{{\ttfamily arXiv:1410.3012 [hep-ph]}}.

\bibitem{Corke:2010yf}
R.~Corke and T.~Sjöstrand, ``{Interleaved Parton Showers and Tuning
  Prospects},'' \href{http://dx.doi.org/10.1007/JHEP03(2011)032}{{\em JHEP}
  {\bfseries 03} (2011) 032},
\href{http://arxiv.org/abs/1011.1759}{{\ttfamily arXiv:1011.1759 [hep-ph]}}.

\bibitem{Skands:2014pea}
P.~Skands, S.~Carrazza, and J.~Rojo, ``{Tuning PYTHIA 8.1: the Monash 2013
  Tune},'' \href{http://dx.doi.org/10.1140/epjc/s10052-014-3024-y}{{\em Eur.
  Phys. J.} {\bfseries C74} (2014) 3024},
\href{http://arxiv.org/abs/1404.5630}{{\ttfamily arXiv:1404.5630 [hep-ph]}}.

\bibitem{Aamodt:2008zz}
{\bfseries ALICE} Collaboration, K.~Aamodt {\em et~al.}, ``{The ALICE
  experiment at the CERN LHC},''
\href{http://dx.doi.org/10.1088/1748-0221/3/08/S08002}{{\em JINST} {\bfseries
  3} (2008) S08002}.

\bibitem{Abelev:2014ffa}
{\bfseries ALICE} Collaboration, B.~B. Abelev {\em et~al.}, ``{Performance of
  the ALICE Experiment at the CERN LHC},''
  \href{http://dx.doi.org/10.1142/S0217751X14300440}{{\em Int. J. Mod. Phys.}
  {\bfseries A29} (2014) 1430044},
\href{http://arxiv.org/abs/1402.4476}{{\ttfamily arXiv:1402.4476 [nucl-ex]}}.

\bibitem{Cortese:2008zza}
{\bfseries ALICE} Collaboration, P.~Cortese {\em et~al.}, ``{ALICE
  electromagnetic calorimeter technical design report},''
CERN-LHCC-2008-014, CERN-ALICE-TDR-014.

\bibitem{Dellacasa:1999kd}
{\bfseries ALICE} Collaboration, G.~Dellacasa {\em et~al.}, ``{ALICE technical
  design report of the photon spectrometer (PHOS)},''
CERN-LHCC-99-04.

\bibitem{Alme:2010ke}
J.~Alme {\em et~al.}, ``{The ALICE TPC, a large 3-dimensional tracking device
  with fast readout for ultra-high multiplicity events},''
  \href{http://dx.doi.org/10.1016/j.nima.2010.04.042}{{\em Nucl. Instrum.
  Meth.} {\bfseries A622} (2010) 316--367},
\href{http://arxiv.org/abs/1001.1950}{{\ttfamily arXiv:1001.1950
  [physics.ins-det]}}.

\bibitem{Aamodt:2010aa}
{\bfseries ALICE} Collaboration, K.~Aamodt {\em et~al.}, ``{Alignment of the
  ALICE Inner Tracking System with cosmic-ray tracks},''
  \href{http://dx.doi.org/10.1088/1748-0221/5/03/P03003}{{\em JINST} {\bfseries
  5} (2010) P03003},
\href{http://arxiv.org/abs/1001.0502}{{\ttfamily arXiv:1001.0502
  [physics.ins-det]}}.

\bibitem{Abeysekara:2010ze}
{\bfseries ALICE EMCal} Collaboration, U.~Abeysekara {\em et~al.}, ``{ALICE
  EMCal Physics Performance Report},''
\href{http://arxiv.org/abs/1008.0413}{{\ttfamily arXiv:1008.0413
  [physics.ins-det]}}.

\bibitem{Cortese:2004aa}
{\bfseries ALICE} Collaboration, P.~Cortese {\em et~al.}, ``{ALICE technical
  design report on forward detectors: FMD, T0 and V0},''
CERN-LHCC-2004-025.

\bibitem{Abelev:2012sea}
{\bfseries ALICE} Collaboration, B.~Abelev {\em et~al.}, ``{Measurement of
  inelastic, single- and double-diffraction cross sections in proton--proton
  collisions at the LHC with ALICE},''
  \href{http://dx.doi.org/10.1140/epjc/s10052-013-2456-0}{{\em Eur. Phys. J.}
  {\bfseries C73} (2013) 2456},
\href{http://arxiv.org/abs/1208.4968}{{\ttfamily arXiv:1208.4968 [hep-ex]}}.

\bibitem{Kral:2012ae}
J.~Kral, T.~Awes, H.~Muller, J.~Rak, and J.~Schambach, ``{L0 trigger for the
  EMCal detector of the ALICE experiment},''
\href{http://dx.doi.org/10.1016/j.nima.2012.06.065}{{\em Nucl. Instrum. Meth.}
  {\bfseries A693} (2012) 261--267}.

\bibitem{Wang:2011zzd}
D.~Wang {\em et~al.}, ``{Level-0 trigger algorithms for the ALICE PHOS
  detector},''
\href{http://dx.doi.org/10.1016/j.nima.2010.11.111}{{\em Nucl. Instrum. Meth.}
  {\bfseries A629} (2011) 80--86}.

\bibitem{Adam2017}
J.~Adam {\em et~al.}, ``{Determination of the event collision time with the
  ALICE detector at the LHC},''
  \href{http://dx.doi.org/10.1140/epjp/i2017-11279-1}{{\em Eur. Phys. J. Plus}
  {\bfseries 132} (2017) 99}, \href{http://arxiv.org/abs/1610.03055}{{\ttfamily
  arXiv:1610.03055 [physics.ins-det]}}.

\bibitem{1748-0221-5-12-C12048}
O.~Bourrion, R.~Guernane, B.~Boyer, J.~L. Bouly, and G.~Marcotte, ``{Level-1
  jet trigger hardware for the ALICE electromagnetic calorimeter at LHC},''
  \href{http://dx.doi.org/10.1088/1748-0221/5/12/C12048}{{\em JINST} {\bfseries
  5} (2010) C12048},
\href{http://arxiv.org/abs/1010.2670}{{\ttfamily arXiv:1010.2670
  [physics.ins-det]}}.

\bibitem{vanderMeer:1968zz}
S.~van~der Meer, ``{Calibration of the Effective Beam Height in the ISR}.''
  {CERN-ISR-PO-68-31}, 1968.

\bibitem{ALICE-PUBLIC-2017-002}
{\bfseries ALICE} Collaboration, ``{ALICE luminosity determination for pp
  collisions at $\sqrt{s}=8$ TeV},''. \url{https://cds.cern.ch/record/2255216}.

\bibitem{Olive:2016xmw}
{\bfseries Particle Data Group} Collaboration, C.~Patrignani {\em et~al.},
  ``{Review of Particle Physics},''
\href{http://dx.doi.org/10.1088/1674-1137/40/10/100001}{{\em Chin. Phys.}
  {\bfseries C40} (2016) 100001}.

\bibitem{Alessandro:2006yt}
{\bfseries ALICE} Collaboration, P.~Cortese {\em et~al.}, ``{ALICE: Physics
  performance report, volume II},''
\href{http://dx.doi.org/10.1088/0954-3899/32/10/001}{{\em J. Phys.} {\bfseries
  G32} (2006) 1295--2040}.

\bibitem{ALICE-PUBLIC-2017-009}
{\bfseries ALICE} Collaboration, ``{Supplemental figures: $\pi^{0}$ and $\eta$
  meson production in proton-proton collisions at $\sqrt{s}=8$ TeV},''.
  \url{http://cds.cern.ch/record/2282851}.

\bibitem{Awes1992130}
T.~Awes, F.~Obenshain, F.~Plasil, S.~Saini, S.~Sorensen, and G.~Young, ``{A
  simple method of shower localization and identification in laterally
  segmented calorimeters},''
  \href{http://dx.doi.org/10.1016/0168-9002(92)90858-2}{{\em Nucl. Instrum.
  Meth.} {\bfseries A311} (1992) 130 -- 138}.

\bibitem{Fruhwirth:1987fm}
R.~Fruhwirth, ``{Application of Kalman filtering to track and vertex
  fitting},''
\href{http://dx.doi.org/10.1016/0168-9002(87)90887-4}{{\em Nucl. Instrum.
  Meth.} {\bfseries A262} (1987) 444--450}.

\bibitem{podolanski1954iii}
J.~Podolanski and R.~Armenteros, ``{III. Analysis of V-events},''
  \href{http://dx.doi.org/10.1080/14786440108520416}{{\em Phil. Mag.}
  {\bfseries 45} (1954) 13--30}.

\bibitem{Matulewicz1990194}
T.~Matulewicz {\em et~al.}, ``{Response of BaF2 detectors to photons of 3-50
  MeV energy},'' \href{http://dx.doi.org/10.1016/0168-9002(90)90259-9}{{\em
  Nucl. Instrum. Meth.} {\bfseries A289} (1990) 194--204}.

\bibitem{CrystalBall:1980}
M.~J. Oreglia, {\em {"A Study of the Reactions $\psi^\prime\to\gamma \gamma
  \psi$"}}.
\newblock PhD thesis, SLAC, Stanford University, Stanford, California 94305,
  1980.
\newblock \url{http://www.slac.stanford.edu/pubs/slacreports/slac-r-236.html}.

\bibitem{Engel:1995sb}
R.~Engel, J.~Ranft, and S.~Roesler, ``{Hard diffraction in hadron hadron
  interactions and in photoproduction},''
  \href{http://dx.doi.org/10.1103/PhysRevD.52.1459}{{\em Phys. Rev.} {\bfseries
  D52} (1995) 1459--1468},
\href{http://arxiv.org/abs/hep-ph/9502319}{{\ttfamily arXiv:hep-ph/9502319
  [hep-ph]}}.

\bibitem{Brun:1987ma}
R.~Brun, F.~Bruyant, M.~Maire, A.~McPherson, and P.~Zanarini, ``{GEANT3},''
  {\em CERN-DD-EE-84-1} (1987) BASE001 1--3.

\bibitem{Aamodt:2011zza}
{\bfseries ALICE} Collaboration, K.~Aamodt {\em et~al.}, ``{Strange particle
  production in proton-proton collisions at $\sqrt{s}=0.9$ TeV with ALICE at
  the LHC},'' \href{http://dx.doi.org/10.1140/epjc/s10052-011-1594-5}{{\em Eur.
  Phys. J.} {\bfseries C71} (2011) 1594},
\href{http://arxiv.org/abs/1012.3257}{{\ttfamily arXiv:1012.3257 [hep-ex]}}.

\bibitem{Abelev:2014laa}
{\bfseries ALICE} Collaboration, B.~B. Abelev {\em et~al.}, ``{Production of
  charged pions, kaons and protons at large transverse momenta in pp and
  Pb–Pb collisions at $\sqrt{s_{\rm NN}}$ =2.76 TeV},''
  \href{http://dx.doi.org/10.1016/j.physletb.2014.07.011}{{\em Phys. Lett.}
  {\bfseries B736} (2014) 196--207},
\href{http://arxiv.org/abs/1401.1250}{{\ttfamily arXiv:1401.1250 [nucl-ex]}}.

\bibitem{Adam:2016emw}
{\bfseries ALICE} Collaboration, J.~Adam {\em et~al.}, ``{Enhanced production
  of multi-strange hadrons in high-multiplicity proton-proton collisions},''
  \href{http://dx.doi.org/10.1038/nphys4111}{{\em Nature Phys.} {\bfseries 13}
  (2017) 535--539},
\href{http://arxiv.org/abs/1606.07424}{{\ttfamily arXiv:1606.07424 [nucl-ex]}}.

\bibitem{Lyons1988rp}
L.~Lyons, D.~Gibaut, and P.~Clifford, ``{How to Combine Correlated Estimates of
  a Single Physical Quantity},''
\href{http://dx.doi.org/10.1016/0168-9002(88)90018-6}{{\em Nucl. Instrum.
  Meth.} {\bfseries A270} (1988) 110}.

\bibitem{Valassi2003mu}
A.~Valassi, ``{Combining correlated measurements of several different physical
  quantities},''
\href{http://dx.doi.org/10.1016/S0168-9002(03)00329-2}{{\em Nucl. Instrum.
  Meth.} {\bfseries A500} (2003) 391--405}.

\bibitem{Lyons1986em}
L.~Lyons, {\em {Statistics For Nuclear And Particle Physicists}}.
\newblock Cambridge, UK: Univ. Pr., 1986.

\bibitem{barlow1989statistics}
R.~J. Barlow, {\em {Statistics: a guide to the use of statistical methods in
  the physical sciences}}, vol.~29.
\newblock John Wiley \& Sons, 1989.

\bibitem{Valassi2013bga}
A.~Valassi and R.~Chierici, ``{Information and treatment of unknown
  correlations in the combination of measurements using the BLUE method},''
  \href{http://dx.doi.org/10.1140/epjc/s10052-014-2717-6}{{\em Eur. Phys. J.}
  {\bfseries C74} (2014) 2717},
\href{http://arxiv.org/abs/1307.4003}{{\ttfamily arXiv:1307.4003
  [physics.data-an]}}.

\bibitem{Lafferty:1995}
G.~Lafferty and T.~Wyatt, ``{Where to stick your data points: The treatment of
  measurements within wide bins},''
  \href{http://dx.doi.org/10.1016/0168-9002(94)01112-5}{{\em Nucl. Instrum.
  Meth.} {\bfseries A355} (1995) 541--547}.

\bibitem{Bylinkin:2015xya}
A.~Bylinkin, N.~S. Chernyavskaya, and A.~A. Rostovtsev, ``{Predictions on the
  transverse momentum spectra for charged particle production at LHC-energies
  from a two component model},''
  \href{http://dx.doi.org/10.1140/epjc/s10052-015-3392-y}{{\em Eur. Phys. J.}
  {\bfseries C75} (2015) 166},
\href{http://arxiv.org/abs/1501.05235}{{\ttfamily arXiv:1501.05235 [hep-ph]}}.

\bibitem{Tsallis1987eu}
C.~Tsallis, ``{Possible Generalization of Boltzmann-Gibbs Statistics},''
\href{http://dx.doi.org/10.1007/BF01016429}{{\em J. Statist. Phys.} {\bfseries
  52} (1988) 479--487}.

\bibitem{Adare:2015ozj}
{\bfseries PHENIX} Collaboration, A.~Adare {\em et~al.}, ``{Inclusive cross
  section and double-helicity asymmetry for $\pi^{0}$ production at midrapidity
  in $p$$+$$p$ collisions at $\sqrt{s}=510$ GeV},''
  \href{http://dx.doi.org/10.1103/PhysRevD.93.011501}{{\em Phys. Rev.}
  {\bfseries D93} (2016) 011501},
\href{http://arxiv.org/abs/1510.02317}{{\ttfamily arXiv:1510.02317 [hep-ex]}}.

\bibitem{PhysRevC.81.034911}
{\bfseries PHENIX} Collaboration, A.~Adare {\em et~al.}, ``{Detailed
  measurement of the ${e}^{+}{e}^{\ensuremath{-}}$ pair continuum in $p+p$ and
  $\mathrm{Au}+\mathrm{Au}$ collisions at $\sqrt{{s}_{\mathit{NN}}}=200$ GeV
  and implications for direct photon production},''
  \href{http://dx.doi.org/10.1103/PhysRevC.81.034911}{{\em Phys. Rev. C}
  {\bfseries 81} (2010) 034911},
  \href{http://arxiv.org/abs/0912.0244}{{\ttfamily arXiv:0912.0244 [nucl-ex]}}.

\bibitem{PhysRevLett.111.012001}
{\bfseries TOTEM} Collaboration, G.~Antchev {\em et~al.},
  ``{Luminosity-Independent Measurement of the Proton-Proton Total Cross
  Section at $\sqrt{s}=8\text{ }\text{ }\mathrm{TeV}$},''
  \href{http://dx.doi.org/10.1103/PhysRevLett.111.012001}{{\em Phys. Rev.
  Lett.} {\bfseries 111} (2013) 012001}.

\bibitem{Abelev:2013bla}
{\bfseries ALICE} Collaboration, B.~B. Abelev {\em et~al.}, ``{Multiplicity
  dependence of the average transverse momentum in pp, p-Pb, and Pb-Pb
  collisions at the LHC},''
  \href{http://dx.doi.org/10.1016/j.physletb.2013.10.054}{{\em Phys. Lett.}
  {\bfseries B727} (2013) 371--380},
\href{http://arxiv.org/abs/1307.1094}{{\ttfamily arXiv:1307.1094 [nucl-ex]}}.

\bibitem{Adam:2015gka}
{\bfseries ALICE} Collaboration, J.~Adam {\em et~al.}, ``{Charged-particle
  multiplicities in proton–proton collisions at $\sqrt{s} = 0.9$ to 8 TeV},''
  \href{http://dx.doi.org/10.1140/epjc/s10052-016-4571-1}{{\em Eur. Phys. J.}
  {\bfseries C77} (2017) 33},
\href{http://arxiv.org/abs/1509.07541}{{\ttfamily arXiv:1509.07541 [nucl-ex]}}.

\bibitem{Adam:2015qaa}
{\bfseries ALICE} Collaboration, J.~Adam {\em et~al.}, ``{Measurement of pion,
  kaon and proton production in proton–proton collisions at $\sqrt{s} = 7$
  TeV},'' \href{http://dx.doi.org/10.1140/epjc/s10052-015-3422-9}{{\em Eur.
  Phys. J.} {\bfseries C75} (2015) 226},
\href{http://arxiv.org/abs/1504.00024}{{\ttfamily arXiv:1504.00024 [nucl-ex]}}.

\bibitem{Adare:2010cy}
{\bfseries PHENIX} Collaboration, A.~Adare {\em et~al.}, ``{Cross section and
  double helicity asymmetry for eta mesons and their comparison to neutral pion
  production in p+p collisions at $\sqrt{s}=200$ GeV},''
  \href{http://dx.doi.org/10.1103/PhysRevD.83.032001}{{\em Phys. Rev.}
  {\bfseries D83} (2011) 032001},
\href{http://arxiv.org/abs/1009.6224}{{\ttfamily arXiv:1009.6224 [hep-ex]}}.

\bibitem{AguilarBenitez:1991yy}
{\bfseries LEBC-EHS} Collaboration, M.~Aguilar-Benitez {\em et~al.},
  ``{Inclusive particle production in 400 GeV/c pp-interactions},''
\href{http://dx.doi.org/10.1007/BF01551452}{{\em Z. Phys.} {\bfseries C50}
  (1991) 405--426}.

\end{thebibliography}\endgroup

\newpage
\appendix
\section{The ALICE Collaboration}
\label{app:collab}

\begingroup
\small
\begin{flushleft}
S.~Acharya\Irefn{org137}\And 
J.~Adam\Irefn{org96}\And 
D.~Adamov\'{a}\Irefn{org93}\And 
J.~Adolfsson\Irefn{org32}\And 
M.M.~Aggarwal\Irefn{org98}\And 
G.~Aglieri Rinella\Irefn{org33}\And 
M.~Agnello\Irefn{org29}\And 
N.~Agrawal\Irefn{org46}\And 
Z.~Ahammed\Irefn{org137}\And 
N.~Ahmad\Irefn{org15}
% ...
\And
\renewcommand\labelenumi{\textsuperscript{\theenumi}~}

\section*{Affiliation notes}
\renewcommand\theenumi{\roman{enumi}}
\begin{Authlist}
\item \Adef{org*}Deceased
\item \Adef{orgI}Dipartimento DET del Politecnico di Torino, Turin, Italy
\item \Adef{orgII}Georgia State University, Atlanta, Georgia, United States
\item \Adef{orgIII}M.V. Lomonosov Moscow State University, D.V. Skobeltsyn Institute of Nuclear, Physics, Moscow, Russia
\item \Adef{orgIV}Department of Applied Physics, Aligarh Muslim University, Aligarh, India
\item \Adef{orgV}Institute of Theoretical Physics, University of Wroclaw, Poland
\end{Authlist}

\section*{Collaboration Institutes}
\renewcommand\theenumi{\arabic{enumi}~}
\begin{Authlist}
\item \Idef{org1}A.I. Alikhanyan National Science Laboratory (Yerevan Physics Institute) Foundation, Yerevan, Armenia
\item \Idef{org2}Benem\'{e}rita Universidad Aut\'{o}noma de Puebla, Puebla, Mexico
\item \Idef{org3}Bogolyubov Institute for Theoretical Physics, Kiev, Ukraine
\item \Idef{org4}Bose Institute, Department of Physics  and Centre for Astroparticle Physics and Space Science (CAPSS), Kolkata, India
\item \Idef{org5}Budker Institute for Nuclear Physics, Novosibirsk, Russia
\item \Idef{org6}California Polytechnic State University, San Luis Obispo, California, United States
\item \Idef{org7}Central China Normal University, Wuhan, China
\item \Idef{org8}Centre de Calcul de l'IN2P3, Villeurbanne, Lyon, France
\item \Idef{org9}Centro de Aplicaciones Tecnol\'{o}gicas y Desarrollo Nuclear (CEADEN), Havana, Cuba
\item \Idef{org10}Centro de Investigaci\'{o}n y de Estudios Avanzados (CINVESTAV), Mexico City and M\'{e}rida, Mexico
\item \Idef{org11}Centro Fermi - Museo Storico della Fisica e Centro Studi e Ricerche ``Enrico Fermi', Rome, Italy
\item \Idef{org12}Chicago State University, Chicago, Illinois, United States
\item \Idef{org13}China Institute of Atomic Energy, Beijing, China
\item \Idef{org14}COMSATS Institute of Information Technology (CIIT), Islamabad, Pakistan
\item \Idef{org15}Department of Physics, Aligarh Muslim University, Aligarh, India
\item \Idef{org16}Department of Physics, Ohio State University, Columbus, Ohio, United States
\item \Idef{org17}Department of Physics, Pusan National University, Pusan, Republic of Korea
\item \Idef{org18}Department of Physics, Sejong University, Seoul, Republic of Korea
\item \Idef{org19}Department of Physics, University of Oslo, Oslo, Norway
\item \Idef{org20}Department of Physics and Technology, University of Bergen, Bergen, Norway
\item \Idef{org21}Dipartimento di Fisica dell'Universit\`{a} 'La Sapienza' and Sezione INFN, Rome, Italy
\item \Idef{org22}Dipartimento di Fisica dell'Universit\`{a} and Sezione INFN, Cagliari, Italy
\item \Idef{org23}Dipartimento di Fisica dell'Universit\`{a} and Sezione INFN, Trieste, Italy
\item \Idef{org24}Dipartimento di Fisica dell'Universit\`{a} and Sezione INFN, Turin, Italy
\item \Idef{org25}Dipartimento di Fisica e Astronomia dell'Universit\`{a} and Sezione INFN, Bologna, Italy
\item \Idef{org26}Dipartimento di Fisica e Astronomia dell'Universit\`{a} and Sezione INFN, Catania, Italy
\item \Idef{org27}Dipartimento di Fisica e Astronomia dell'Universit\`{a} and Sezione INFN, Padova, Italy
\item \Idef{org28}Dipartimento di Fisica `E.R.~Caianiello' dell'Universit\`{a} and Gruppo Collegato INFN, Salerno, Italy
\item \Idef{org29}Dipartimento DISAT del Politecnico and Sezione INFN, Turin, Italy
\item \Idef{org30}Dipartimento di Scienze e Innovazione Tecnologica dell'Universit\`{a} del Piemonte Orientale and INFN Sezione di Torino, Alessandria, Italy
\item \Idef{org31}Dipartimento Interateneo di Fisica `M.~Merlin' and Sezione INFN, Bari, Italy
\item \Idef{org32}Division of Experimental High Energy Physics, University of Lund, Lund, Sweden
\item \Idef{org33}European Organization for Nuclear Research (CERN), Geneva, Switzerland
\item \Idef{org34}Excellence Cluster Universe, Technische Universit\"{a}t M\"{u}nchen, Munich, Germany
\item \Idef{org35}Faculty of Engineering, Bergen University College, Bergen, Norway
\item \Idef{org36}Faculty of Mathematics, Physics and Informatics, Comenius University, Bratislava, Slovakia
\item \Idef{org37}Faculty of Nuclear Sciences and Physical Engineering, Czech Technical University in Prague, Prague, Czech Republic
\item \Idef{org38}Faculty of Science, P.J.~\v{S}af\'{a}rik University, Ko\v{s}ice, Slovakia
\item \Idef{org39}Faculty of Technology, Buskerud and Vestfold University College, Tonsberg, Norway
\item \Idef{org40}Frankfurt Institute for Advanced Studies, Johann Wolfgang Goethe-Universit\"{a}t Frankfurt, Frankfurt, Germany
\item \Idef{org41}Gangneung-Wonju National University, Gangneung, Republic of Korea
\item \Idef{org42}Gauhati University, Department of Physics, Guwahati, India
\item \Idef{org43}Helmholtz-Institut f\"{u}r Strahlen- und Kernphysik, Rheinische Friedrich-Wilhelms-Universit\"{a}t Bonn, Bonn, Germany
\item \Idef{org44}Helsinki Institute of Physics (HIP), Helsinki, Finland
\item \Idef{org45}Hiroshima University, Hiroshima, Japan
\item \Idef{org46}Indian Institute of Technology Bombay (IIT), Mumbai, India
\item \Idef{org47}Indian Institute of Technology Indore, Indore, India
\item \Idef{org48}Indonesian Institute of Sciences, Jakarta, Indonesia
\item \Idef{org49}INFN, Laboratori Nazionali di Frascati, Frascati, Italy
\item \Idef{org50}INFN, Laboratori Nazionali di Legnaro, Legnaro, Italy
\item \Idef{org51}INFN, Sezione di Bari, Bari, Italy
\item \Idef{org52}INFN, Sezione di Bologna, Bologna, Italy
\item \Idef{org53}INFN, Sezione di Cagliari, Cagliari, Italy
\item \Idef{org54}INFN, Sezione di Catania, Catania, Italy
\item \Idef{org55}INFN, Sezione di Padova, Padova, Italy
\item \Idef{org56}INFN, Sezione di Roma, Rome, Italy
\item \Idef{org57}INFN, Sezione di Torino, Turin, Italy
\item \Idef{org58}INFN, Sezione di Trieste, Trieste, Italy
\item \Idef{org59}Inha University, Incheon, Republic of Korea
\item \Idef{org60}Institut de Physique Nucl\'eaire d'Orsay (IPNO), Universit\'e Paris-Sud, CNRS-IN2P3, Orsay, France
\item \Idef{org61}Institute for Nuclear Research, Academy of Sciences, Moscow, Russia
\item \Idef{org62}Institute for Subatomic Physics of Utrecht University, Utrecht, Netherlands
\item \Idef{org63}Institute for Theoretical and Experimental Physics, Moscow, Russia
\item \Idef{org64}Institute of Experimental Physics, Slovak Academy of Sciences, Ko\v{s}ice, Slovakia
\item \Idef{org65}Institute of Physics, Academy of Sciences of the Czech Republic, Prague, Czech Republic
\item \Idef{org66}Institute of Physics, Bhubaneswar, India
\item \Idef{org67}Institute of Space Science (ISS), Bucharest, Romania
\item \Idef{org68}Institut f\"{u}r Informatik, Johann Wolfgang Goethe-Universit\"{a}t Frankfurt, Frankfurt, Germany
\item \Idef{org69}Institut f\"{u}r Kernphysik, Johann Wolfgang Goethe-Universit\"{a}t Frankfurt, Frankfurt, Germany
\item \Idef{org70}Institut f\"{u}r Kernphysik, Westf\"{a}lische Wilhelms-Universit\"{a}t M\"{u}nster, M\"{u}nster, Germany
\item \Idef{org71}Instituto de Ciencias Nucleares, Universidad Nacional Aut\'{o}noma de M\'{e}xico, Mexico City, Mexico
\item \Idef{org72}Instituto de F\'{i}sica, Universidade Federal do Rio Grande do Sul (UFRGS), Porto Alegre, Brazil
\item \Idef{org73}Instituto de F\'{\i}sica, Universidad Nacional Aut\'{o}noma de M\'{e}xico, Mexico City, Mexico
\item \Idef{org74}IRFU, CEA, Universit\'{e} Paris-Saclay, Saclay, France
\item \Idef{org75}iThemba LABS, National Research Foundation, Somerset West, South Africa
\item \Idef{org76}Joint Institute for Nuclear Research (JINR), Dubna, Russia
\item \Idef{org77}Konkuk University, Seoul, Republic of Korea
\item \Idef{org78}Korea Institute of Science and Technology Information, Daejeon, Republic of Korea
\item \Idef{org79}KTO Karatay University, Konya, Turkey
\item \Idef{org80}Laboratoire de Physique Subatomique et de Cosmologie, Universit\'{e} Grenoble-Alpes, CNRS-IN2P3, Grenoble, France
\item \Idef{org81}Lawrence Berkeley National Laboratory, Berkeley, California, United States
\item \Idef{org82}Moscow Engineering Physics Institute, Moscow, Russia
\item \Idef{org83}Nagasaki Institute of Applied Science, Nagasaki, Japan
\item \Idef{org84}National and Kapodistrian University of Athens, Physics Department, Athens, Greece
\item \Idef{org85}National Centre for Nuclear Studies, Warsaw, Poland
\item \Idef{org86}National Institute for Physics and Nuclear Engineering, Bucharest, Romania
\item \Idef{org87}National Institute of Science Education and Research, HBNI, Jatni, India
\item \Idef{org88}National Nuclear Research Center, Baku, Azerbaijan
\item \Idef{org89}National Research Centre Kurchatov Institute, Moscow, Russia
\item \Idef{org90}Niels Bohr Institute, University of Copenhagen, Copenhagen, Denmark
\item \Idef{org91}Nikhef, Nationaal instituut voor subatomaire fysica, Amsterdam, Netherlands
\item \Idef{org92}Nuclear Physics Group, STFC Daresbury Laboratory, Daresbury, United Kingdom
\item \Idef{org93}Nuclear Physics Institute, Academy of Sciences of the Czech Republic, \v{R}e\v{z} u Prahy, Czech Republic
\item \Idef{org94}Oak Ridge National Laboratory, Oak Ridge, Tennessee, United States
\item \Idef{org95}Petersburg Nuclear Physics Institute, Gatchina, Russia
\item \Idef{org96}Physics Department, Creighton University, Omaha, Nebraska, United States
\item \Idef{org97}Physics department, Faculty of science, University of Zagreb, Zagreb, Croatia
\item \Idef{org98}Physics Department, Panjab University, Chandigarh, India
\item \Idef{org99}Physics Department, University of Cape Town, Cape Town, South Africa
\item \Idef{org100}Physics Department, University of Jammu, Jammu, India
\item \Idef{org101}Physics Department, University of Rajasthan, Jaipur, India
\item \Idef{org102}Physikalisches Institut, Eberhard Karls Universit\"{a}t T\"{u}bingen, T\"{u}bingen, Germany
\item \Idef{org103}Physikalisches Institut, Ruprecht-Karls-Universit\"{a}t Heidelberg, Heidelberg, Germany
\item \Idef{org104}Physik Department, Technische Universit\"{a}t M\"{u}nchen, Munich, Germany
\item \Idef{org105}Purdue University, West Lafayette, Indiana, United States
\item \Idef{org106}Research Division and ExtreMe Matter Institute EMMI, GSI Helmholtzzentrum f\"ur Schwerionenforschung GmbH, Darmstadt, Germany
\item \Idef{org107}Rudjer Bo\v{s}kovi\'{c} Institute, Zagreb, Croatia
\item \Idef{org108}Russian Federal Nuclear Center (VNIIEF), Sarov, Russia
\item \Idef{org109}Saha Institute of Nuclear Physics, Kolkata, India
\item \Idef{org110}School of Physics and Astronomy, University of Birmingham, Birmingham, United Kingdom
\item \Idef{org111}Secci\'{o}n F\'{\i}sica, Departamento de Ciencias, Pontificia Universidad Cat\'{o}lica del Per\'{u}, Lima, Peru
\item \Idef{org112}SSC IHEP of NRC Kurchatov institute, Protvino, Russia
\item \Idef{org113}Stefan Meyer Institut f\"{u}r Subatomare Physik (SMI), Vienna, Austria
\item \Idef{org114}SUBATECH, IMT Atlantique, Universit\'{e} de Nantes, CNRS-IN2P3, Nantes, France
\item \Idef{org115}Suranaree University of Technology, Nakhon Ratchasima, Thailand
\item \Idef{org116}Technical University of Ko\v{s}ice, Ko\v{s}ice, Slovakia
\item \Idef{org117}Technical University of Split FESB, Split, Croatia
\item \Idef{org118}The Henryk Niewodniczanski Institute of Nuclear Physics, Polish Academy of Sciences, Cracow, Poland
\item \Idef{org119}The University of Texas at Austin, Physics Department, Austin, Texas, United States
\item \Idef{org120}Universidad Aut\'{o}noma de Sinaloa, Culiac\'{a}n, Mexico
\item \Idef{org121}Universidade de S\~{a}o Paulo (USP), S\~{a}o Paulo, Brazil
\item \Idef{org122}Universidade Estadual de Campinas (UNICAMP), Campinas, Brazil
\item \Idef{org123}Universidade Federal do ABC, Santo Andre, Brazil
\item \Idef{org124}University of Houston, Houston, Texas, United States
\item \Idef{org125}University of Jyv\"{a}skyl\"{a}, Jyv\"{a}skyl\"{a}, Finland
\item \Idef{org126}University of Liverpool, Liverpool, United Kingdom
\item \Idef{org127}University of Tennessee, Knoxville, Tennessee, United States
\item \Idef{org128}University of the Witwatersrand, Johannesburg, South Africa
\item \Idef{org129}University of Tokyo, Tokyo, Japan
\item \Idef{org130}University of Tsukuba, Tsukuba, Japan
\item \Idef{org131}Universit\'{e} Clermont Auvergne, CNRS/IN2P3, LPC, Clermont-Ferrand, France
\item \Idef{org132}Universit\'{e} de Lyon, Universit\'{e} Lyon 1, CNRS/IN2P3, IPN-Lyon, Villeurbanne, Lyon, France
\item \Idef{org133}Universit\'{e} de Strasbourg, CNRS, IPHC UMR 7178, F-67000 Strasbourg, France, Strasbourg, France
\item \Idef{org134}Universit\`{a} degli Studi di Pavia, Pavia, Italy
\item \Idef{org135}Universit\`{a} di Brescia, Brescia, Italy
\item \Idef{org136}V.~Fock Institute for Physics, St. Petersburg State University, St. Petersburg, Russia
\item \Idef{org137}Variable Energy Cyclotron Centre, Kolkata, India
\item \Idef{org138}Warsaw University of Technology, Warsaw, Poland
\item \Idef{org139}Wayne State University, Detroit, Michigan, United States
\item \Idef{org140}Wigner Research Centre for Physics, Hungarian Academy of Sciences, Budapest, Hungary
\item \Idef{org141}Yale University, New Haven, Connecticut, United States
\item \Idef{org142}Yonsei University, Seoul, Republic of Korea
\item \Idef{org143}Zentrum f\"{u}r Technologietransfer und Telekommunikation (ZTT), Fachhochschule Worms, Worms, Germany
\end{Authlist}
\endgroup
\end{document}